\documentclass[]{aa}
\usepackage{graphicx}
\usepackage[varg]{txfonts}
\usepackage{breqn}
\usepackage{float}
\usepackage{longtable,lscape}
\usepackage{amsmath}
\usepackage{afterpage}
\usepackage{amssymb}
\usepackage[colorlinks=true,citecolor=blue]{hyperref}
\usepackage{multirow}
\usepackage{upgreek}

\begin{document} 

\title{Perturbers: SPHERE detection limits to planetary-mass companions in protoplanetary disks}



\author{R. Asensio-Torres \inst{\ref{MPIA}}, 
        Th. Henning \inst{\ref{MPIA}},
        F. Cantalloube \inst{\ref{MPIA}},
       P. Pinilla \inst{\ref{MPIA}},
       D. Mesa\inst{\ref{padova}},
       A. Garufi\inst{\ref{firenze}},
       S. Jorquera \inst{\ref{unichile}},
        R. Gratton \inst{\ref{padova}},
        G. Chauvin\inst{\ref{grenoble}, \ref{unichile}},
        J. Szul\'agyi\inst{\ref{zurich}},
        R. van Boekel\inst{\ref{MPIA}},
        R. Dong\inst{\ref{victoria}},
        G.-D.~Marleau\inst{{\ref{tubingen},\ref{bern},\ref{MPIA}}},
        M. Benisty\inst{\ref{unichile2},\ref{grenoble}},
        M. Villenave \inst{\ref{grenoble}},
      C. Bergez-Casalou\inst{\ref{MPIA}},
        C. Desgrange\inst{\ref{MPIA}, \ref{lyon2}},
        M. Janson\inst{\ref{stockholm}},
        M. Keppler\inst{\ref{MPIA}},
       M. Langlois\inst{\ref{lyon}},
        F. Ménard \inst{\ref{grenoble}},
      E. Rickman \inst{\ref{geneve}, \ref{ESA}},
         T. Stolker \inst{\ref{leiden}},
         M. Feldt\inst{\ref{MPIA}},
         T. Fusco\inst{\ref{marseille}, \ref{onera}},
         L. Gluck\inst{\ref{grenoble}}
         A. Pavlov\inst{\ref{MPIA}},
         J. Ramos\inst{\ref{MPIA}}
}

\authorrunning{R. Asensio-Torres et al.}
\titlerunning{Perturbers}

\institute{Max-Planck-Institut f\"{u}r Astronomie (MPIA), K\"{o}nigstuhl 17, D-69117 Heidelberg, Germany \email{asensio-torres@mpia.de}\label{MPIA}
    \and INAF-Osservatorio Astronomico di Padova, Vicolo dell'Osservatorio 5, 35122 Padova, Italy \label{padova}
    \and INAF, Osservatorio Astrofisico di Arcetri, Largo E. Fermi 5, 50125 Firenze, Italy \label{firenze}
    \and Unidad Mixta Internacional Franco-Chilena de Astronomía, CNRS/INSU UMI 3386 and Departamento de Astronomía, Universidad de Chile, Casilla 36-D, Santiago, Chile \label{unichile}
    \and Université Grenoble Alpes, CNRS, IPAG, 38000, Grenoble, France \label{grenoble}
    \and ETH Z\"urich, Institute for Particle Physics and Astrophysics, Wolfgang-Pauli-Strasse 27, CH-8093, Z\"urich, Switzerland \label{zurich}
    \and Department of Physics $\&$ Astronomy, University of Victoria, Victoria, BC, V8P 1A1, Canada \label{victoria}
    \and Institut f\"ur Astronomie und Astrophysik, Universit\"at T\"ubingen, Auf der Morgenstelle 10, D-72076 T\"ubingen, Germany \label{tubingen}
    \and Physikalisches Institut, Universit\"{a}t Bern, Gesellschaftsstr.~6, CH-3012 Bern, Switzerland \label{bern}
    \and Unidad Mixta Internacional Franco-Chilena de Astronomía (CNRS, UMI 3386), Departamento de Astronomía, Universidad de Chile, Camino El Observatorio 1515, Las Condes, Santiago, Chile \label{unichile2}
    \and Department of Physics, Ecole Normale Supérieure de Lyon, 69364, Lyon, France \label{lyon2}
    \and Department of Astronomy, Stockholm University, AlbaNova University Center, SE-106 91 Stockholm, Sweden \label{stockholm}
    \and Université de Lyon, CRAL/CNRS, CRAL, UMR 5574, CNRS, saint genis laval, France \label{lyon}
    \and Observatoire de Genève, Chemin Pegasi 51, 1290 Versoix, Switzerland \label{geneve}
    \and European Space Agency (ESA), ESA Office, Space Telescope Science Institute, 3700 San Martin Dr, Baltimore, MD 21218, USA\label{ESA}
    \and Leiden Observatory, Leiden University, Niels Bohrweg 2, 2333 CA Leiden, The Netherlands \label{leiden}
    \and Aix Marseille Universit\'e, CNRS, CNES,  LAM, Marseille, France\label{marseille}
    \and DOTA, ONERA, Universit\'{e} Paris Saclay, F-91123, Palaiseau France \label{onera}}


 \abstract{The detection of a wide range of substructures such as rings, cavities and spirals has become a common outcome of high spatial resolution imaging of protoplanetary disks, both in the near-infrared scattered light and in the thermal millimetre continuum emission. The most frequent interpretation of their origin is the presence of planetary-mass companions perturbing the gas and dust distribution in the disk (perturbers), but so far the only bona-fide detection has been the two giant planets carving the disk around PDS 70. Here, we collect a sample of 15 protoplanetary disks showing substructures in SPHERE scattered light images and present a homogeneous derivation of planet detection limits in these systems. To obtain mass limits we rely on different post-formation luminosity models based on distinct formation conditions, which are critical in the first Myrs of evolution. We also estimate the mass of these perturbers through a Hill radius prescription and a comparison to ALMA data. Assuming that one single planet carves each substructure in scattered light, we find that more massive perturbers are needed to create gaps within cavities than rings, and that we might be close to a detection in the cavities of RX J1604.3-2130A, RX J1615.3-3255, Sz Cha, HD 135344B and HD 34282. We reach typical mass limits in these cavities of 3--10\,$M\rm_{Jup}$. For planets in the gaps between rings, we find that the detection limits of SPHERE high-contrast imaging are about an order of magnitude away in mass, and that the gaps of PDS 66 and HD 97048 seem to be the most promising structures for planet searches. The proposed presence of massive planets causing spiral features in HD 135344B and HD 36112 are also within SPHERE's reach assuming hot-start models.These results suggest that current detection limits are able to detect hot-start planets in cavities, under the assumption that they are formed by a single perturber located at the centre of the cavity. More realistic planet mass constraints would help to clarify whether this is actually the case, which might point to perturbers not being the only way of creating substructures.}


\maketitle 

\let\oldpageref\pageref
\renewcommand{\pageref}{\oldpageref*}

%

\section{Introduction}

Protoplanetary disks (PPDs) are the by-product of the star formation process, and the place where giant planets form before all the gas is accreted onto the star or dispersed over a period of $\sim$\,3-10\,Myr \citep{fedele2010}. In the last few years, high-resolution observations have opened a new era in our understanding of the gas and dust around young stars. These observations show a plethora of complex substructures in PPDs that are remarkably common when imaged with sufficient angular resolution, including gaps, cavities, rings, vortices, asymmetries, and spiral arms \citep[e.g.,][]{vanboekel2017,avenhaus2018, andrews2018, long2019, garufi2020}.

The origin of these morphologies remains unclear, and different mechanisms have been proposed to explain them \citep[e.g., ][]{flock2015, okuzumi2016, takahashi2016, cieza2016, gonzalez2017, dullemond2018, riols2020}. A common interpretation is to describe them as signposts of planetary-mass companions interacting with the disk, which requires a formation of the giant planets and their location within the gaps in less than a few Myrs. In this scenario, the massive planet creates a pressure bump in the gas that stops the radial drift of the dust, which gets trapped in the pressure maxima creating a ring-like structure. In general, mm-sized dust is affected by the drag with the gas and it is easily trapped in pressure bumps \citep{drazkowska2016, taki2016}. Small $\upmu$m-sized dust is however coupled to the gas, following its distribution and possibly populating the gap. As a result, spatial segregation is expected in the distribution of small and large dust particles  \citep{rice2006,  dejuanovelar2013, pinilla2015,hendler2018}. The exact morphology and structure of these regions will eventually depend on planet mass and PPD properties, such as viscosity and temperature \citep[e.g.,][]{whipple1972, pinilla2012}.

To understand how these structures are formed and to shed light on the planet formation mechanism, SPHERE high-resolution scattered light observations probe the surface layers of the optically thick dust using the polarized differential imaging technique \citep[PDI;][]{deboer2020a, vanholstein2020}. This mode detects the polarized light scattered off $\upmu$m-sized grains, and is thus an effective way of removing the stellar halo without altering the underlying disk morphology with post-processing techniques. To this date, there are $\sim$\,90 disks imaged in this mode, from individual large disks around intermediate mass stars \citep[e.g.,][]{benisty2015,ginski2016, stolker2016, pohl2017} to surveys designed to alleviate observational biases \citep[DARTTS;][]{avenhaus2018}. These results show a ubiquity of substructures in scattered light in IR bright disks, unless the disks are small or too faint.

The indirect detection of protoplanets creating these observed PPD substructures is particularly hindered by the presence of circumstellar material, while stellar activity also limits the performances of radial velocity surveys. As of today, direct imaging is the best technique to detect and characterize at the same time the planet and the disk morphology, although the companion luminosity might also be affected by extinction due to the surrounding disk material \citep{szulagyi2018, szulagyi2019,szulagyigarufi, sanchis2020}. So far, besides the exception of the two planets around PDS 70 \citep{keppler2018, mueller2018, haffert2019, mesa2019}, only not-confirmed planetary-mass candidates have been found with direct imaging in PPDs. A few kinematic detections of embedded protoplanets have also been proposed \citep{pinte2018, teague2018, pinte2019}, but there is not yet a final confirmation of these objects either.  

Detection limits to low-mass companions via total intensity high-contrast imaging data are available for only a handful of targets that show hints of substructures. These data, however, are analysed with different algorithms that remove the stellar halo, using different mass-luminosity relationships, and sometimes outdated and heterogeneous stellar parameters. 

In this paper we present a homogeneous study to determine planet detection limits within the substructures of a sample of PPDs observed in $\upmu$m-sized dust with SPHERE. Our goal is to perform a systematic analysis to planet sensitivities in these systems, and to understand how far off we are from detecting perturbers with direct imaging. In Section \ref{sec:data} we motivate the PPD sample and present the high contrast imaging reduction of the data. In Section \ref{sec:sensitivities} we convert the obtained detection limits to mass sensitivities, and assess the effect of different formation luminosities on the detectability of low-mass companions. In Section \ref{sec:perturbers} we estimate the mass and location of the potential companions carving these substructures, and compare them to the SPHERE detection limits. Finally, in Section \ref{sec:discuss} we discuss the results and caveats of this paper, and in Section \ref{sec:conclusion} we present the conclusions.

\begin{figure*}
\centering
\setlength{\unitlength}{\textwidth}
\begin{picture}(1,0.55)
  \put(0.,0.00){\includegraphics[width=0.48\textwidth]{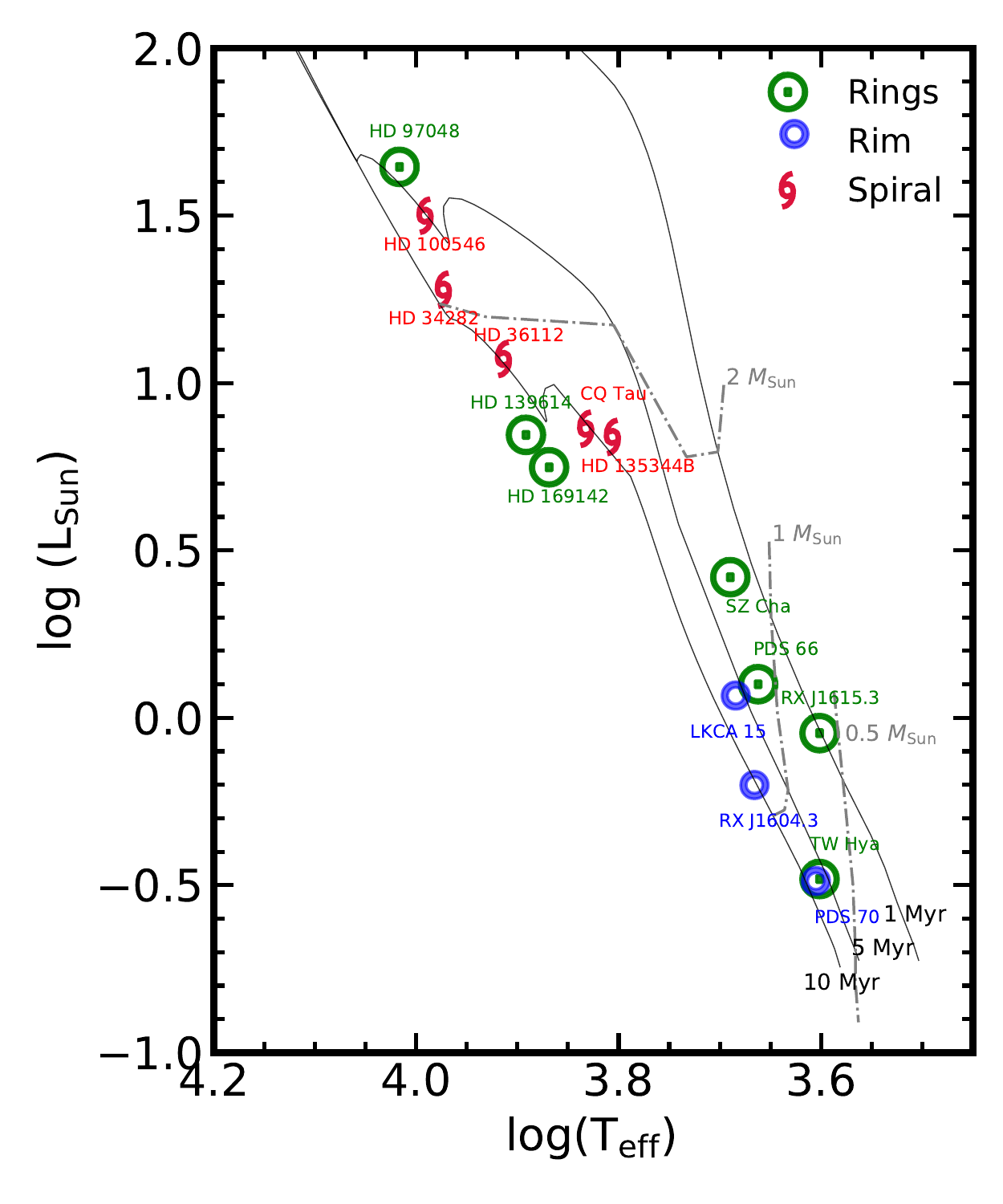}}
 
  \put(0.5,-0.005){\includegraphics[width=0.48\textwidth]{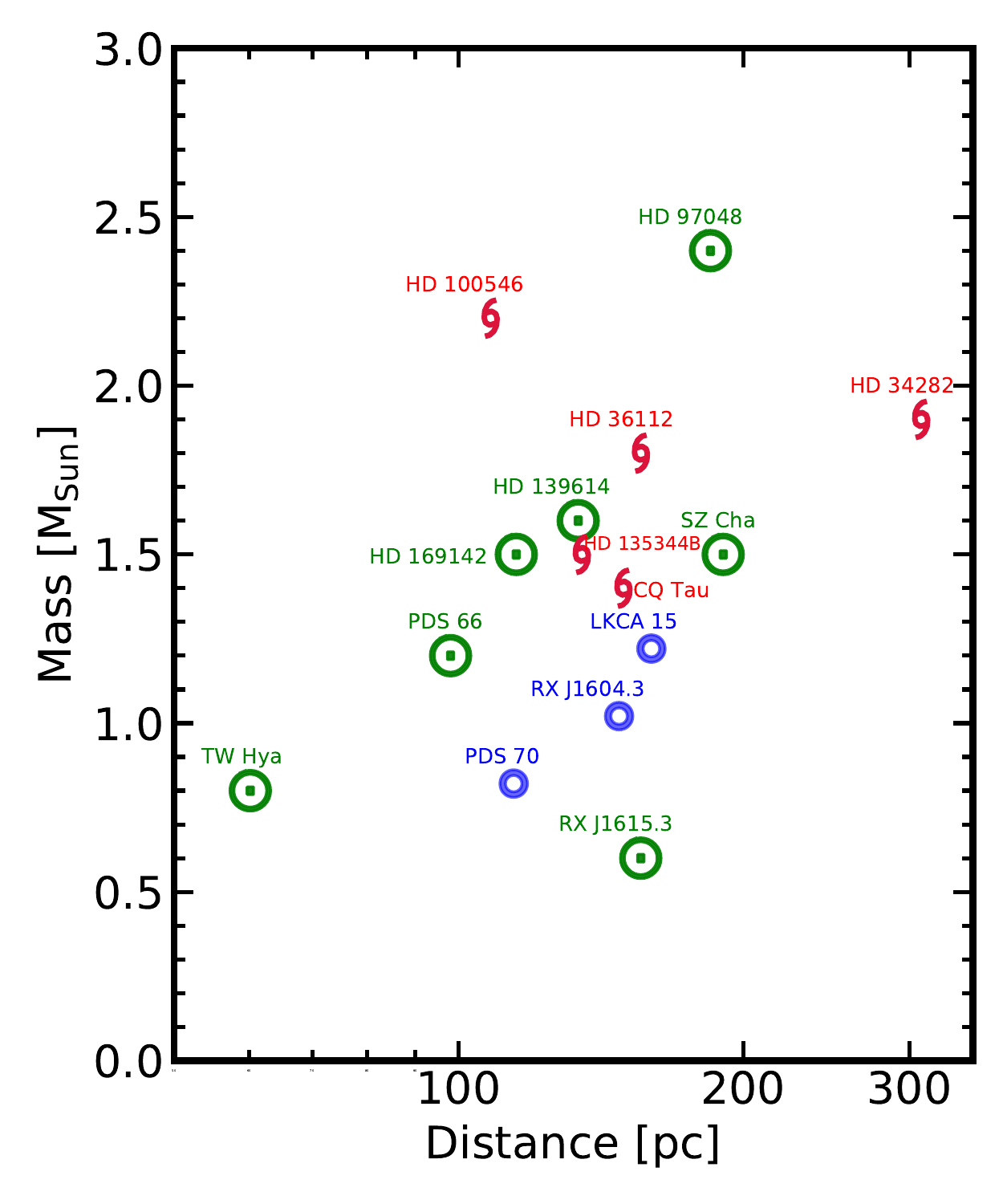}}
  
\end{picture}
\caption{(Left) Hertzsprung-Russell diagram of the PPD hosts in our sample. They are classified into \textit{Rings}, \textit{Rim} or \textit{Spiral}, based on their main detected substructures, and the MIST tracks for young PMS stars are overplotted \citep{choi2016}. The resulting stellar parameters are shown in Table \ref{tab:host_param} and have been derived as explained in Section \ref{sec:stellar_param} (Right) Mass-distance diagram of the same population.  }
\label{fig:HR}
\end{figure*}

\renewcommand{\arraystretch}{1.5}
\begin{table*}
\caption{\label{tab:host_param}Stellar parameters of the targets}
\small
\centering
\begin{tabular}{lccccccccccc}
\hline\hline
        Target
        & RA (J2000)
        & DEC (J2000)
        & SpT
        & V
        & d
        & Teff
        & L$_{*}$
       &  $A\rm_{V}$
        & M$_{*}$
       & Age

       \\
       
       & (h:m:s)
       & (deg:m:s)
       & 
       &(mag)
       & (pc)
       & (K)
       & (L$\rm_{Sun}$)
       & (mag)
    & ($M\rm_{Sun}$)
       & (Myr)
      
       \\
       
\hline
LkCa 15 & 04:39:17.79 & +22:21:03.38  & K5 & 12.0 & 157.2\,$\pm$\,0.6 & 4800 & 1.11\,$\pm$\,0.04 & 0.6 &   1.2\,$\pm$\,0.1 & 6.3$^{+0.7}_{-1.9}$  \\
HD 34282  & 05:16:00.47 &  -09:48:35.39 & B9.5 & 9.8 & 318.6\,$\pm$\,2.2 & 9400 & 19.0\,$\pm$\,1.8 &  0.6 &  1.9\,$\pm$\,0.1 & 8.9$^{+0.7}_{-1.9}$    \\
HD 36112 & 05:30:27.52  & +25:19:57.08 & A8V & 8.3 & 155.9\,$\pm$\,0.8 & 8200 & 11.8\,$\pm$\,0.5 & 0.2 & 1.8\,$\pm$\,0.1 & 8.9$^{+2.1}_{-1.0}$   \\
CQ Tau  & 05:35:58.46 & +24:44:54.08  & F5IV & 10.0 & 149.4\,$\pm$\,1.3 & 6800 & 7.3\,$\pm$\,0.6 & 0.4 &  1.4\,$\pm$\,0.1 & 9.9$^{+2.1}_{-2.9}$   \\
Sz Cha  & 10:58:16.74 & -77:17:17.18 &  K0 & 12.7 & 190.2\,$\pm$\,0.9 &  4900 & 2.6\,$\pm$\,0.2  &  1.5 & 1.5\,$\pm$\,0.2 & 1.9$^{+0.9}_{-0.7}$ &  \\
TW Hya  & 11:01:51.90 & -34:42:17.03 & K6V & 10.5 & 60.14\,$\pm$\,0.05 & 4000 & 0.33\,$\pm$\,0.02 &  0.0 & 0.8\,$\pm$\,0.1 & 6.3$^{+3.7}_{-1.9}$    \\
HD 97048 & 11:08:03.31 & -77:39:17.49 & A0V & 9.0 & 184.4\,$\pm$\,0.8 & 10400 & 44.2\,$\pm$\,5.7 &  1.1 &  2.4\,$\pm$\,0.2 & 3.9$^{+1.7}_{-0.4}$   \\
HD 100546 & 11:33:25.44 & -70:11:41.23 & A0V & 6.3 & 108.1\,$\pm$\,0.4 & 9800 & 31.6\,$\pm$\,0.7 &  0.1 & 2.2\,$\pm$\,0.2 & 5.0$^{+1.3}_{-0.6}$   \\
PDS 66 & 13:22:07.54 &  -69:38:12.21 & K1V & 10.4 & 97.9\,$\pm$\,0.1 & 4600 & 1.26\,$\pm$\,0.04 & 0.8 &  1.2\,$\pm$\,0.2 &  3.1$^{+0.8}_{-0.9}$   \\
PDS 70 & 14:08:10.15 & -41:23:52.57 & K7IV & 12.2 & 112.4\,$\pm$\,0.2 &  4000 & 0.31\,$\pm$\,0.02 & 0.0 & 0.8\,$\pm$\,0.1 &  7.9$^{+3.1}_{-2.9}$   \\
HD 135344B & 15:15:48.44 & -37:09:16.02 & F8V & 8.7 & 135.0\,$\pm$\,0.4 & 6400 & 6.9\,$\pm$\,0.4 &  0.4 & 1.5\,$\pm$\,0.1 & 8.9$^{+2.1}_{-1.1}$   \\
HD 139614 & 15:40:46.38 & -42:29:53.53  &  A9V & 8.2 & 133.6\,$\pm$\,0.5 & 7800 & 7.0\,$\pm$\,1.6 & 0.0 &  1.6\,$\pm$\,0.1 & 12.2$^{+4.1}_{-1.1}$   \\
RX J1604.3-2130A & 16:04:21.65 & -21:30:28.54 & K2 & 12.3 & 145.3\,$\pm$\,0.6 & 4600 & 0.60\,$\pm$\,0.03 &  1.1 & 1.0\,$\pm$\,0.1 & 11.1$^{+3.3}_{-3.1}$   \\
RX J1615.3-3255 & 16:15:20.23 & -32:55:05.09  & K5  & 12.0 & 155.6\,$\pm$\,0.6 & 4000  & 0.90\,$\pm$\,0.02 & 0.6 & 0.6\,$\pm$\,0.1 &  1.0$^{+0.4}_{-0.2}$    \\
HD 169142 & 18:24:29.77 &  -29:46:49.32 & F1V & 8.2 & 114.0\,$\pm$\,0.8 & 7400 & 5.6\,$\pm$\,1.2 &  0.0 & 1.5\,$\pm$\,0.2 &  12.3$^{+6.4}_{-1.2}$    \\

\hline
\end{tabular}
\tablefoot{Stellar parameters of the PPD hosts considered in this work. References for the stellar temperature and optical extinction can be found in Table F.1 of \citet{garufi2018}. Mass, luminosity and age have been derived here using \textit{Gaia} EDR3 parallaxes \citep{gaia2020}.\\
}
\end{table*}

\section{Sample selection and data reduction}
\label{sec:data}

We collected PPD systems observed with SPHERE that comply with the following criteria:

\begin{enumerate}
    
    \item The PPD shows rings, cavities or spirals in SPHERE/PDI observations.
    
    \item The host star is single or has no stellar companions closer than 3\,$\arcsec$.
    
    \item The PPD has an inclination below $\sim$\,60\,deg.
    
     \item The PPD has been observed in coronagraphic angular differential imaging mode \citep[ADI,][]{marois2006} with SPHERE, suitable for the detection of low-mass companions.
    
\end{enumerate}

The resulting list of targets includes disks that have substructures observed in the small dust component, and allows us to obtain sensitivities to planets that are not critically affected by projection effects and extinction towards the line of sight \citep[e.g.,][]{szulagyigarufi}. Moreover, our selection removes close binaries to optimize that the physical and morphological properties of the PPDs are not influenced by the presence of a stellar companion; this allows us to attribute the substructures to intrinsic physical processes related to the disk or host star \citep[e.g.,][]{durisen2007}, or otherwise to the presence of planetary-mass objects. 

The target selection was mostly based on the compilation by \citet{garufi2018} and the latest results of the DARTTS-S survey \citep{avenhaus2018, garufi2020}, followed by a cross-match with the ESO archive of ADI data. The final sample is comprised of 15 PPDs, whose substructures are characterized based on their visual appearance in the PDI images. \citet{garufi2018} classified them in \textit{Spirals}, \textit{Ring}, \textit{Rim}, \textit{Giant}, \textit{Faint}, \textit{Small} and \textit{Inclined}. This classification is rather subjective and can be affected by projection effects and the observing conditions of the different datasets; this implies that some PPDs will have some overlap, i.e., rims and/or rings (e.g., PDS 66), or giants with spirals (e.g. HD 100546). In this section we further simplify the classification and assign one class to each PPD depending on which feature is most prominent; \textit{Spiral} when spiral arms in the $\upmu$m-sized dust are seen, \textit{Rings} when the PPD shows signs of a series of resolved rings with gaps in between those, and \textit{Rim} when there is a detection of a large central cavity with a clear bright rim (typically known as transition disks). Some systems reveal different substructures when observed with ALMA in the gas or mm-dust \citep[e.g.,][]{teague2019}, but we do not consider those images for the classification. HD 100546 and HD 34282 are complicated systems with the presence of arcs, spirals and cavities. Here we put them in the \textit{Spiral} group, but the presence of rims or rings in these systems will also be taken into account (see Appendix \ref{appendix_A}). For instance, HD 34282 hosts an inner cavity with a rim-like structure at $\sim$\,88\,au, and a potential single-armed spiral feature farther out better observed after deprojection \citep{deboer2020b}. These individual substructures within a given PPD (not the global classification made in this section) will be taken into account separately to derive potential planet masses and locations in Section \ref{sec:perturbers}.

\subsection{Stellar parameters}\label{sec:stellar_param}

In Table \ref{tab:host_param} we show the derived parameters for the stellar hosts present in our sample. We follow the method outlined in \citet{garufi2018}, but in this case we obtain updated values using the new {\it Gaia} Early Data Release 3 parallaxes \citep[{\it Gaia} ED3,][]{gaia2016, gaia2020}. We build the stellar SEDs using {\it VizieR} and fit the wavelength range 0.4\,$\upmu$m\,--\,1.3\,mm to the PHOENIX photospheric models \citep{hauschildt1999}, which allows us to calculate the stellar luminosity using literature extinction $A\rm_{V}$ and effective temperature $T\rm_{eff}$. On the left panel of Figure \ref{fig:HR} we show a Hertzsprung-Russell diagram of the sample, where we estimate each individual mass and age using the MIST pre-main-sequence (PMS) tracks \citep{choi2016}. These isochrones are consistent with other PMS tracks such as Parsec \citep{bressan2012}, Baraffe \citep{baraffe2015} and Dartmouth \citep{Dotter2008}. To calculate uncertainties on the age and mass, the error bars on the luminosity (in turn derived from a 20\,$\%$ uncertainty on the optical extinction) and on $T\rm_{eff}$ ($\pm$\,200\,K) is propagated through the tracks. The right panel shows the location of the targets in the stellar mass-distance space. Distances to the targets range between 60--300\,pc, but most of them lie around 150\,pc, as members of star-forming regions such as Sco-Cen and Taurus. Spirals tend to appear in disks around massive stars, while objects classified as \textit{Rim}, with prominent cavities, are only resolved in the scattered light in $<$\,1.5\,$M\rm_{Sun}$ stellar systems (although some of the most massive objects classified as \textit{Spiral} also count with resolved cavities, see Appendix \ref{appendix_A}).

We note here that the derived isochronal parameters for young PMS stars carry a moderate uncertainty. Stellar ages is the most critical parameter to estimate reliable sensitivities to planet mass; deviations in $L$/$T\rm_{eff}$ values, the use of different evolutionary tracks, the effect of magnetic fields and the initial position of the star at $t$\,$=$\,0 can all contribute to dubious estimations of individual ages \citep[e.g.,][]{ruben2019}. Our approach allows us to obtain a homogeneous and consistent classification of the different PPDs, while the adopted ages (Table \ref{tab:host_param}) also tend to agree within error bars with other individual studies. For instance, PDS 70 is found to have an age of 5.4\,Myr \citep{mueller2018}, and TW Hya is most consistent with 8\,Myr \citep{sokal2018}. HD 139614 has an age of 10.75\,$\pm$\,0.77\,Myr, recently derived via astereoseismology \citep{murphy2020}, in close agreement with our values. For HD 169142, an estimation of 6$^{+6}_{-3}$\,Myr was found for the M-type wide binary companion 2M1824 \citep{grady2007}, for which only the upper uncertainty value is consistent with our estimation, although isochronal ages for low-mass stars seem to be underestimated by a factor of $\sim$\,2 \citep{ruben2019}.

\begin{table*}
\caption{\label{tab:adi}Perturbers' ADI epochs}
\small
\centering
\begin{tabular}{lccccccccc}
\hline\hline
        Target
        & Date  
        & Program ID
        & $t\rm_{int}$
        & Sky rot
        & Inc
        & PA 
       &  Mode
       & Global substructure

       \\
       
       &(UT)
       &
       & (min)
       & (deg)
       & (deg)
       & (deg)
       & 
       & 
       \\
\hline
LkCa 15 & 2015-11-29 & 096.C-0241(B) & 84.3 & 26.4 & 50 & 60 & IRDIFS_EXT & Rim \\
HD 34282  & 2015-10-26 & 096.C-0241(A) & 67.2 & 53.5 & 56 & 119 & IRDIFS & Spiral \\
HD 36112 & 2018-12-17 & 1100.C-0481(K) & 102.4 & 29.1 & 21 & 62 & IRDIFS & Spiral \\
CQ Tau  & 2016-12-21 & 298.C-5014(A) & 126.7 & 44.4 & 35 & 55 & IRDIFS_EXT & Spiral \\
Sz Cha & 2017-02-10 & 198.C-0209(E) & 102.4 & 34.3 & 48.9 & 158.3 & IRDIFS & Ring \\
TW Hya  & 2015-02-04 & 095.C-0298(H)& 67.2 & 76.7 & 7 & 320 & IRDIFS & Rim \\
HD 97048 & 2016-03-28 & 096.C-0241(E) & 84.3 & 24.5 & 39.9 & 2.8 & IRDIFS & Ring \\
HD 100546 & 2015-05-29 & 095.C-0298(B) & 102.1 & 34.3 & 42.5 & 139.1 & IRDIFS &Spiral \\
PDS 66 & 2017-02-11 &  198.C-0209(E) & 116.3 & 38.5 & 30.1 & 189.2 &  IRDIFS & Ring  \\
PDS 70 & 2018-02-25 & 1100.C-0481(D) & 142.4 &  95.7 & 49.7 & 158.6 & IRDIFS_EXT & Rim \\
HD 135344B & 2015-05-15 & 095.C-0298(A) & 67.2 &  63.6  & 11 & 62 & IRDIFS_EXT &  Spiral \\
HD 139614 & 2017-06-12 & 198.C-0209(H) & 100.8 & 69.6 & 17.6 & 276.5 & IRDIFS  & Ring \\
RX J1604.3-2130A & 2015-06-10 & 095.C-0673(A) & 93.3 & 142.8 & 6 & 80 & IRDIFS_EXT & Rim \\
RX J1615.3-3255 & 2015-05-15 & 095.C-0298(A) & 67.2 & 74.3 & 47 & 146 & IRDIFS & Ring \\
HD 169142 & 2019-05-18 & 1100.C-0481(N) & 76.8 & 118.4 & 13 & 5 & IRDIFS_EXT & Ring \\

\hline
\end{tabular}
\tablefoot{ADI observations of the target list in Table \ref{tab:host_param}. The IRDIS side of the observations were obtained in dual-band mode (H2-H3 and K1-K2 for IRDIFS and IRDIFS-EXT, respectively), except for CQ Tau, for which the full $K\rm_{s}$ band was used. References for the inclination (Inc) and position angle (PA) of the disks can be found in Appendix \ref{appendix_A}.}
\end{table*}

\begin{table}
\caption{\label{tab:weather}Weather conditions of the ADI epochs}
\small
\centering
\begin{tabular}{lccc}
\hline\hline
        Target
        & Wind speed  
        & $\tau_{0}$
        & Seeing
       
       \\
       & (m/s)
       & (ms)
       & ($\arcsec$)

       \\
\hline
LkCa 15  & 1.0 & 19.5 & 1.53 \\
HD 34282 &  1.8 & 1.3 & 1.18 \\
HD 36112 & 4.1 & 6.9 & 1.02 \\
CQ Tau &  5.0 & 8.3 & 0.69 \\
Sz Cha &  4.9 & 3.7 & 0.69 \\
TW Hya &  4.5 & 11.5 & 0.58 \\
HD 97048 &  8.5 & 3.6 & 0.95 \\
HD 100546 &  8.2 & 1.7 & 0.81 \\
PDS 66 &  11.4 & 2.8 & 0.99 \\
PDS 70 &  2.9 & 7.0 & 0.42 \\
HD 135344B &  6.5 & 8.5 & 0.50  \\
HD 139614 &  13.9 & 1.5 &  1.81 \\
RX J1604.3-2130A &  11.0 & 2.0 & 1.40\\
RX J1615.3-3255 &  6.9 & 6.7 & 0.64\\
HD 169142 & 12.6 & 4.1 & 0.43\\

\hline
\end{tabular}
\tablefoot{Median wind speed, turbulence coherence time and seeing during the ADI observations of Table \ref{tab:adi}. Low wind effects are usually avoided with a wind speed $>$3\,m/s, and a $\tau_{0}$\,$>$\,3\,ms effectively eradicates the wind-driven halo in SPHERE observations.}
\end{table}

\subsection{Reduction of the pupil-tracking data}

For all the objects in the target list of Table \ref{tab:host_param}, we performed high-contrast imaging ADI reductions to obtain detection limits to the presence of perturbers creating the substructures in the small dust distribution. Table \ref{tab:adi} shows the corresponding pupil-tracking epochs. When more than one ADI observation of the same system was available, we analysed the epoch that provided a better contrast based on published results, collected sky rotation and weather conditions (see Table \ref{tab:weather}). The ADI observing modes are split about equally between IRDIFS and IRDIFS-EXT; IRDIFS was the preferred mode by the SpHere INfrared Exoplanets survey \citep[SHINE; e.g.,][]{chauvin2017}, which mainly targetted 10-200\,Myr-old nearby stars. This mode results in dual-band imaging in the $H$ band with the infrared dual-band imager and spectrograph \citep[IRDIS;][]{dohlen2008}, and $YJ$ spectrophotometry with the near-infrared integral field spectrograph \citep[IFS;][]{claudi2008} in the Y–J range with a FoV of 1.7\,$\arcsec$\,$\times$\,1.7\,$\arcsec$. With the inclusion of younger stars located beyond 100\,pc in the survey (particularly from Sco-Cen) or dedicated follow-up of PPDs, the IRDIFS-EXT mode was used, which extends the dual-band imaging to the $K$ band and the IFS spectrophotometry to $H$ band, as this mode is more adapted for the detection of young red L-type planets.

The reduction and analysis have been conducted homogeneously for the entire dataset. For the IRDIS side we first used the vlt-sphere repository \citep{vltpf}\footnote{\url{https://github.com/avigan/SPHERE}}, which is a python-based pipeline to pre-reduce SPHERE data \citep{beuzit2019}. We sorted out the files and performed static calibrations, such as flat field reduction, dark and sky subtraction, and bad pixel correction. This was followed by star registration, for which we used the waffle pattern to find the location of the host star behind the coronagraphic mask, and flux calibration to get the difference in flux between the unsaturated star and the science frames. After this pre-reduction step we obtained a cube of centred saturated frames, collapsed unsaturated stellar frames for flux calibration, a sky rotation file and the wavelength in which the observations were carried out. To treat the quasi-static speckle noise affecting high-contrast imaging observations, further PSF-removal was conducted with the ANgular DiffeRential OptiMal Exoplanet Detection Algorithm \citep[ANDROMEDA;][]{cantalloube2015}. This algorithm is based on the maximum likelihood estimation. It first attenuates extended structures by applying a high-pass Fourier filtering that removes 25\,$\%$ of the lowest frequencies \citep[with an energy loss of $\sim$\,18\,$\%$, see Fig. 1 in ][]{cantalloube2015}. Subsequently, ANDROMEDA performs a model cross-correlation of the signature that a point source leaves after two frames at different rotation angles are subtracted. This processing is thus sensitive to the presence of point sources in the field of view. Before feeding the vlt-sphere output to ANDROMEDA, we performed a frame selection to remove those that deviate more than $n$ standard deviations of the mean of the 3D cube in a 20\,px radius centred on the star. The process is repeated for n$=$1,2,3... until the rejection fraction is lower than 20\,$\%$. Finally, ANDROMEDA provides two 2D maps: (i) the estimated contrast that a point source would have at every location in the image, and (ii) the corresponding uncertainty on this estimated contrast. The limiting contrast is therefore given by the uncertainty map, previously normalised empirically to account for the non-gaussian residual noise at small separation \citep{cantalloube2015}. This map can be used to derive a 1D 5$\sigma$ projected detection limit by taking its azimuthal median.

IFS data was pre-reduced via the SPHERE Data Center \citep{delorme2017}, and the resulting cubes went through the the SpeCal software \citep{galicher2018}, which further removes the stellar halo by applying an ADI-based method. We selected an Angular and Spectral Differential Imaging (ASDI) reduction following the principal component analysis approach in \citet{mesa2015} for the spatial and spectral channels. This method is aggressive in terms of speckle subtraction and seems to reach better contrasts than IRDIS-ADI within the IFS FoV \citep{langlois2021sphere}. Moreover, 1D IFS-ASDI projected contrast curves were computed as the azimuthal standard deviation of the noise residuals, corrected for self-subtraction via fake planet injections. To obtain 2D IFS-ASDI limiting magnitude maps that account azimuthally for the presence of disk residuals, we evaluated, around each pixel, the standard deviation within a ring of inner and outer radius of 0.7 and 2.5\,FWHM, respectively.

\setlength{\unitlength}{\textwidth}
\begin{figure*}
\includegraphics[width=1\textwidth]{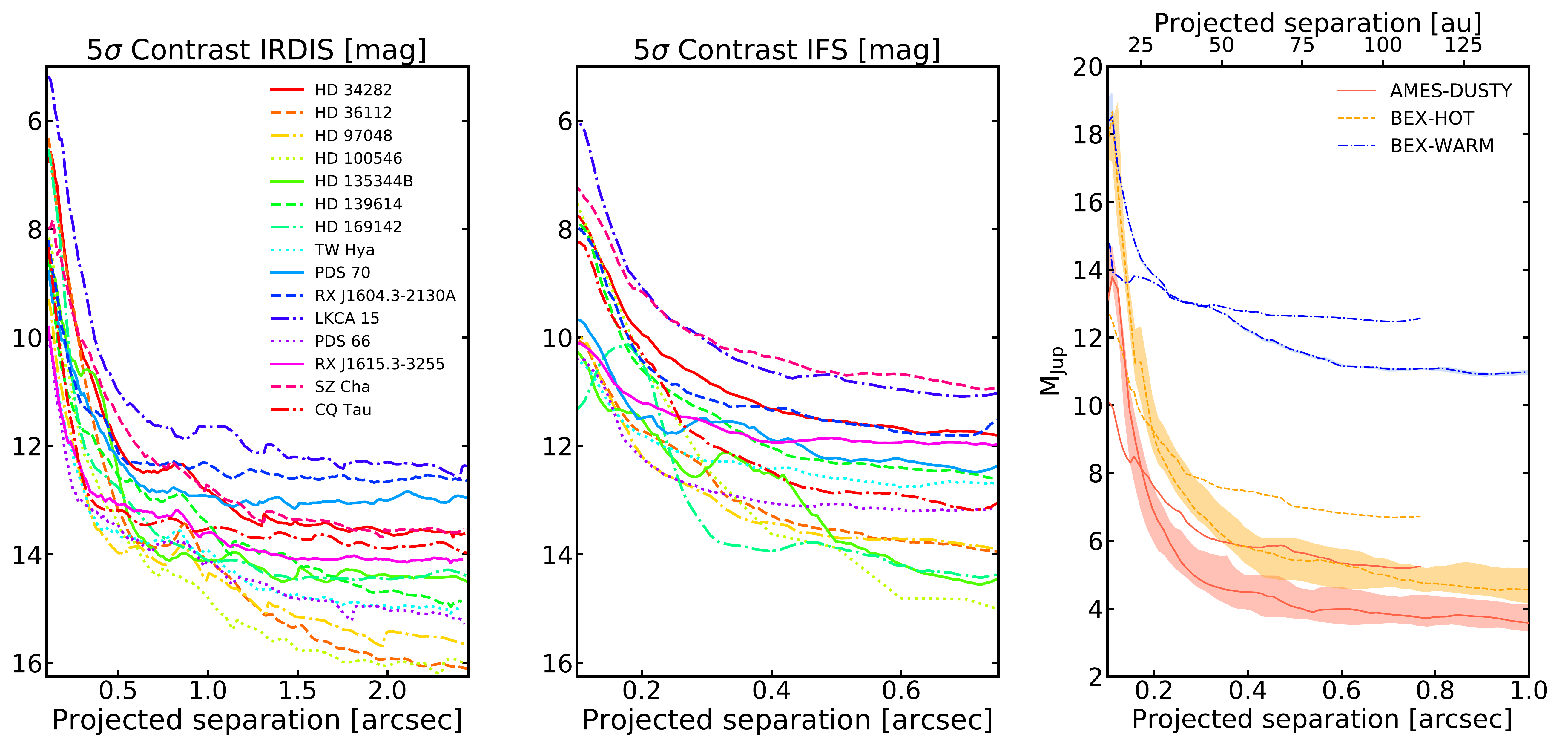}
\caption{(Left) SPHERE 1D projected contrast curves for ANDROMEDA/IRDIS data in the $H2$, $K1$ or $K\rm_{s}$ filter, depending on the observing mode as stated in Table \ref{tab:adi}. We show the results down to the edge of the coronagraph ($\sim$\,100\,mas) (Centre) Corresponding ASDI/IFS magnitude limits for the inner regions. (Right) Median mass detection limits assuming the initial luminosities given by the AMES-DUSTY, BEX-HOT and BEX-WARM models. To convert semi-major axis to au, we have adopted the median distance in our sample of $\sim$\,145\,pc. Thick curves correspond to the ANDROMEDA/IRDIS performance, while the shaded areas show the median sensitivities for the upper and lower limits on the age of the companions. ASDI/IFS mass limits are shown in the same curve style up to $\sim$\,0.8\,$\arcsec$. Individual deprojected detection limits are given in Appendix \ref{appendix_B}. }
\label{fig:CCs}
\end{figure*}

\section{Detection limits}\label{sec:sensitivities}

In Figure \ref{fig:CCs}  we show the projected 1D contrast limits achieved by ANDROMEDA/IRDIS and SpeCal-ASDI/IFS for the entire PPD sample. In general, ANDROMEDA/IRDIS achieves magnitude differences of $\sim$\,11--14\,mag at 0.5\,$\arcsec$ in the contrast-limited regime, and down to 16\,mag at larger separations limited by background sensitivity. With the use of ASDI, IFS contrasts seem to improve the IRDIS limits at close separations, within $\lesssim$\,0.4\,$\arcsec$. Besides the presence of disk residuals, variations in Strehl ratio, weather conditions, collected sky rotations and magnitudes of the host stars all contribute to the final detection limits. 

To convert these contrast to detectable companion masses, a post-formation luminosity of the perturber needs to be assumed. This is known as the `initial luminosity' with respect to the cooling phase, and is set by the radiative transfer and thermodynamics during the accretion phase \citep[e.g., ][]{mordasini2017,szulagyimordasini, marleau2019}. These differences can lead to divergences between the hot (more luminous) and cold (fainter) starts of about 3 orders of magnitude in luminosity during the first few Myrs \citep[e.g.,][]{marley2007, spiegel2012}. Although it is likely that a smooth range in planet formation luminosities between the extreme hot and cold models exist, the coldest starts do not seem to be a valid representation of the physical conditions of the current population of imaged planets; for instance, dynamical masses of the two planets around $\beta$ Pic seem to fall close to the predictions of a hot start, either formed through disk instability or a hot core accretion process  \citep{brandt2020, nowak2020}. Likewise, cold starts are unable to reproduce the dynamical masses of the four giant planets in the HR 8799 system \citep{wang2018}. Although these results could be affected by an observational bias, where we simply have not yet detected cold-start planets because they are fainter (see Section \ref{sec:detection_rates}), theoretical studies of the accretion shock also favour hotter start models \citep{marleau2019b}.

Here we rely on the AMES-DUSTY atmospheric models \citep{chabrier2000} to reproduce the hottest, most luminous outcome of planet formation, usually associated with disk instability. This model includes dust absorption and scattering, since young planets seem to have condensates in their atmospheres \citep[e.g.,][]{mueller2018}. AMES-DUSTY has been widely used by the community, and starts with a completely formed planet at arbitrary initial entropy for a given mass.

We also use the Bern EXoplanet cooling curves (BEX), which are based on the population synthesis models by \citet{mordasini2017}. They provide luminosities of young giant planets formed through the core accretion paradigm. Depending on whether the accretion shock luminosity is radiated away or deposited into the planet, the initial luminosities of the newborn planets are classified in diminishing order of brightness from `hottest' to `coldest'. These are then expanded into evolutionary tracks at constant mass, reproducing the cooling under different atmospheric conditions \citep{marleau2019}. We make use of the BEX-COND tracks, where the initial planet luminosities are coupled to the boundary conditions given by the condensate-free AMES-COND atmospheric models \citep{allard2001}, and use these resulting BEX-COND gravities and temperatures as a function of time to calculate the magnitudes via DUSTY. We adopt the `hot' and `warm'-start initial conditions from BEX-COND, which we call BEX-HOT and BEX-WARM. These two relations correspond to a fit extension of the cold nominal (BEX-HOT) and cold classical (BEX-WARM) population in \citet{marleau2019}; see Equation~1 in \citet{vigan2020}.

On the right panel of Figure \ref{fig:CCs}, we see the effect of these different post-formation luminosities on the median projected sensitivity to planet masses around a stellar host located at the typical distance of our sample (145\,pc). Down to $\sim$\,15\,au (0.1\,$\arcsec$), assuming that the AMES-DUSTY models are valid, the SPHERE observations would be sensitive to planets below 10\,$M\rm_{Jup}$, and down to 4\,$M\rm_{Jup}$ at $>$\,0.5\,$\arcsec$. This is different by a factor $\sim$\,$\times$\,1.5 if the BEX-HOT start models are considered, and planets below 11\,$M\rm_{Jup}$ would not be detectable if the cold BEX-WARM models are a good representation of planet formation. 

There are however a few caveats in the use of evolutionary models. If the perturber is formed via core accretion, there will be a \textit{delay time} with respect to the formation of its host star \citep[e.g.,][]{fortney2005, bonnefoy2014}. For gas giants the delay cannot be longer than the typical PPD lifetime of $\sim$\,3\,Myr \citep[e.g.][]{ribas2015}; most giant planets around an A-type star will have acquired their final mass after 2\,Myr \citep{williams2011, marleau2019}. Moreover, if the planet is still forming, its intrinsic luminosity could be augmented by accretion onto the planet surface and/or shock and thermal emission from a potential circumplanetary disk (CPD) \citep{szulagyi2014, zhu2015, szulagyimordasini, aoyama2018, aoyama2019, aoyama2021}, while at the same time also reddened by the presence of material in the circumplanetary environment \citep{szulagyi2018, szulagyi2019, szulagyi2020}. In these cases where formation is ongoing, these evolutionary models of post-formation luminosities are probably not a very solid representation of the physical parameters of the perturbers. However, without comprehensive spectroscopic analysis it is extremely difficult to derive the mass of a forming protoplanet from the expected magnitudes, as the contribution of accretion can be very important, creating a spread in luminosity of up to $\sim$\,2\,dex for a given planet mass (see Figure 2 in \citet{mordasini2017}). Here we suppose that the perturber starts its post-formation cooling at the same time as the star reaches its pre-main sequence phase, and assume that any potential time delay would be included within the estimated error bars on the age of the host star. Extinction effects are discussed in Section \ref{subsec:extinction}.

\begin{figure}
\centering
\setlength{\unitlength}{\textwidth}
\includegraphics[width=0.5\textwidth]{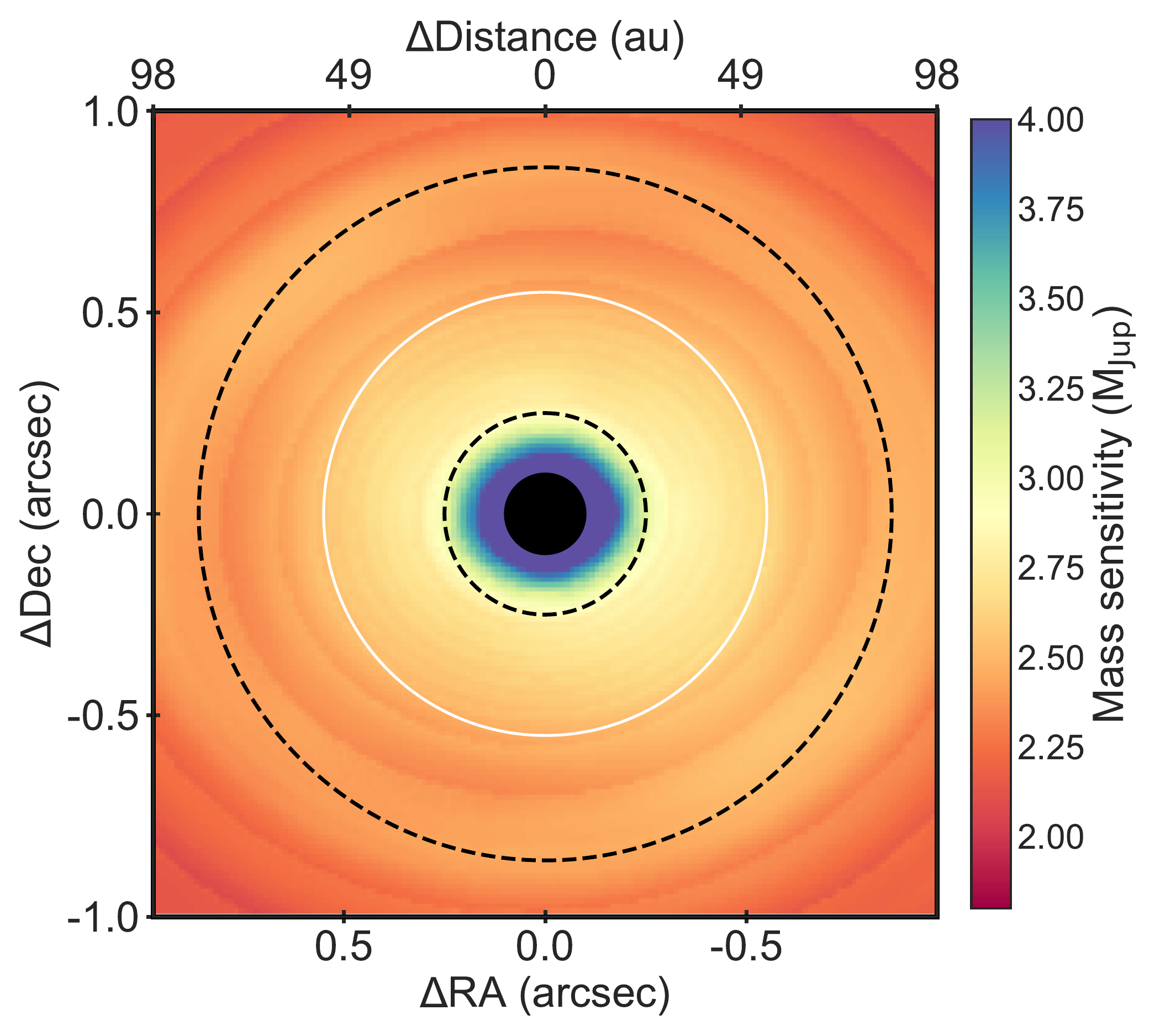}
\caption{5$\sigma$ mass sensitivity to companions in the PDS 66 system and the IRDIS $H2$ filter. The removal of the stellar halo has been performed with ANDROMEDA, and AMES-DUSTY initial conditions have been assumed for the perturber's luminosity. The black dashed circumferences show the outer radius of the bright compact region at 25\,au and the location of the outer ring at 85\,au, respectively, as seen in scattered light by \citet{avenhaus2018}. The white thick circumference shows the proposed location of a perturber, at 55\,au, creating the ring-like substructure, which coincidentally corresponds to a further depleted region in the scattered light data. The black circle shows the location of the star and has a radius of 0.1\,$\arcsec$. The 2D map has been deprojected based on the disk parameters of Table \ref{tab:host_param}.}
\label{fig:example_PDS66}
\end{figure}

\begin{figure}
\centering
\setlength{\unitlength}{\textwidth}
\includegraphics[width=0.5\textwidth]{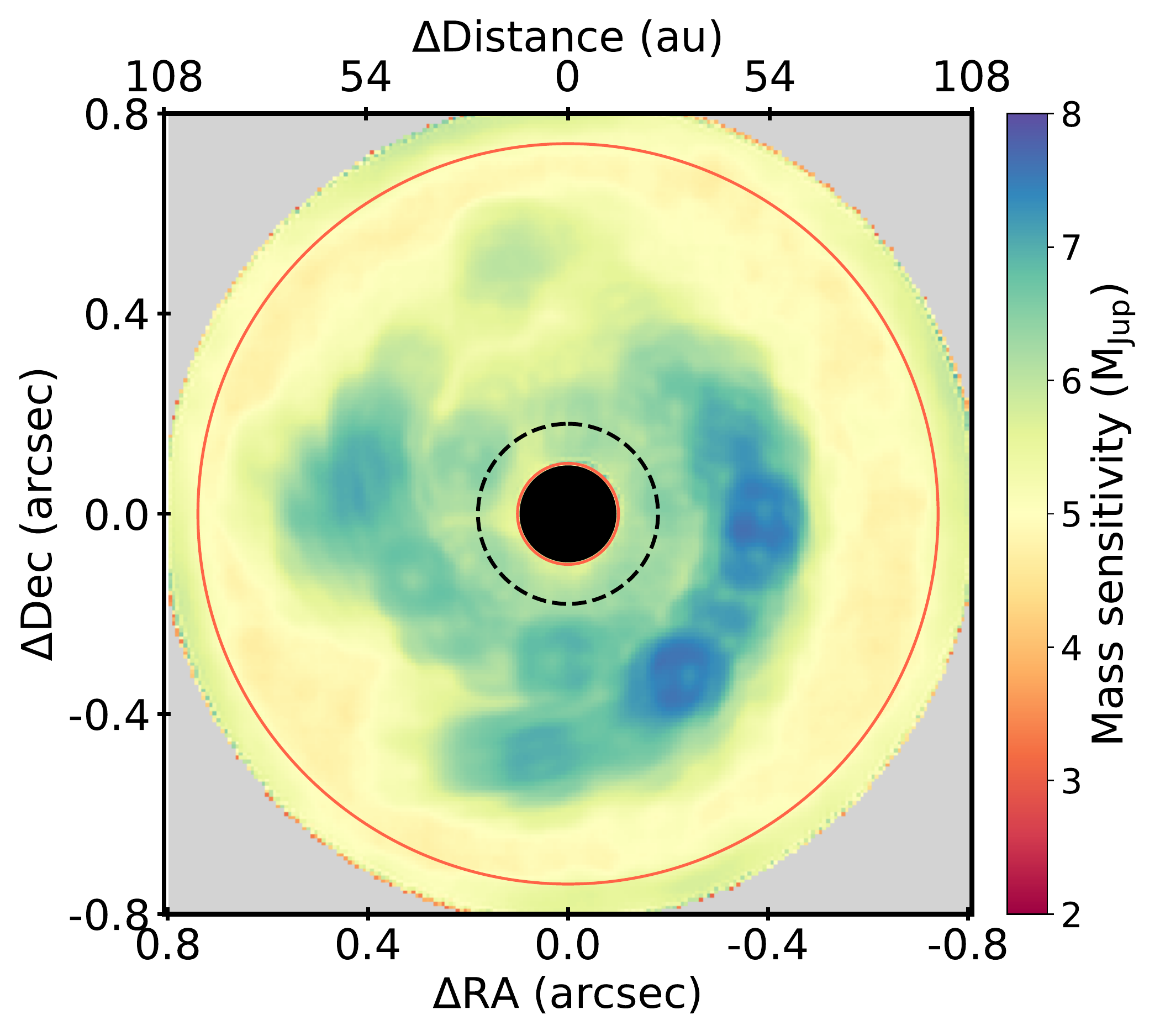}
\caption{5$\sigma$ mass sensitivity to companions in the HD 135344B system achieved by the ASDI/IFS reduction. As in Figure \ref{fig:example_PDS66}, AMES-DUSTY initial conditions have been assumed for the perturber's luminosity, and the 2D map has been deprojected based on the disk parameters of Table \ref{tab:host_param}. Disk residuals caused by the spiral pattern are clearly seen. The black dashed circumference shows the location of the cavity in scattered light as observed by SPHERE/PDI \citep{stolker2016}. The inner red thick circumference shows the proposed location of a perturber in the middle of the cavity, very close to the coronagraphic mask, and the outer one represents the proposed location of a 5--10\,$M\rm_{Jup}$ companion creating the spirals \citep{dongfung2017a}. }
\label{fig:example_hd135344b}
\end{figure}

Finally, we convert the projected 2D limiting contrasts to mass limits for each pixel in the image. These are then used to (i) obtain a detection probability map for each target (included in Appendix \ref{appendix_C}) via the Multi-purpose Exoplanet Simulation System \citep[MESS, ][]{bonavita2012} and (ii) obtain 1D deprojected mass sensitivities. For the latter, we deprojected the 2D maps by inclination and position angle to be in the disk plane (see an ANDROMEDA/IRDIS example in Figure \ref{fig:example_PDS66}, and in \ref{fig:example_hd135344b} for the ASDI/IFS mode), and at each radial separation we take the median of the limiting mass values to construct 1D sensitivities, which are shown for each individual system in Appendix \ref{appendix_B}. We also overplot the location of the substructures seen in scattered light and in the mm continuum as explained in Appendix \ref{appendix_A}. In the following section we will use this method to constrain the detectability of the perturbers that may be creating the PPD substructures.

\section{Population of perturbers in SPHERE PPDs}\label{sec:perturbers}

To approximate the mass and location of the perturbers causing the substructures in the PPDs of our sample, we first rely on the simulated results in the literature where gap morphology in the $\upmu$m-sized dust is taken into account \citep{pohl2017,dongfung2017b,vanboekel2017}, together with the simulated planets that may reproduce the spirals in scattered light \citep{dongfung2017a, baruteau2019}. However, these derivations make up only a subset of the substructures seen in our target list. Given the intricate disk-planet interaction and the large numbers of gap structures we are dealing with, in addition to the simulated perturbers obtained from the literature we also consider two different ways of estimating their masses: 1) Gap width proportional to the planet Hill radius and 2) Difference in cavity size in near infrared scattered light images and ALMA mm continuum.

\begin{figure}
\centering
\setlength{\unitlength}{\textwidth}
\includegraphics[width=0.48\textwidth]{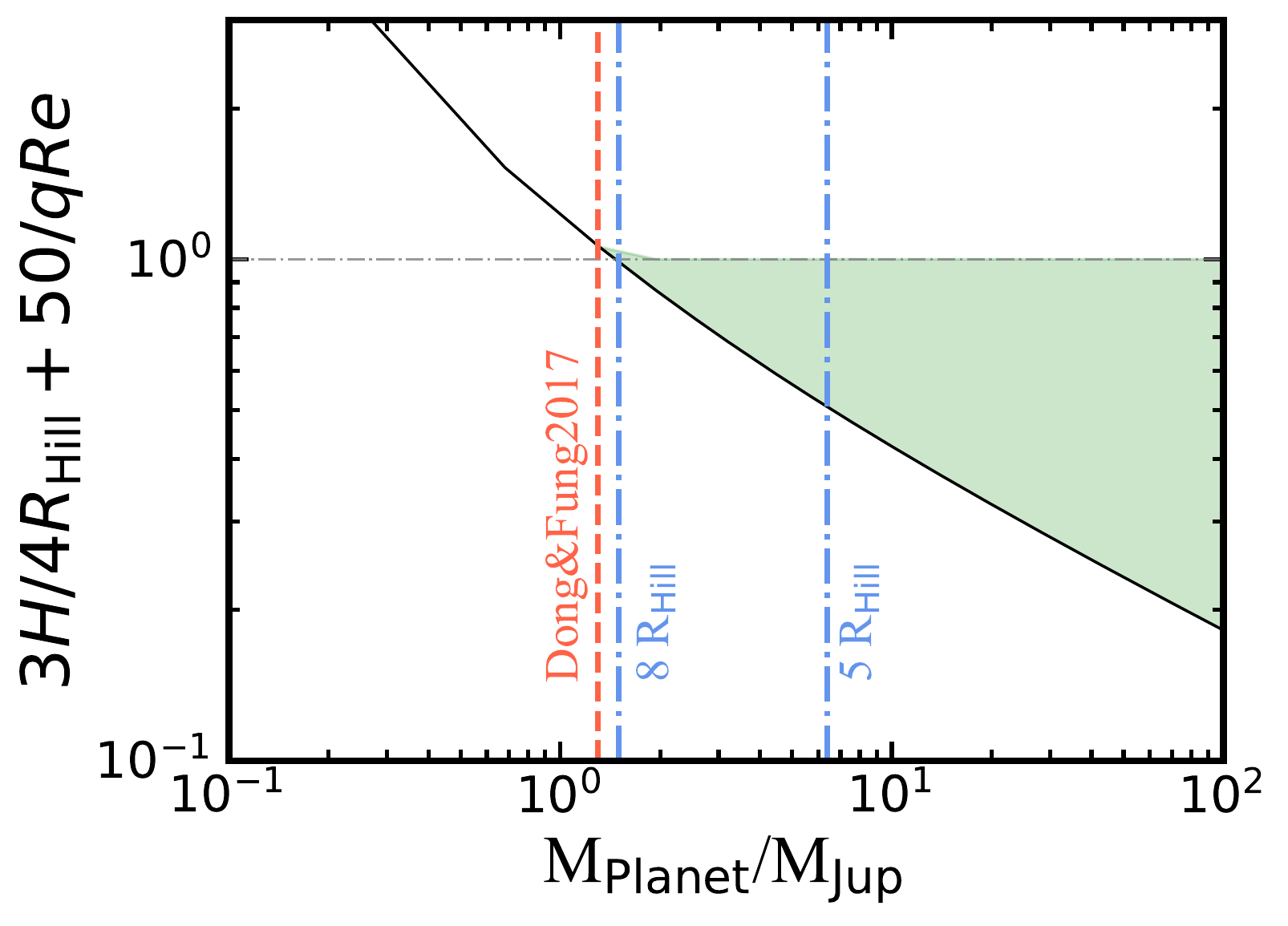}
\caption{\citet{crida2006} criterion of Equation \ref{eq:crida} applied to the gap-opening planet mass in the PPD around HD 97048. The green-shaded area shows the masses at which a planet would carve a deep gap in the gas at the observed location of 127\,au \citep{ginski2016}, assuming a disk viscosity parameter of $\alpha = 10^{-3}$. The vertical blue dashed-dotted lines correspond to a planet creating a depleted region of width 5\,R$\rm_{Hill}$\ and 8\,R$\rm_{Hill}$\, respectively, following Equation \ref{eq:hill}. The red dashed line indicates the estimated perturber mass by \citet{dongfung2017b}, using a detailed hydrodynamical + radiative transfer simulation of the gap morphological parameters.}
\label{fig:crida_HD97048}
\end{figure}

\subsection{Gap Opening Mass from Hill Radii}
\label{sec:Hill}

The gap width that a planet carves in the disk can be approximated by a proportionality factor applied to its Hill sphere of influence, defined as $ R\rm_{Hill} = R\rm_{p} (M\rm_{p}/ 3M\rm_{\star})^{1/3}$. A simple estimation that can relate the observed width in scattered light to the perturber's mass is given by:

\begin{equation}
 \mathit{R}\rm_{PDI,peak} - \mathit{R}\rm_{PDI,gap} = \mathit{k}\,\mathit{R}\rm_{Hill}
    \label{eq:hill}
\end{equation}

where the width is defined as the distance between the gap minimum $R\rm_{PDI,gap}$ and the peak of the ring $R\rm_{PDI,peak}$ seen in PDI, and $k$ is a scaling factor. This formulation is similar to the latest estimations of gap-carving planet masses in the millimetre continuum \citep[e.g.,][]{lodato2019}, which are in agreement with the individually-modelled DSHARP planet population and the broader collection from \citet{bae2018}.

To estimate the scaling factor in the $\upmu$m-sized dust, we consider the gap located at $\sim$\,127\,au in the PPD around HD 97048. The edges of this gap are well resolved in SPHERE/PDI and total intensity, and the surface brightness of the bottom is detected at 5$\sigma$ background sensitivity \citep{ginski2016}. First, we apply the analytical gap-opening criterion from \citet{crida2006}, which does not take into account gap morphology nor dust dynamics and evolution. This solution gives the minimum planet mass that clears 90\,$\%$ of the gas disk when the following criterion is met:

\begin{equation}
    \frac{3}{4}\frac{H}{R\rm_{Hill}} + \frac{50}{q\textit{Re}} 	\lesssim 1
    \label{eq:crida}
\end{equation}

where $H$ is the local scale height and \textit{Re} the Reynolds number defined as $r\rm_{p}$2$\Omega\rm_{p}$/$\nu$, with $r\rm_{p}$ being the radius of the planet orbit, $\Omega\rm_{p}$ the angular velocity and $\nu$ the viscosity $\nu$ = $\alpha$\,c$\rm_{s}^{2}$/$\Omega\rm_{p}$. The shaded area of Figure \ref{fig:crida_HD97048} shows the allowed mass values according to this approach for $\alpha$ = 10$^{-3}$ and assuming $H$/$r\rm_{p}$ = 0.056 (same value as found by \citet{dongfung2017b} at the gap location through Equation \ref{eq:width}, see later). A perturber with a mass $\geq$\,1.3\,$M\rm_{Jup}$ is necessary to create a well-depleted gap in the gas. 

To take into consideration the morphological features that a planet causes in the disk, hydrodynamical simulations have investigated the gap parameters that a planet embedded in a PPD with a given physical properties would impose on the gas \citep[e.g.,][]{fung2014, duffell2015, kanagawa2015, kanagawa2016}. The associated analytical criteria approximate these simulations and link planet masses to gap depth and width. Following \citet{dongfung2017b}, the gap depth in gas is given by:

\begin{equation}\label{eq:depth}
    \frac{\sum_{0}(r\rm_{min, \sum})}{\sum(r\rm_{min, \sum})} = 0.043\,q^{2}\,\left(\frac{H}{r}\right)^{-5}\,\alpha^{-1} + 1
\end{equation}

and the normalised and dynamical gap width is found to scale only with the local disk aspect ratio as:

\begin{equation}\label{eq:width}
    \frac{r\rm_{in, \sum} - r\rm_{out, \sum}}{ r\rm_{min, \sum}} = 5.8\,\frac{H}{r}
\end{equation}

where $q$, $\alpha$ and $H/r$ are the mass ratio between the companion and the host star, the accretion disk viscosity parameter \citep{shakura1973}, and the disk aspect ratio, respectively. The locations (r$\rm_{in, \sum}$ and r$\rm_{out, \sum}$) are the edges of the gap, defined as the positions where the surface density reaches the mean of the minimum $\sum(r\rm_{min, \sum})$ and undepleted $\sum_{0}(r\rm_{min, \sum})$ surface density. These morphological parameters also need to be connected to the observed ones in scattered-light images. 2D and 3D hydrodynamical simulations coupled to radiative transfer calculations show that this translation depends mostly on the inclination of the disk and the angular resolution and sensitivity of the observations \citep{szulagyi2018}. Using this formulation, \citet{dongfung2017b} obtained a value for the mass of the perturber of 1.3\,$M\rm_{Jup}$ for $\alpha$ = 10$^{-3}$ and $H$/$r$ = 0.056 (see Figure \ref{fig:crida_HD97048}). This is in agreement with the derivation using Crida's criterion, although it does not always have to be the case, as equations \ref{eq:depth} and \ref{eq:width} are accommodated to each specific gap morphology, unlike equation \ref{eq:crida}. When comparing the two criteria at the depth assumed by Crida, this later prescription tends to derive a higher planet mass \citep[see, e.g., Figure 12 in ][]{camille2020}.

We finally apply the gap-opening Hill radius criterion of equation \ref{eq:hill} to this gap. A scaling factor of $k = 8$ is necessary to simulate the mass derived by \citet{dongfung2017b}. We find that for LkCa 15, the $\sim$90\,au gap in TW Hya and HD 169142, we are also able to approximate the simulations of \citet{dongfung2017b} with $k$\,$\sim$\,8--10. A $k$\,$>$10 scaling is however necessary to account for their derived masses in the $\sim$\,78\,au gap around RX J1615.3-3255 and the $\sim$\,22\,au gap in TW Hya, which is related to the individual gap morphology that the Hill radius approximation does not account for. To give an insight on the uncertainties involved when applying this basic Hill radius approximation, variations in the $k$ factor from 8 to 12 can be translated to a perturber's mass estimate between $\sim$1.5 to 0.5\,$M\rm_{Jup}$ for the gap we are considering in HD 97048. Likewise, another source of uncertainty is the viscosity $\alpha$ assumed by the models; \citet{dongfung2017a} find that the mass of the perturber changes by a factor of 10 between the two extreme values they considered, from $\alpha = 10^{-4}$ to $\alpha = 10^{-2}$.

Here we take from the literature the location of all the gaps between rings and within cavities found in the SPHERE scattered light images of our PPD sample, and assume that one planet at that location opens a gap that scales with $k\,=\,8$. This also includes PDS 70, for which we ignore the two giant planets detected in its cavity for the rest of the analysis in order to be used as a point of reference and comparison with other systems. An individual brief discussion of each system and its substructures is provided in Appendix \ref{appendix_A}, together with a very simple representation of their scattered light appearance and the locations at which we locate the perturbers (Figure \ref{fig:drawing}). If the gap minimum location is not explicitly calculated in the literature, we place the putative planet in the centre between a pair of rings. When the region between the star and the inner ring is depleted down to the coronagraphic IWA in the $\upmu$m-sized dust, we treat it as a cavity. In these cases, we locate the planet at half the distance between the star and the rim, which does not necessarily need to be the case in a real scenario. We also treat the outer rings of PDS 70 and LkCa 15 as cavities, although non-resolved inner disk signals in the small dust are detected close to the mask. With this prescription we seek to account for a global estimate of planet masses in real observations of non fully-depleted gaps. 

\begin{figure}
\centering
\setlength{\unitlength}{\textwidth}
\hbox{\hspace{-0.25cm}\includegraphics[width=0.5\textwidth]{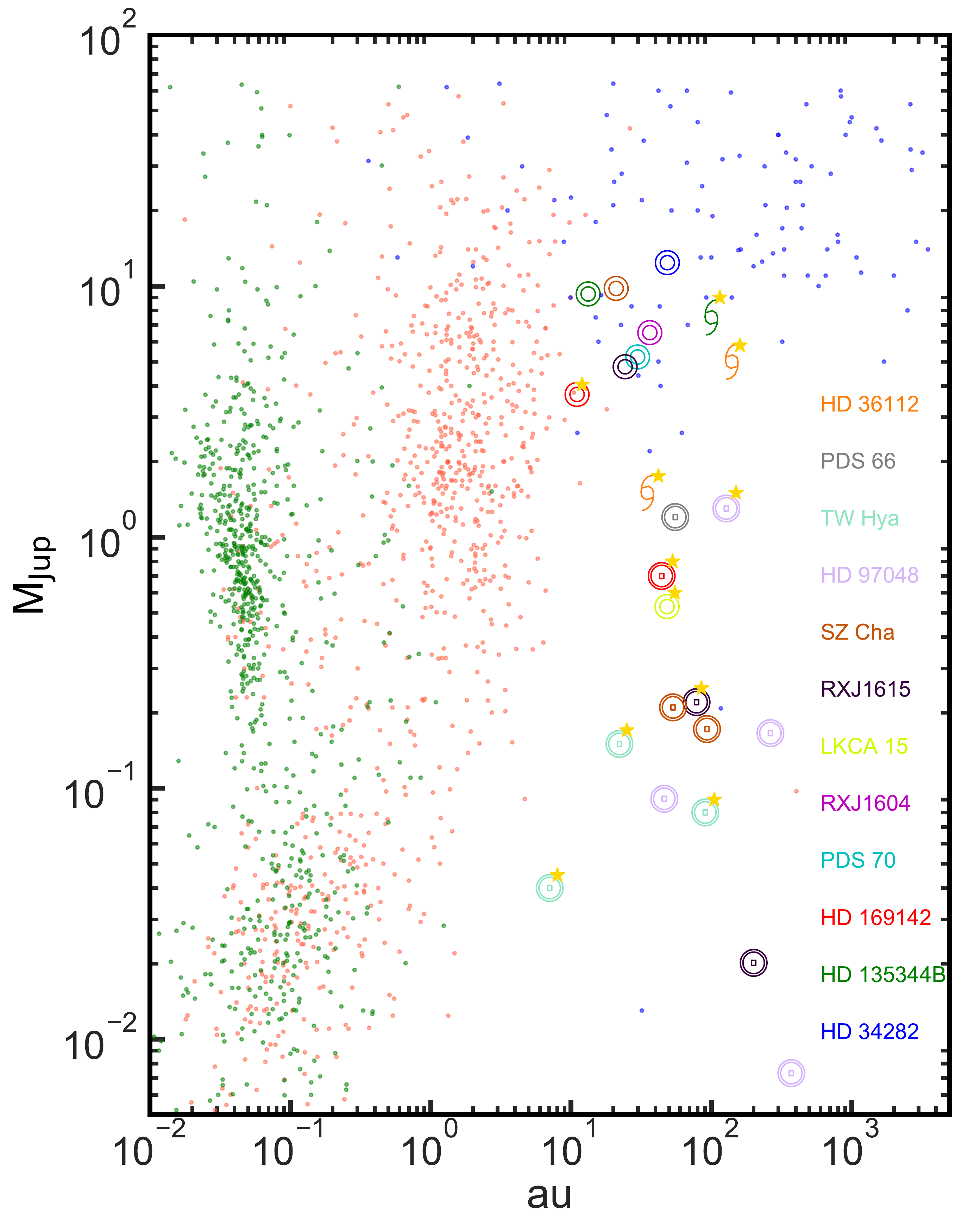}}
\caption{Underlying population of potential perturbers derived from PPD substructures in the SPHERE scattered light images. For planets in the gaps between rings (circles with inner dots) and planets in cavities (circles), these values are estimated using a proportionality factor of 8\,$R\rm_{Hill}$ to account for the mass that would create the observed width in the scattered light gap (Equation \ref{eq:hill}). When the gap morphology in scattered light was taken into account, planet masses are taken from detailed simulations in the literature for $\alpha$ $=$ 10$^{-3}$, and are marked with an asterisk. Simulated planets creating spiral features are also included as spiral symbols (see references in the Appendix \ref{appendix_A} and see also Figure \ref{fig:drawing}). The distribution of confirmed planets in the exoplanets.eu catalogue are shown as green (radial velocity discoveries), orange (transits) and blue (imaging) small dots.}
\label{fig:hill}
\end{figure}

\subsubsection{Underlying population of perturbers}

\begin{figure}
\centering
\setlength{\unitlength}{\textwidth}
\hbox{\hspace{-0.5cm}\includegraphics[width=0.54\textwidth]{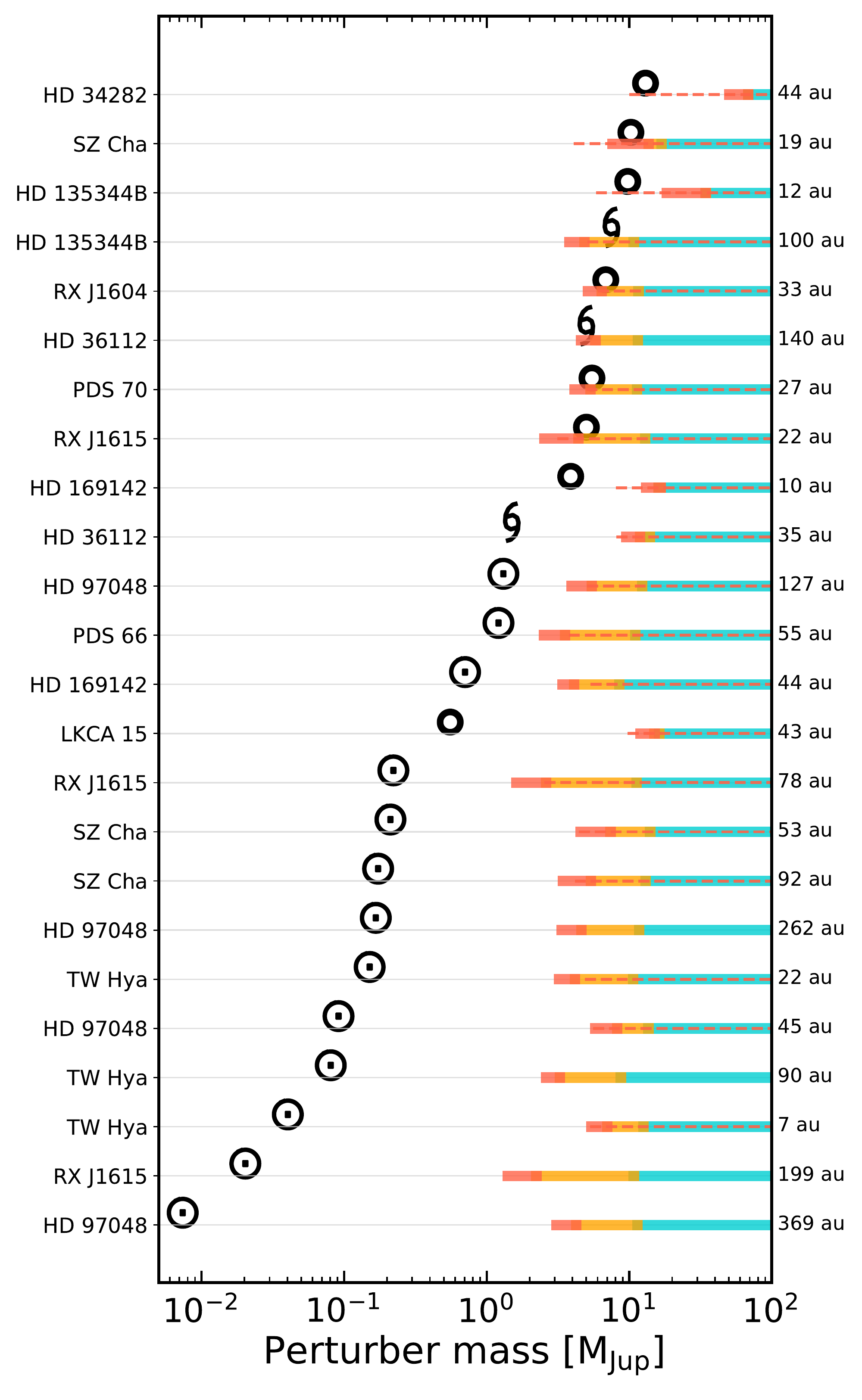}}
\caption{Population of perturbers from Figure \ref{fig:hill} compared to our derived 5$\sigma$ deprojected mass sensitivities at their location, which is shown on the right side of the figure. SPHERE/IRDIS with AMES-DUSTY, BEX-HOT and BEX-WARM initial starts are shown as red, orange and blue bars, respectively. The dashed lines correspond to SPHERE/IFS-ASDI contrasts and AMES-DUSTY post-formation luminosity. }
\label{fig:hill_sensitiv}
\end{figure}

Figure \ref{fig:hill} shows the planet masses derived through Hill radius and $k = 8$. The perturbers are located at the gap minimum within the cavities and in between rings of the PPDs in our sample. When derivations from simulations exist in the literature, we adopt those ones. They are marked with an asterisk in the figure. We also include the latest simulated planet masses and locations that explain the spirals seen in HD 36112 \citep{baruteau2019} and HD 135344B \citep{dongfung2017a}. We do not consider in this section three systems: CQ Tau, HD 100546 and HD 139614, because they have complicated structures with no clear gaps, or no simulated planets creating the spirals exist. From the figure, two different populations of perturbers seem to arise; more massive 3--10\,$M\rm_{Jup}$ objects that create big cavities, and a more sparse population of planets creating gaps within rings in the $\sim$\,0.01--1\,$M\rm_{Jup}$ regime. The latter are all massive enough to potentially create detectable gaps in PPDs \citep[e.g.,][]{bitsch2018}. We note that these derived perturbers are objects that could migrate and evolve \citep[e.g.,][]{lodato2019}, and so the mass-position diagram of Figure \ref{fig:hill} is not directly comparable to the population observed by the indirect methods. In Figure \ref{fig:hill_sensitiv} these masses are compared to the derived 5$\sigma$ mass detection limits at the proposed location of the perturber, computed for the different starting luminosities after deprojecting the 2D mass sensitivity maps. 

These two figures suggest that:

\begin{figure}
\setlength{\unitlength}{\textwidth}
\includegraphics[width=0.49\textwidth]{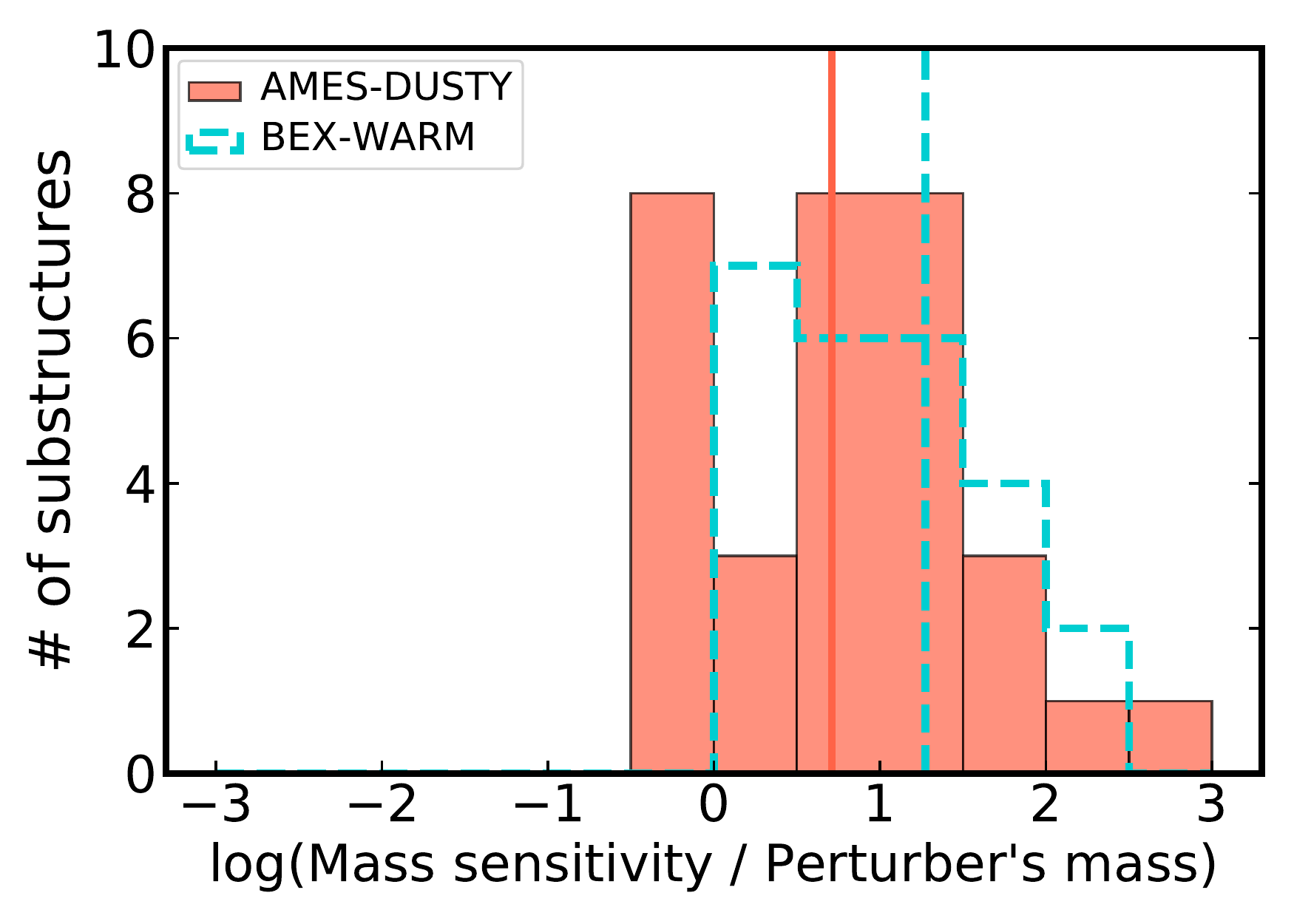}
\caption{Histogram of the difference between estimated planet masses that create substructures in our PPD sample and the derived contrast mass limits at their location. Vertical lines mark the median value for each mass-luminosity relationship. Units are in orders of magnitude in mass.}
\label{fig:histo_hill}
\end{figure}

\begin{itemize}

    \item Our sensitivities do not reach masses for planets creating the gaps in the disk population with rings.\\

    \item Planet masses needed to create the observed cavities are in general larger than those in the gaps between rings, unless the assumption that only one planet induces the cavity is incorrect. \\
    
    \item If hot-start models are a good representation of the initial planet luminosities, we could be close to a detection in several systems, omitting the extinction due to the potential surrounding circumplanetary and circumstellar material. So far, two giant planets in PDS 70 are the only bona-fide detections, which might imply the presence of currently undetected lower-mass companions in these systems, or that other physical mechanisms may be responsible for the formation of the cavity, such as dead-zones, MHD winds, instabilities or grain growth at snow-lines \citep{flock2015,pinilla2016, takahashi2016, cieza2016}. \\

    \item Planet masses creating gaps within scattered-light rings are consistent with the population of undetected companions that is emerging from the high-resolution ALMA data \citep[see Figure 21 in][]{zhang2018}.

\end{itemize}

Figure \ref{fig:histo_hill} shows a histogram version of the results in Figure \ref{fig:hill_sensitiv}. We can see that there are 8 cases (i.e., the first bin) where these putative planets with hottest starts could be detectable; 6 in cavities (HD 34282, Sz Cha, HD 135344B, RX J1604, PDS 70 and RX J1615) and 2 creating spiral patterns (HD 135344B and HD 36112). Additionally, 3 perturbers might be close to a detection (defined here as less than half an order of magnitude away in mass, forming the second bin) in the cavity of HD 169142 and the gaps between rings in HD 97048 and PDS 66. Table \ref{tab:promising} shows these most promising systems where to look for perturbers based on this analysis. For the perturbers in cavities very close to the coronagraphic mask, spatial resolution and projection effects can also affect the detection. Overall, these results might point towards the fact that planets may not be the only way of creating substructures, given the current number of detections, unless cold initial conditions represent a better description of their luminosity, which is unlikely according to \citet{marleau2019b}.

\begin{table*}
\caption{\label{tab:promising}Perturbers close to detection based on the mass estimates of Section \ref{sec:Hill}}
\small
\centering
\begin{tabular}{lccccc}
\hline\hline
  ID &   Gap &  Location  &  Planet mass & AMES-DUSTY &   BEX-HOT  \\
    &    &   (au) &  ($M\rm_{Jup}$) & ($M\rm_{Jup}$) &   ($M\rm_{Jup}$)  \\

\hline
        HD 34282 &      in-cavity     &   44&     11.7    &   10.0 &  11.8 \\
        SZ Cha &           in-cavity &      19 &   9.3  & 4.1 &  6.2 \\
        HD 135344B &         in-cavity &      12 &  8.8 & 5.8 &  7.0 \\
      HD 135344B &       Spiral &     100 &   5--10 & 3.8 &  4.9 \\
        RX J1604&         in-cavity &      33 &  6.2 & 5.1 &  6.4 \\
          HD 36112 &        Spiral &     140 &    5.0    &   4.6 &  5.8 \\
        PDS 70 &         in-cavity&      27 &   4.9 & 4.1 &  5.3 \\
         RX J1615 &         in-cavity &      22 & 4.5 & 2.5 &  4.3 \\
        HD 169142 &         in-cavity &    10 &    3.5 & 8.1 & 9.5 \\
         HD 97048 &           in-rings &     127 &   1.3 & 3.9 &  5.5 \\
        PDS 66 &           in-rings &      55 &   1.2 & 2.5 &  3.5 \\

\hline
\end{tabular}
\tablefoot{Perturbers that, according to the estimations of Figure \ref{fig:hill_sensitiv}, are the most promising to be found. PDS 70 treated as planet-less here. Planet masses in gaps and creating the spiral patterns are taken as described in Figure \ref{fig:hill}. The columns AMES-DUSTY and BEX-HOT are our derived SPHERE detection limits (best of ANDROMEDA/IRDIS and ASDI/IFS) using the AMES-DUSTY and BEX-HOT models (see Section \ref{sec:sensitivities}). References for the locations of the perturbers can be found in Appendix \ref{appendix_A}. See also Figure \ref{fig:drawing} for a sketch of these morphologies.}
\end{table*}

\subsection{Gap opening mass from scattered light-mm continuum comparison}

In this section we derive planet masses for the systems in which resolved rings and/or cavities exist both in SPHERE scattered light and in ALMA mm continuum observations (see Table \ref{tab:ALMA}). SPHERE PDI observations trace the small dust coupled to the gas in a PPD, while ALMA data probe mm-sized dust grains settled in the disk midplane. For this reason, the pressure bump created by the presence of a planetary-mass companion can leave its imprint in the different appearance of the disk at different wavelengths. In general, large grains will accumulate more effectively at the peak of the gas pressure, as no drag force exists in the pressure bump, making the gas to rotate near or exactly at Keplerian velocity, while $\upmu$m-sized dust is well-coupled to the gas and moves along with it. This effect creates a spatial difference in the cavity sizes observed in the mm continuum with ALMA and in polarised scattered light, which is dependent on the planet mass and disk location.

\citet{dejuanovelar2013} simulated the radial distribution of the dust due to the presence of 1, 9, and 15\,$M\rm_{Jup}$ giant planets located at 20, 40 and 60\,au around a solar-mass star, after 3\,Myr of evolution and $\alpha$ viscosity of 10$^{-3}$. We use their functional form of the ratio in the cavity size seen in $\upmu$m and mm dust for a given planet mass:

\begin{equation}
\centering
f(M\rm_{p})=c\left(\frac{M\rm_{p}}{{M\rm_{Jup}}}\right)^{\Gamma}
\label{eq:ovelar}
\end{equation}

with $f(M\rm_{p})$ defined as the ratio between the location of the scattered light inner edge of the cavity seen with SPHERE/ZIMPOL, and the peak in 850\,$\upmu$m flux with ALMA, $c\sim0.85$ and $\Gamma\sim[-0.22,-0.18,-0.16]$ for planet orbital radii $R_{\rm{p}}=[20,40,60]$, respectively.

\begin{figure}
\setlength{\unitlength}{\textwidth}
\hbox{\hspace{-0.3cm}\includegraphics[width=0.51\textwidth]{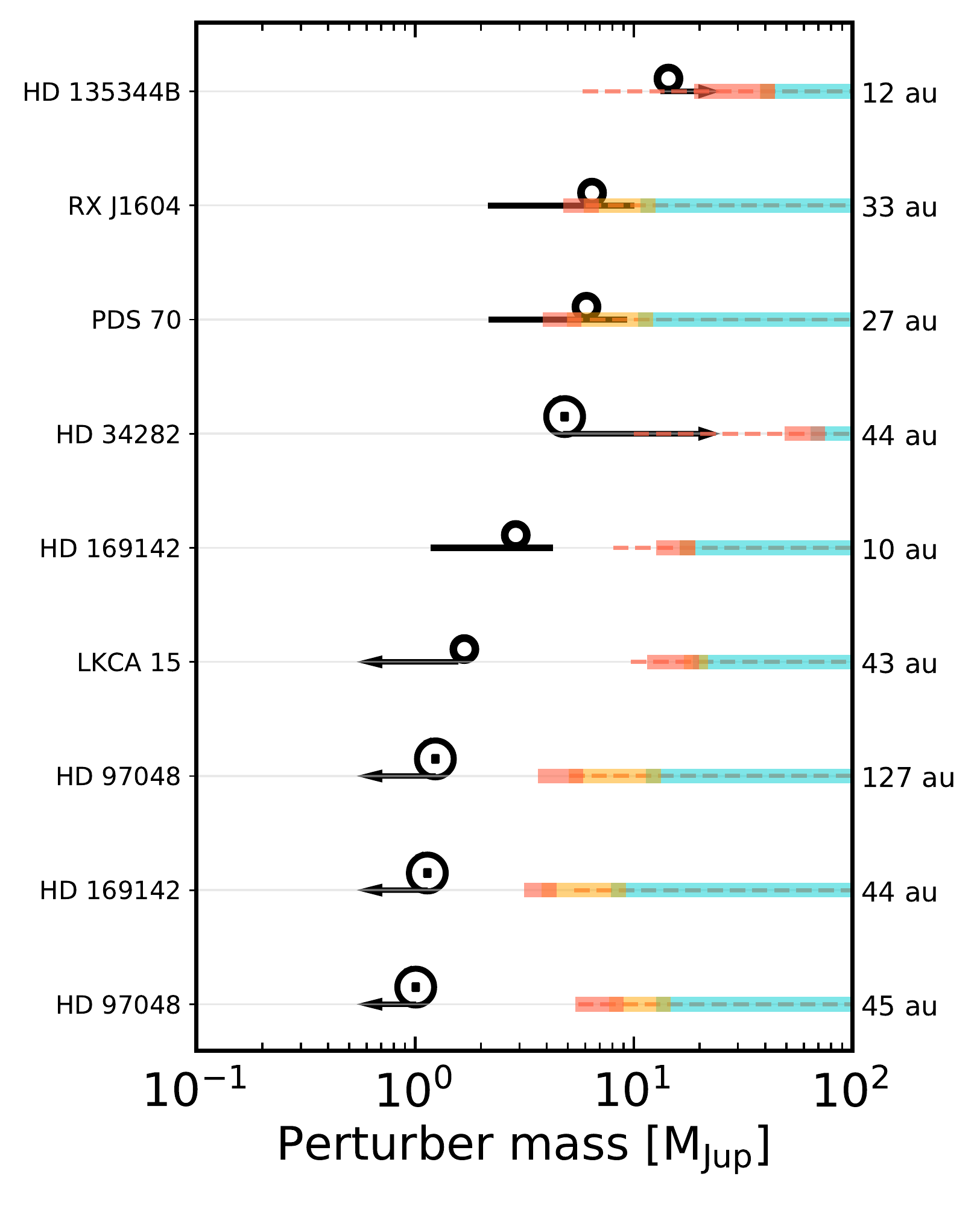}}
\caption{Predicted masses from the \citet{dejuanovelar2013} prescription (black) using SPHERE scattered light images and ALMA mm peak emission. Arrows are used when the estimated mass limit is either lower than 1\,$M\rm_{Jup}$ or higher than 15\,$M\rm_{Jup}$, i.e., outside the regime of Equation \ref{eq:ovelar}. ANDROMEDA/IRDIS detection limits with AMES-DUSTY, BEX-HOT and BEX-WARM initial starts are shown as red, orange and blue bars, respectively. The dashed lines correspond to ASDI/IFS AMES-DUSTY detection limits.}
\label{fig:juanovelar}
\end{figure}

The location of the scattered light inner edge is usually not provided from the observations. Following \citet{villenave2019}, we derive for each gap a minimum planet mass (M$\rm_{p}$ (min)), assuming that the scattered light edge is at the half distance between the peak of the disk in PDI and the gap minimum, and a maximum planet mass (M$\rm_{p}$ (max)), taking the scattered light edge as the position of the peak in PDI. We use as $R\rm_{p}$ the location of the gap in scattered light. In the cases where $f(M\rm_{p})$ is close to 1, the predicted mass is below 1\,$M\rm_{Jup}$. This regime is outside the range of the simulations, and observational uncertainties in the derivation of the peak locations would be of critical importance. We thus limit ourselves to the 1--15\,$M\rm_{Jup}$ mass range, and treat these values as the upper and lower thresholds of the resulting planet masses outside these limits. We also extrapolate the value of $\Gamma$ to the location of the gaps in our sample, assuming that it behaves linearly at distances >60\,au within the 1--15\,$M\rm_{Jup}$ mass range.

\begin{table*}
\caption{\label{tab:ALMA}Perturbers' masses creating SPHERE/PDI-ALMA substructures}
\small
\centering
\begin{tabular}{lccccccc}
\hline\hline
        Host ID &    Min PDI &    Peak PDI  &    Peak mm   &  M$\rm_{p}$ (min) &  M$\rm_{p}$ (max) & AMES-DUSTY &  BEX-HOT \\
            &    (au) &    (au) &      (au)  &   ($M\rm_{Jup}$)     &  ($M\rm_{Jup}$)   &  ($M\rm_{Jup}$) &  ($M\rm_{Jup}$) \\
\hline
       HD 135344B &   12 &   24 &   51 &  12.5 &  $>$15 &   5.8 &  7.0 \\
        RX J1604        &   33 &   66 &   90 &   2.2 &   9.5 &  5.1 &  6.4 \\
           PDS 70 &   27 &   54 &   75 &   2.1 &   8.5 &  4.1 &  5.3 \\
    HD 34282 &   44 &   89 &  138 &   4.8 &  $>$15 &    10.0 &  11.8\\
        HD 169142 &   10 &   20 &   25 &   1.2 &   4.1 &  8.1 &  9.5\\
          LKCA 15 &   43 &   59 &   66 &   $<$1 &   1.6 &   9.7 &  12.3 \\
         HD 97048 &  127 &  188 &  189 &   $<$1 &   1.2    &  3.9 &  5.5 \\
        HD 169142 &   44 &   63 &   64 &   $<$1 &   1.1 &   3.4 &  4.1 \\
         HD 97048 &   45 &   54 &   55 &   $<$1 &   $<$1 &   5.6 & 7.5\\

\hline
\end{tabular}
\tablefoot{Minimum and maximum mass of the perturbers, M$\rm_{p}$ (min) and M$\rm_{p}$ (max), respectively, calculated using Eq. \ref{eq:ovelar}. The derived detection limits at the perturber's location using AMES-DUSTY and BEX-HOT tracks (best of ANDROMEDA/IRDIS and ASDI/IFS) are shown in the last two columns. References for the location of the substructures can be found in Appendix \ref{appendix_A}.}
\end{table*}

\begin{figure}
\centering
\setlength{\unitlength}{\textwidth}
\hbox{\hspace{-0.15cm}\includegraphics[width=0.5\textwidth]{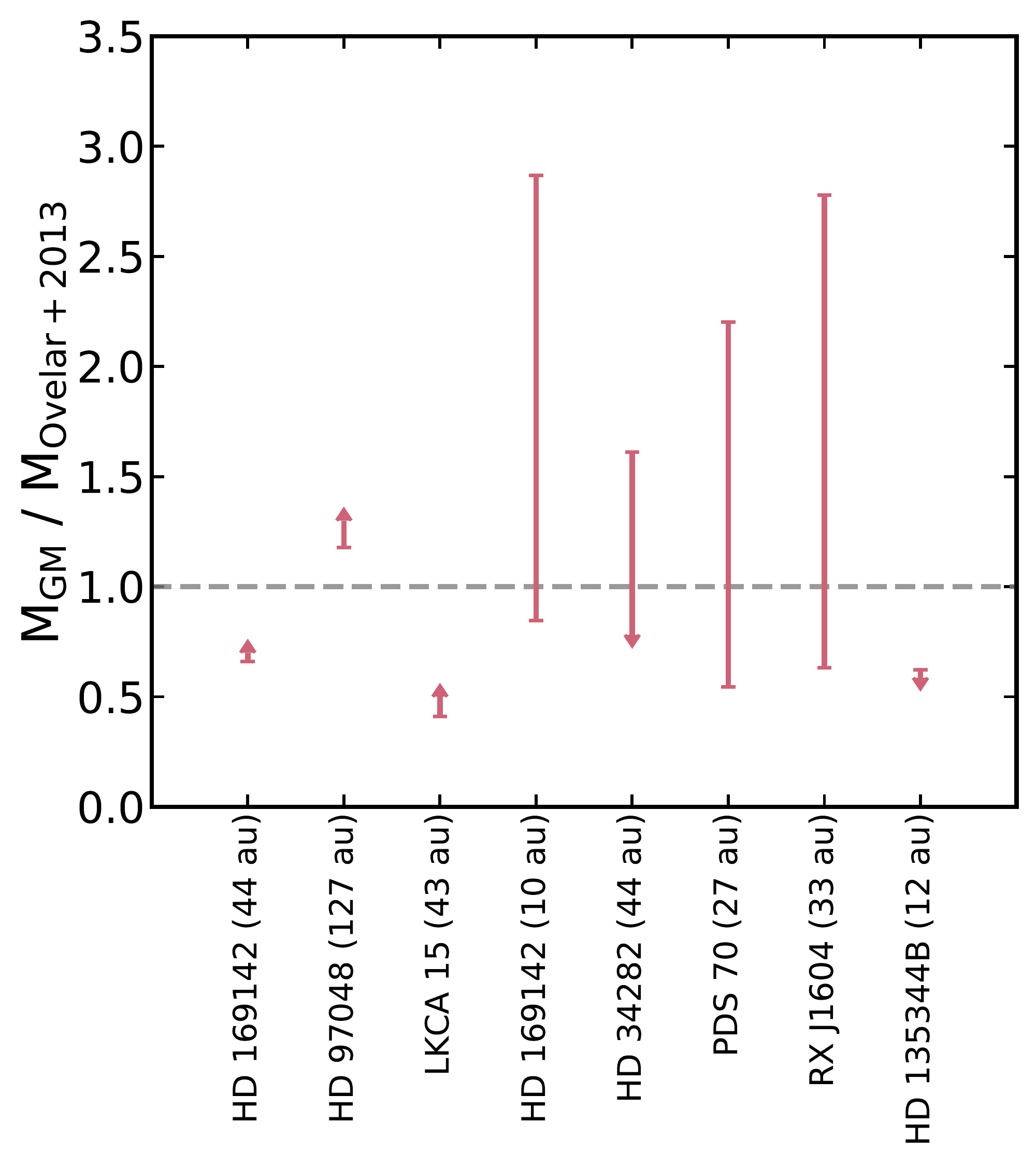}}
\caption{Difference in the estimated mass of the perturbers of Figure \ref{fig:hill} (M$\rm_{GM}$), calculated using the gap morphological features observed only in scattered light, and those calculated via \citet{dejuanovelar2013}, which compares the cavity sizes in scattered light and the mm continuum (M$\rm_{Ovelar+2013}$). The only uncertainties that have been assumed are those linked to the location of the scattered light inner edge of equation \ref{eq:ovelar}.  }
\label{fig:ovelar-hill}
\end{figure}

From this approximation, we see in Figure \ref{fig:juanovelar} that perturbers within large cavities are close to detection under AMES-DUSTY conditions (RX J1604, PDS 70 and HD 135344B). Again, planets in the gaps between a pair of rings such as HD 97048 and HD 169142 seem to be at least about half an order of magnitude lower in mass than currently detectable. The derived masses for these systems are also in good agreement to those derived in the last section from Hill radii or estimated from simulations that relate the mass to the morphological features of the gaps (see Tables \ref{tab:promising} and \ref{tab:ALMA}). In Figure \ref{fig:ovelar-hill} we show this difference for those substructures carved by 1--15\,$M\rm_{Jup}$ planets. For most of these perturbers, the mass derived using the ALMA observations is typically within a factor of $\sim$\,0.4 and 1.5 of that obtained via the Hill radius criterion. We do not show here the potential planet creating the gap at $\sim$\,45\,au around HD 97048, as only an upper limit can be constrained with Equation \ref{eq:crida}, but both approaches yield a planet mass well below 1\,$M\rm_{Jup}$.

\section{Discussion}\label{sec:discuss}

\subsection{Why is the current detection rate so low?} \label{sec:detection_rates}

These homogeneously-derived detection limits in mass have clear implications on the detectability of potential companions in very young PPDs. Indeed, our results suggest that, if planets creating substructures in PPDs have a very hot start, the majority of them are probably less massive than $\sim$\,4\,$M\rm_{Jup}$  (see Figure \ref{fig:CCs}). The mass detection limits are however affected by a factor $\sim$\,3 if planets were formed via cold core-accretion models, which is a consequence of the very young age of our sample. If we compare our sensitivities to the SPHERE/SHINE survey (see Figure B.1 in \citet{vigan2020}), we see that their limits are moved towards higher masses, as SHINE is comprised mostly of $\sim$\,50-Myr old stars which are members of nearby young moving groups \citep{desidera2021sphere}. Moreover, the luminosities at the beginning of formation have less of an impact on their final results because the hot and cold populations tend to converge to almost identical luminosities at ages $>$\,50\,Myr \citep{baraffe2003,marley2007, spiegel2012, mordasini2017, linder2019}, contrary to our results for $<$10\,Myr targets.

The detection limits derived in this work are in line with the non-detection of sub-Jupiter-mass planets in gaps within rings, but our derivation of planet masses in cavities seem to suggest the presence of $\sim$\,5\,$M\rm_{Jup}$ companions, which would be close to a detection in several systems. Despite of this, only two bona-fide giant protoplanets have been imaged, both of them in the cavity of PDS 70 \citep{mesa2019}, although some other candidates have been proposed in HD 100546 \citep{brittain2014, quanz2015}, LkCa 15 \citep[see][]{ sallum2015} and HD 169142 \citep{gratton2019}, but remain uncertain \citep[e.g.,][]{rameau2017,currie2019}.

An explanation for the low detection rate of giant planets in PPDs could be that the perturbers' masses are overestimated. Indeed, the problem of estimating masses from the observed morphologies in the dust is still complex; current analytical equations describing the gap can only be used after making several assumptions, such as the disk temperature, viscosity, gas accretion effects, dust evolution or migration. Simulations of planet-disk interactions also need to account for more complex and realistic thermodynamics that account for the cooling efficiency of the disk, which would impact the morphology and the number of gaps that a planet can produce \citep[e.g, ][]{szulagyi2017, miranda2020}. Including more physics in the theoretical simulations in fact suggests that the derived masses by locally isothermal relations might be underestimated. Under the isothermal assumption, for a given planet mass the gap is wider than in the case where the disk does not cool rapidly. This implies that the derived masses by current isothermal relations might be underestimating the real mass of the perturbers, as even small planets could open up wide gaps. The same conclusion is found if gas accretion is taken into account \citep{camille2020}. 

If the planet is not on a circular or coplanar orbit, the dynamical mass estimates (as Equations \ref{eq:depth} and \ref{eq:width}) would also be modified, e.g., a planet on an eccentric orbit may open a wider but shallower gap \citep{muley2019}. This assumption on the planet orbit is mostly justified on the basis of parameter restriction on the theoretical simulations, and it is not clear how this could affect when applied to a large sample. For instance, PDS 70b is found to be on an eccentric orbit of $e \sim 0.17$ \citep{wang2021}, and the presence of misaligned inner disks, such the one around J1604 in this sample, might point to the presence of a companion in a highly-inclined orbit \citep{mayama2012, pinilla2018}.

Nevertheless, our masses derived via gap morphology and those calculated with ALMA data seem to agree, and are probably representative of the order of magnitude of the perturbers' mass. For the special case of PDS 70, which we treated as a planet-less PPD, the derived mass of a single perturber carving its cavity at 27\,au has a mass of 4.9\,$M\rm_{Jup}$ using the Hill radii prescription, or between 2.1--8.5\,$M\rm_{Jup}$ following equation \ref{eq:ovelar}. This is comparable to the mass of PDS 70b or c (both of $\sim$\,5\,$M\rm_{Jup}$) at 23 and 30\,au, respectively \citep{mesa2019}. If accretion luminosity is taken into account as a contributor to the observed luminosity, the mass of PDS 70b decreases to about 1\,$M\rm_{Jup}$ \citep{stolker2020}. Assuming that PDS 70c follows a similar pattern, the combined mass of both planets will be in the range 2--3\,$M\rm_{Jup}$. This fact may indeed be affecting the current number of detections in cavities; as claimed by \citet{dodson2011}, several undetected lower-mass planets of mass $\geq$\,1\,$M\rm_{Jup}$ might be creating the large holes in optically thin transition disks, instead of a single very massive companion. Assuming the presence of more than one perturber inside of a cavity would greatly affect the mass estimates of Equation \ref{eq:hill}, which lead to the assumption that \textit{Rim} systems have more massive planets than \textit{Ring} systems, with the only exception of LkCa 15. There is not enough information in scattered light yet to derive the mass of multi-planet systems in cavities from their observed morphology, but simulated scattered light images of cavities opened by multiple Jovian planets seem to resemble the observations \citep[see Figures 5 and 7 in][]{dong2015}. 

Another possibility is that some (or most) of these planets reradiate the energy influx generated during the gas accretion shock, and consequently are better described by cold starts, which is not theoretically supported \citep{szulagyimordasini,marleau2019b}. In this scenario, in direct imaging surveys we would have detected only those that are born with high entropies, but the bulk of the planets that create substructures, which would be formed by a cold process, remains undetected after the PPD phase \citep{vigan2020}. However, planets of $<$10\,Myr may still be actively accreting the material that resides within the gap, inside the circumplanetary disk from which the planet feeds. In this case of a cold formation pathway, the resulting accretion luminosity radiated away by these planets would be easily detectable, and could be even brighter than the planet irradiation itself, especially at near-IR wavelengths \citep{zhu2015, szulagyi2019, sanchis2020}. \citet{brittain2020} circumvent this apparent inconsistency by proposing a episodical accretion of circumplanetary disks, in which accretion outbursts where planets could be detected only occur during 2\,$\%$ of the runaway accretion phase. 

Nonetheless, the number of systems with resolved depleted cavities or spiral patterns in scattered light which also have high-contrast total intensity observations is still low, as shown in this work. Detection probabilities are also hampered by projection effects in PPDs that are not seen face-on, especially at close semimajor axes (see the detection probability maps in Appendix \ref{appendix_C}), and the emission of disk residuals in the ADI data, which are sometimes difficult to distinguish from real companions \citep{sissa2018, gratton2019}. Further observations of young cavity-hosting disks in both polarised and ADI modes would help constrain all these different scenarios, and to understand whether other theoretical interpretations not related to the presence of perturbers might be necessary to explain the various morphologies, such as effective dust growth at snowlines \citep{pinilla2017}, variations in the ionisation level of the gas \citep{flock2015}, or photoevaporation winds \citep{owen2012}.

\subsection{Extinction effects}
\label{subsec:extinction}

Throughout this study we have assumed that the observed emission of these young planets will not be affected by the presence of disk material along the line of sight. However, all forming planets have to be surrounded by circumplanetary material regardless of their mass, since they still accrete. As long as they do it, there has to be material around to accrete from, which means that planetary atmospheres can well be extincted, shifting the emission to longer wavelengths. If an extended cloud of material around the planet exists, observations might not even be detecting the planet itself, but the reprocessed emission of the dust \citep[e.g.,][]{szulagyi2019, stolker2020}. As seen for PDS 70b, even with 1--5\,$\upmu$m photometry and spectroscopy, the protoplanet nature of these objects makes it difficult to disentangle the intrinsic planet emission from that of the circumplanetary environment, and whether the dust reddening the spectrum is located in the planet's atmosphere, in the circumplanetary envelope or in the circumstellar disk \citep{mueller2018, haffert2019, mesa2019, chris2019, stolker2020, wang2020}.

If the bulk of the perturbers creating the scattered-light gaps between rings have masses of a fraction of that of Jupiter as the results suggest, they might not open deep enough gaps to clear all the disk material, and the extinction could indeed completely impede their detection in the near-infrared. According to 3D high-resolution hydrodynamical simulations of disk-planet interaction by \citet{sanchis2020}, their emission (intrinsic and accretion shock luminosity) can be highly attenuated by the surrounding dust (of about $\sim$13 mag in $H$ band and $\sim$4 in $L$ at locations of $\sim$100\,au for a 1\,$M\rm_{Jup}$ companion), depending on disk aspect ratio, surface densities, viscosities, dust processing and planet formation parameters. However, when the undepleted disk surface density is low, even low-mass objects seem to suffer little extinction at any distance in the PPD. For instance, in PDS 70 the infrared extinction has an incidence in $H$ band only for planets $<$\,2\,$M\rm_{Jup}$ at distances of $<$\,40\,au, and due to the very low surface density of the PPD around CQ Tau \citep{ubeira2019}, extinction is only somewhat relevant for 1 Jupiter-mass planets within 20\,au, with an $L$-band extinction of 0.3\,mag \citep{sanchis2020}. For the three potential sub-Jupiter planets carving the gaps in TW Hya, \citet{vanboekel2017} estimated an extinction in $H$ band of about 2 magnitudes during the late detached phase, and for the same system \citet{sanchis2020} increased the mass upper limits obtained by \citet{ruane2017} only from 1 to 2\,$M\rm_{Jup}$, when extinction is taken into account. \citet{maire2017} also found that the detection mass limits beyond the spirals in HD 135344B might be underestimated by not more than 2\,$M\rm_{Jup}$. These results suggest that, depending on the individual system, the derived mass sensitivities to low-mass perturbers depleting the $\upmu$m-sized dust may be too optimistic, with a critical effect in dense PPDs, and a low or moderate influence in PPDs with a lower surface density.

Perturbers with masses higher than 4\,$M\rm_{Jup}$ will not suffer substantially from extinction in any band or distance, according to \citet{sanchis2020}, even for very dense disks, because the material is cleared within the gap. However, those simulations do not count with a well-resolved planet vicinity, which means that the density of material is lower than it would be with a fully-resolved CPD \citep[see][]{szulagyi2019}. In addition, they assume a face-on configuration of the disk, whereas the circumstellar disk could distort the vision of the planet for inclined systems \citep[e.g.,][]{szulagyigarufi}. For these reasons, these values can be better considered as lower limits for extinction. 

In summary, we expect that the emission originating in the potential $>$\,4\,$M\rm_{Jup}$ planets carving the cavities will not be heavily obscured, as it has been the case for PDS 70b \citep{mesa2019}. Hence, in the majority of situations perturbers susceptible of being importantly extincted ($>$2\,mag) in the $H$ and $K$ bands are in any case too low mass ($\leq$\,2\,$M\rm_{Jup}$) to be detected by the SPHERE detection limits, while more massive planets that are within SPHERE's reach would probably suffer less obscuration. A more detailed luminosity estimate of the forming planets would consist in coupling evolutionary models of `naked' planets (as in Figure \ref{fig:hill_sensitiv}) with individual high-resolution hydrodynamical simulations of the disk.

\subsection{Potential for future detection}

\begin{figure}
\centering
\setlength{\unitlength}{\textwidth}
\includegraphics[width=0.5\textwidth]{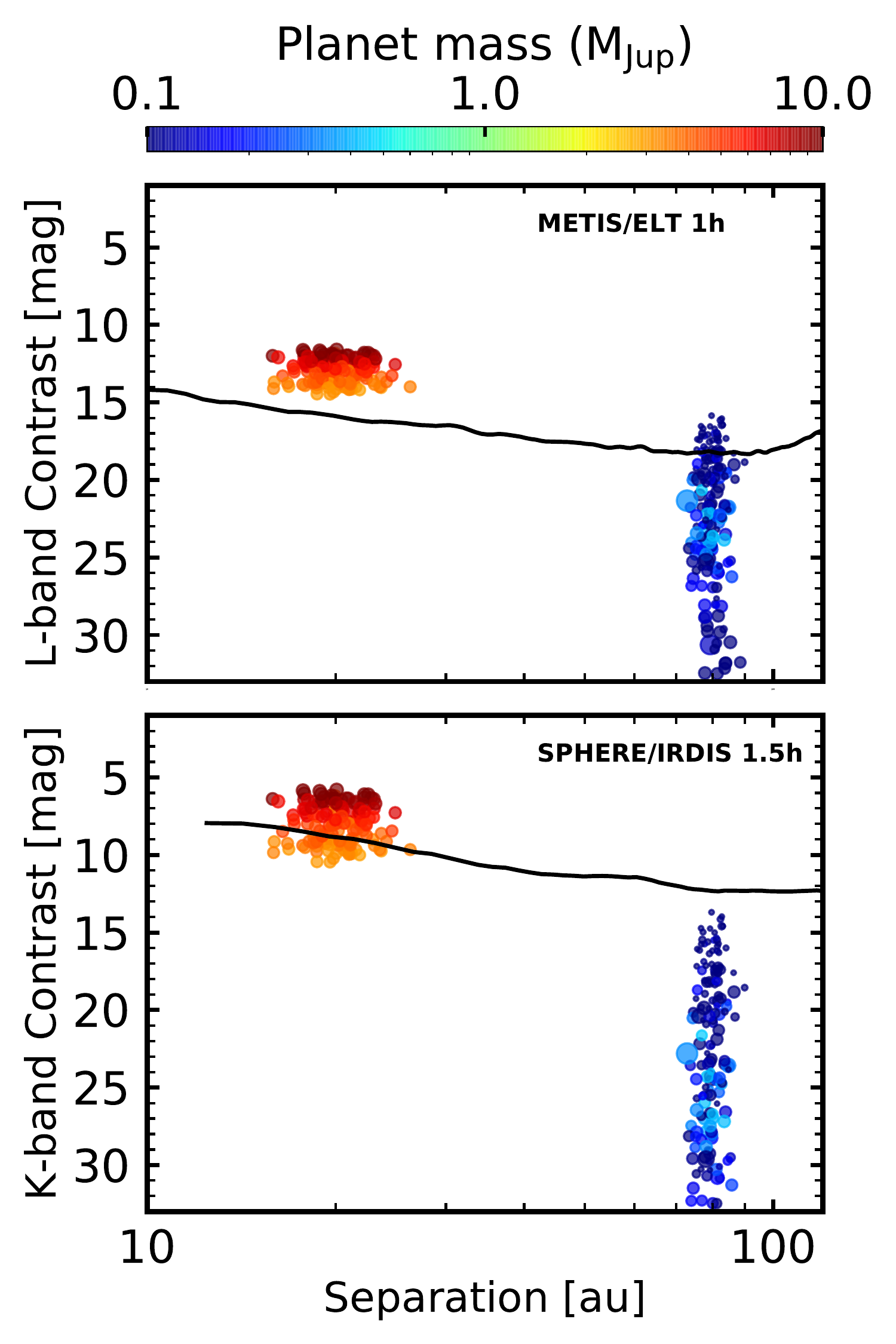}
\caption{Contrast magnitudes for 8\,Myr-old planets in the NG76 population synthesis model \citep{emsenhuber2020a,emsenhuber2020b}. Two populations of 0.05--0.5\,$M\rm_{Jup}$ and 3--10\,$M\rm_{Jup}$ planets are artificially located at 80 and 20\,au, respectively. Estimated extinction coefficients have been taken from \citet{sanchis2020}, see text. The METIS contrast curve has been obtained from \citet{carlomagno2020}, and the SPHERE limits correspond to the derived contrast in this work for RX J1604, which we adopt as a representative host star. Small random shifts in separation are applied for a better visibility of the populations, and dot sizes are connected to planet radius.  }
\label{fig:metis}
\end{figure}

Based on the results of the two previous sections, giant planets in cavities seem to be within the reach of current high-contrast imagers such as SPHERE, and constitute the most promising places where to image a giant planet. Planets carving the gaps between rings however have masses of about $\sim$0.1\,$M\rm_{Jup}$, similar to those of the undetected planetary-mass companions responsible for the mm-dust gaps seen in high-resolution ALMA observations in the DSHARP survey \citep{zhang2018} and in the Taurus star-forming region \citep{long2018}. Even without considering the presence of extinction, this level of contrast is unattainable for current instruments (see Figures \ref{fig:hill_sensitiv} and \ref{fig:histo_hill}).

If the surface density depletions are indeed caused by the presence of embedded sub-Jupiter mass companions, their detection might be more favorable at longer wavelengths. A future instrument that would enable their detection is the Mid-Infrared E-ELT Imager and Spectrograph \citep[METIS; ][]{brandl2014}. The top panel of Figure \ref{fig:metis} shows the detection limit in $L$ band of METIS after 1\,h of integration, compared to two different populations of synthetic planets; 0.05--0.5\,$M\rm_{Jup}$ companions that might potentially be creating gaps in the outer disks, which we locate at 80\,au, and more massive 3--10\,$M\rm_{Jup}$ objects opening up cavities at 20\,au. Planet parameters have been obtained from the synthetic population model NG76\footnote{\url{https://dace.unige.ch/populationAnalysis/?populationId=ng76}} \citep{emsenhuber2020a, emsenhuber2020b} for a median age of our sample of 8\,Myr.
We considered objects with an atmosphere and in the detached evolutionary phase or later, and calculated their approximate brightnesses in $L$ band assuming that the planets radiate as blackbodies, following \citet{vanboekel2017}. The simulated contrast curve was taken from \citet{carlomagno2020}, and uses a ring-apodized vortex coronagraph (RAVC). To calculate the contrast between the host star and the synthetic population, we have assumed a host star such as RX J1604, an early K dwarf located at the median distance of our sample. We also include extinction coefficients as derived by \citet{sanchis2020} for an unperturbed surface density of $\Sigma$ = 127\,g/cm$^{2}$, consistent with the densest disks in star forming regions; we take their values for a 5\,$M\rm_{Jup}$ planet at 20\,au (see their Table A.2) and for a 1\,$M\rm_{Jup}$ planet at 100\,au (see their Table A.4), and apply them to the in-cavity and in-rings planets, respectively. The sizes of the dots are correlated with the planet radius.

We see that METIS will be able to detect giant planets forming the resolved cavities in scattered light with masses $>$\,3\,$M\rm_{Jup}$, and possibly also a small fraction of those in the gaps between rings in the systems where extinction is low. On the lower panel of Figure \ref{fig:metis} the same calculations are made for SPHERE/IRDIS in $K$ band. As expected, it shows that the current instrumentation is in the limit of providing giant planet images within cavities, but the bulk of the sub-Jupiter-mass planets is out of reach.

This implies that the use of complementary pipelines and the weather conditions during the observations may now have a critical effect in detecting giant planets for which instruments such as SPHERE are close to the detection limis \citep[see, e.g., Fig A.1 in][]{keppler2018}. Obtaining the best possible observing conditions for the most promising systems in Table \ref{tab:promising}, together with detailed and varied reduction schemes, could be the most favorable case for a new protoplanet detection in a PPD in the near future.

Direct coronagraphic imaging at long wavelengths ($>$\,3\,$\upmu$m) with the James Webb Telescope (JWST) will also be useful for detecting and characterizing young Saturn-like planets at projected distances $>$\,1\,$\arcsec$ \citep[e.g.,][]{beichman2010}. For nearby and large PPDs, JWST may be able to provide the most constringent mass limits of planets creating substructures at hundreds of au, down to $\sim$\,0.2\,$M\rm_{Jup}$ around dwarf K and M host stars \citep{carter2020}, and also to characterize their atmospheric compositions and planet parameters, which would in turn shed light on their formation pathway \citep{linder2019}.

Finally, another way to look for these perturbers is to detect hydrogen emission lines, such as H$\alpha$, originating in the shock-heated gas during the accretion of disk material \citep{aoyama2019}. If detected, this emission constitutes the main observational signature of the ongoing formation of a substellar object \citep{eriksson2020}. Observations of PPDs with SPHERE/ZIMPOL and MUSE have so far been unable to detect planetary-mass companions around PPDs hosts other than the pair around PDS 70, despite reaching potential sensitivities down to 1\,$M\rm_{Jup}$ \citep{haffert2019, zurlo2020}.

\section{Conclusions}\label{sec:conclusion}

In this work we have presented a homogeneous ADI data reduction of 15 PPDs observed with SPHERE which are characterised by rings, cavities or spirals in scattered light. We provided detection limits to low-mass companions creating these substructures, and studied the difference between the current SPHERE mass limits and the estimated sensitivity that would be needed to achieve a detection in these systems. The main results of this paper can be summarised as:

\begin{enumerate}

    \item  We find that the detection limits in our PPD sample (K and earlier-type host stars of $\lesssim$\,10\,Myr) are greatly affected by the assumed initial luminosity of the perturber, with median sensitivities that vary for the hottest AMES-DUSTY starts from $\sim$\,9 to 4\,$M\rm_{Jup}$, respectively for perturbers in close ($\sim$10\,au) and wide ($\sim$100s of au) orbits. If a cold start based on core accretion formation is considered, these detection limits are deteriorated by a factor of $\sim$\,3, and sensitivities to perturbers barely reach the planetary-mass regime. An assessment of the detection limits in mass for each individual system can be found in Appendix \ref{appendix_B}.\\

    \item We performed an estimation of the mass of the currently undetected perturbers via three approaches: 1) Literature simulations based on the scattered light morphology of the disk; 2) Gap width in scattered light proportional to the Hill radius of the planet; 3) Location of the scattered light cavity compared to its radius in ALMA mm continuum observations. Assuming that one perturber is responsible for creating one substructure, the underlying population of planets is located between $\sim$\,10--400\,au and within a mass range of 0.01–1 $M\rm_{Jup}$ for gaps in between rings, and more massive companions of up to $\sim$\,3–10 $M\rm_{Jup}$ in cavities and spirals. They represent the potential objects that would be needed to create the current substructures seen in young disks, which have so far no direct correspondence to any detected population of exoplanets.\\
    
    \item We compared the estimated masses to the obtained detection limits at the perturbers' positions. We find that SPHERE is about one order of magnitude away in mass from detecting the majority of planets creating the gaps in between scattered-light rings. If the presence of material attenuating the planet emission is important, this difference could be more pronounced. Perturbers creating cavities tend to be more massive, and current SPHERE detection limits seem to be good enough for potential detections, consistent with the discovery of two giant planets in PDS 70.\\
    
    \item  Besides PDS 70, we find that the cavities in RX J1604, RXJ1615, Sz Cha, HD 135344B and HD 34282 are the most promising systems for a future direct detection of giant planets. Perturbers of the order of a Jupiter mass might also be found in between the rings of PDS 66 and HD 97048.\\
    
    \item Future imaging instruments such as ELT/METIS and the JWST may be able to explore the hottest population of planets depleting the $\upmu$m-sized dust between rings. These surveys will help to constrain the initial luminosities of the perturbers and link planet masses to gap morphologies.\\
    
    \item Current dynamical mass constraints assume the presence of a single planet on a coplanar and circular orbit in an isothermal disk. This premise might not be universally realistic, which could lead to overestimated mass derivations. Further work in this area exploring a larger parameter space in the properties of planets would help reduce these uncertainties on the planet masses creating substructures in the scattered light.

\end{enumerate}

\begin{acknowledgements}
R.A-T and T.H acknowledge support from the European Research Council under the 
Horizon 2020 Framework Program via the ERC Advanced Grant Origins 83 24 28. P.P. acknowledges support provided by the Alexander von Humboldt Foundation in the framework of the Sofja Kovalevskaja Award endowed by the Federal Ministry of Education and Research. G-DM acknowledges the support of the DFG priority program SPP 1992 `Exploring the Diversity of Extrasolar Planets' (KU 2849/7-1) and from the Swiss National Science Foundation under grant BSSGI0$\_$155816 `PlanetsInTime'. Parts of this work have been carried out within the framework of the NCCR PlanetS supported by the Swiss National Science Foundation. We also thank the referee, whose comments significantly improved the manuscript. This work has made use of data from the European Space Agency (ESA) mission
{\it Gaia} (\url{https://www.cosmos.esa.int/gaia}), processed by the {\it Gaia}
Data Processing and Analysis Consortium (DPAC,
\url{https://www.cosmos.esa.int/web/gaia/dpac/consortium}). Funding for the DPAC
has been provided by national institutions, in particular the institutions
participating in the {\it Gaia} Multilateral Agreement. This research made use of Astropy,\footnote{http://www.astropy.org} a community-developed core Python package for Astronomy \citep{astropy:2013, astropy:2018}.

\end{acknowledgements}


\bibliographystyle{aa_url} 
\bibliography{biblio} 

\begin{appendix}
\section{Notes on individual targets}
\label{appendix_A}

\noindent
\textbf{HD 135344B:} Disk with spiral and inner cavity. PDI resolved in \citet{stolker2016}, showing an inner cavity at $\sim$\,24\,au and a clear spiral pattern, from which we take the inclination and PA. To explain the spiral arms, many different predictions exist involving the presence of a massive companion. For instance, \citet{dongfung2017a} modelling suggests a giant $\sim$\,5--10\,$M\rm_{Jup}$ on a wide orbit $\sim$\,100\,AU to explain the arm contrast in scattered light, which is in agreement with the results of the symmetry-based method by \citet{fung2015}. \citet{vandermarel2016} proposes another possibility with a planet located inside the inner gap at 30\,au that produces a vortex further out, which gives rise to the spiral pattern. Here we simply consider the outer massive planet from \citet{dongfung2017a} (calculated with a distance to the target of 140\,au, instead of our updated 135\,au value from GAIA-EDR3). We also consider a planet located in the middle of the 24\,au cavity using the Hill radius approach. Although very close to the coronagraphic mask, this distance corresponds roughly to 0.1\,$\arcsec$, at which position we obtain the sensitivity limit. ALMA observations from \citet{cazzoletti2018} show a mm-dust cavity at 50\,au and a crescent peaking at $\sim$\,80\,au.  \\

\noindent
\textbf{HD 139614:} This is a complicated system in scattered light, with shadows and a system of misaligned inner rings, probably due to the presence of a misaligned companion. A bright ring is seen down to $\sim$\,16\,au or 0.12\,$\arcsec$ \citep{muroarena2020}, together with three arcs at larger separations that might be caused by variations in the scale height profile. No clear gaps in the scattered light are detected, and for this reason no perturber is derived for this system. Mass detection limits can be found in Appendix \ref{appendix_B}. We adopt the inclination and position angle of the outer disk. ALMA data are not available yet.
\\

\noindent
\textbf{HD 97048:} Four gaps and rings are seen in scattered light by \citet{ginski2016}, whose values we adopt after GAIA-EDR3 distance correction. For gap 2 we assume the derivation from \citet{dongfung2017b} of 1.3\,$M\rm_{Jup}$ for $\alpha$ = 10$^{-3}$ (although we note that they used a distance to the system of 158\,pc), which is in line with the kinematic detection of a 2--3$M\rm_{Jup}$ planet candidate in that same gap at $\sim$\,130\,au \citep{pinte2019}. Inclination and position angle are from the same study. Ring 1 and 2 have direct correspondence in the mm at 55\,au \citep[peak at 0.3\,$\arcsec$,][]{vanderplas2017a} and 189\,au \citep{francis2020}. \\

\noindent
\textbf{PDS 66:} Scattered light disk is part of the DARTSS-S survey \citep{avenhaus2018}. A clear ring-like structure can be seen at a best-fit value of 85\,au. There is no significant deficiency in polarized signal close to the coronagraph; actually a bright compact region out to 25\,au is detected, which rules out the presence of a highly depleted inner cavity of $\upmu$m-sized grains. We thus classify this object as a ring and put the putative planet in the faint region between the compact inner structure and the ring, i.e., at 55\,au, which in turn corresponds to an apparent discontinuity seen in the data. Inclination and position angle from this work. To our knowledge, there is no high-resolution ALMA data for this system yet. \\

\noindent
\textbf{LkCa 15:} This is a transitional disk resolved by SPHERE in  \citet{thalmann2016}, with the outer disk located at $\sim$\,58\,au and a gap minimum at 43\,au, where we locate the planet. Inclination and position angle from this work. A perturber of mass 0.5\,$M\rm_{Jup}$ for $\alpha = 10^{-3}$ seems necessary to carve the cavity in $\upmu$m-sized dust \citep{dongfung2017b}. ALMA observations show a dust cavity of 66\,au \citep{jin2019}.\\

\noindent
\textbf{RX J1615.3-3255:} This is a transitional disk showing three different rings in PDI at 44, 166 and 232\,au \citep{deboer2016, avenhaus2018}. Ellipses are fit to these three substructures, providing locations and geometry. We classify this disk as a ring, but treat the inner rim as a cavity in terms of planet-carving structure in Hill radii. The gap at $\sim$78\,au (labelled as G in \citet{deboer2016}) can be created by a planet of mass 0.2\,$M\rm_{Jup}$ for $\alpha = 10^{-3}$, according to \citet{dongfung2017b}. The third planet is put in the middle of the location between the two outer rings. There is also an additional ring-like structure between the two inner rings seen by \citet{avenhaus2018}, only seen on the northeastern side, that we do not consider here. ALMA observations also show three rings at closer separations than seen in the small dust (Benisty, M. priv. comm), so we do not use their position to derive planet masses.  \\

\noindent
\textbf{HD 169142:} The $J$ band PDI data by \citet{pohl2017} reveal a close to face-on disk with a double-ring structure with a central cavity of $\sim$\,20\,au. The two rings are fitted with an ellipse, whose parameters we adopt here. Using dust evolution models, they found that two giant planets of 3.5 and 0.7\,$M\rm_{Jup}$ can reproduce the disk structure in SPHERE/PDI and the location and width of the gap, which we adopt. A dedicated SPHERE ADI study of \citet{gratton2019} finds a potential accreting planet candidate between the two rings at 38\,au, detected in reflected polarized light and with a mass of $\sim$\,2\,$M\rm_{Jup}$ for an age of 5\,Myr. A point-like structure is also at $\sim$\,10\,au by \citet{ligi2018}.This disk is classified as Ring, even though we treat the inner ring as a cavity. The system has been observed by ALMA at high resolution, showing the inner cavity and the ring at 25\,au, but also the outer ring as composed of three narrow components. In this comparison between scattered light and mm emission, we link the outer ring in scattered light to the central narrow ring component resolved by ALMA and peaking at 64\,au \citep{perez2019}. \\

\noindent
\textbf{HD 100546:} This PPD presents a complex structure of rings, arms and spirals. First imaged in polarimetric scattered light with the ZIMPOL instrument of SPHERE in the optical by \citet{garufi2016}, and later with IRDIS in the near-infrared by \citet{sissa2018}. A $\sim$\,13\,au (0.12\,$\arcsec$) cavity close to the coronagraphic mask is detected, together with a complex structure of wrapped arms beyond 1\,$\arcsec$ that may form a spiral, and three additional small arm-like structures closer in. The presence of two planetary-mass companions has also been suggested \citep[e.g.,][]{brittain2014, quanz2015}, but their nature is still debated \citep{rameau2017}. We derive no perturber masses for this system given its intrincate morphology in scattered light, but present detection limits in Appendix \ref{appendix_B}. High-resolution ALMA observations by \citet{pineda2019} detect a ring-like structure between 20--40\,au, which can be reproduced with a vortex and two rings peaking at 21 and 30\,au. An unresolved, central compact emission is also detected. We take the inclination and position angle from that study. \\

\noindent
\textbf{PDS 70:} This transitional disk hosts two giant planets carving a wide gap of 54\,au \citep{keppler2018, mueller2018, haffert2019}. For our calculations, in order to provide a comparison, we do not consider that this system host the planets at those positions. An inner bright structure below 17\,au is observed, but not resolved. We thus put our potential planet in the centre of the outer ring at 27\,au. Inclination and PA are taken from \citet{hashimoto2012}. Follow-up high resolution observations with ALMA emission found the outer cavity peaking at 75\,au in the dust continuum, and an inner ring at 10\,au \citep{keppler2019, francis2020}.\\

\noindent
\textbf{Sz Cha:} This disk has been observed in IRDIS $H$-band DPI mode. It is characterised by three different rings at semi-major axis between 0.2\,$\arcsec$ and 0.6\,$\arcsec$. We adopt here the ellipse fitting parameters derived by Hagelberg et al. (priv. comm.), and classify the disk as Ring, treating the inner one as a cavity. We are not aware of current observations of this object in the mm emission with enough resolution to observe substructures.  \\

\noindent
\textbf{HD 34282:} Scattered light image from \citet{deboer2020b}. Two rings seems to be well-fitted, the inner one (rim R2) at semi-major axis of $\sim$\,89\,au. The second one after scale height deprojection seems to be more coincident with a single-armed spiral. There seems to be no simulations of planets causing the spiral in scattered light yet. Inclination and PA are from this paper. ALMA data from \citet{vanderplas2017b} see a ring in the continuum extending from 75\,au to 359\,au, but peaking at the deprojected radius of 138\,au. According to their work, a 50 $M\rm_{Jup}$ object at 0.1\,$\arcsec$ ($\sim$\,30\,au) could explain the ALMA ring. We thus classify this object as \textit{Spiral}, and use the inner rim to derive a planetary mass inside the scattered-light cavity via Hill radius, which will be similar to the ALMA planet candidate. \\

\noindent
\textbf{RX_J1604.3-2130A:} This transition disk shows variable dips and a ring peaking at 66\,au in scattered light \citep{pinilla2018}. The dips might be originated by the presence of an inner misaligned ring. The cavity is wider in ALMA observations, which reveal an elliptical fit to the outer disk peaking at 90\,au  in the dust continuum \citep{mayama2018}.\\

\noindent
\textbf{HD 36112:} This transitional disk with an inner eccentric cavity is resolved in the mm wavelengths out to $\sim$\,50\,au in deprojected distance \citep{dong2018}, but not in PDI with SPHERE down to $\sim$\,15\,au \citep{benisty2015}. Scattered light observations however show a pair of spirals with a large opening angle from $\sim$0.25--0.45\,$\arcsec$. The object is thus classified as Spiral, and the inclination and position angle are taken from \citet{isella2010}. \citet{baruteau2019} have simulated planet masses and locations that create these substructures. They take into account previous upper limits on planet sensitivities \citep[e.g.,][]{reggiani2018}, and reproduce the spirals in scattered light and the sub-mm crescent substructure with two planets of mass 1.5\,$M\rm_{Jup}$ and 5\,$M\rm_{Jup}$ located at 35\,au and 140\,au (for a distance of 160\,pc), inside and outside the spirals, respectively. We assume these simulated planets as the ones creating the substructures and compare them to our sensitivities. As no resolved inner cavity is seen with SPHERE, we do not use this object for the ALMA-Scattered light comparison. \\

\noindent
\textbf{TW Hya:}  SPHERE shows a face on disk with three wide albeit shallow gaps in the $\upmu$m-sized dust at approximately 7, 22 and 90\,au \citep{vanboekel2017}. This work derives the planet masses carving these non fully-depleted gaps from their depths following \citet{duffell2015}. We scale these values for $\alpha$ = 10$^{-3}$ and obtain 0.04\,$M\rm_{Jup}$ for the innermost gap. We rely on \citet{dongfung2017b} for the masses of the perturbers causing the two outer gaps, 0.15 and 0.08\,$M\rm_{Jup}$, respectively. We treat the three gaps as rings, as ALMA observations in the sub-mm show a ring at $\sim$\,2\,au. The $\sim$22\,au depression is also observed by ALMA in the surface brightness of the continuum \citep[e.g.,][]{huang2018}, but no clear correspondence with brightness peaks can be done between the millimetre and scattered light. Even though \citet{vanboekel2017} assumed a face-on disk, here we take the inclination and position angle as derived by \citet{qi2004} from CO imaging of the outer disk.\\

\noindent
\textbf{CQ Tau:} SPHERE PDI data shows a spiral pattern extending up to $\sim$0.4\,$\arcsec$, but a cavity is not resolved outside the coronagraphic mask (Benisty et al. in prep.). NIRC2 observations in $L$ band excluded the presence of giant planetary-mass companions down to 5\,$M\rm_{Jup}$ outside the spiral region \citep{uyama2020}. \citet{ubeira2019} reported ALMA data showing a depleted cavity with a rim at $\sim$\,57\,au. Inclination and position angle are taken from this work.

\clearpage

\begin{figure*}
\centering
\setlength{\unitlength}{\textwidth}
\hbox{\hspace{-0.cm}\includegraphics[width=0.9\textwidth]{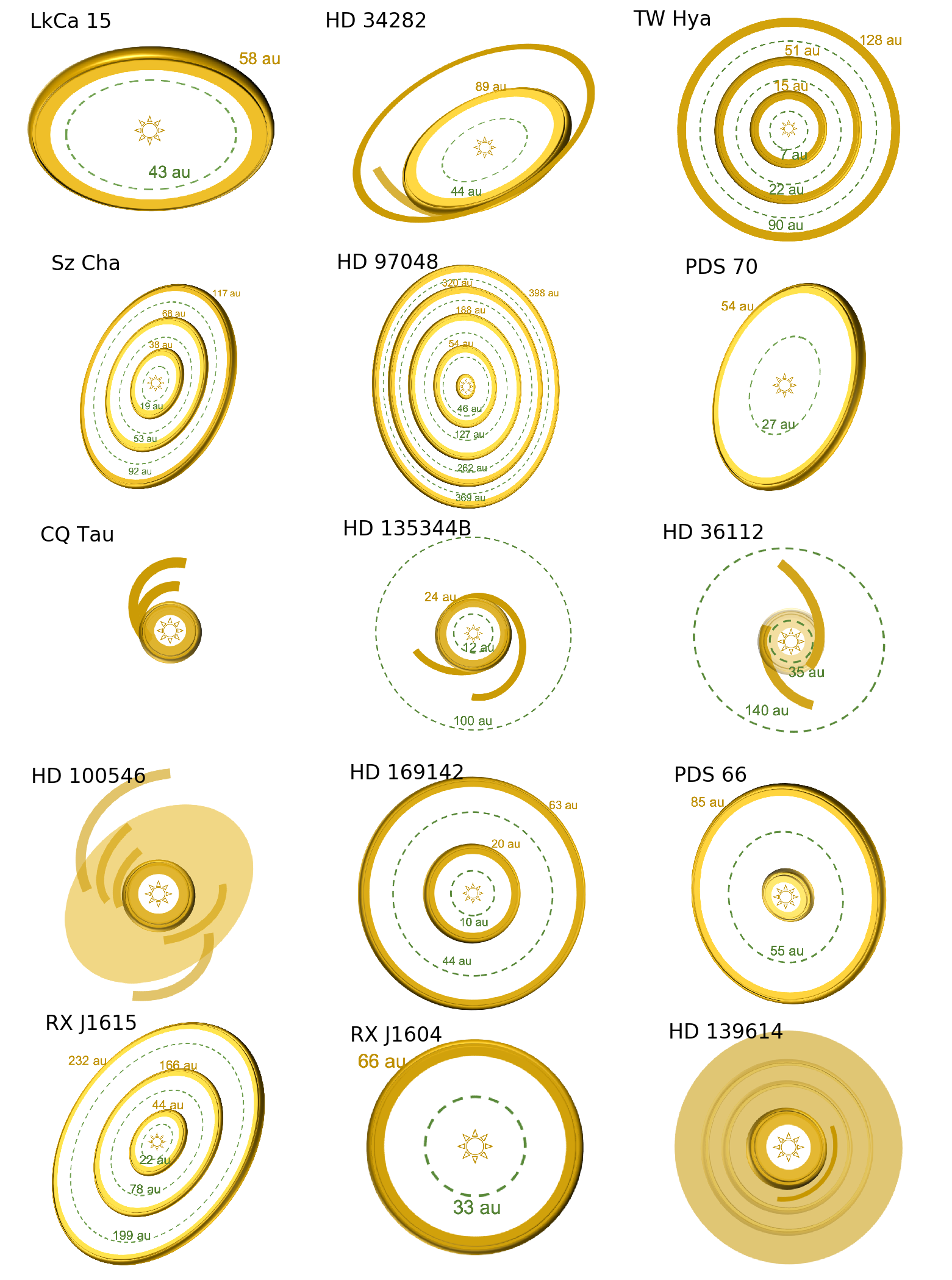}}
\caption{Sketch representation of the scattered light appearance of the PPD sample. We include only the gaps and substructures that were considered as to be created by perturbers in Section \ref{sec:perturbers}. The adopted locations of these perturbers are shown as green dashed curves. Scattered light rings are illustrated with thick rings, while arms and spiral patterns are shown as flat curled curves. The disks are projected according to their inclination and position angle in Table \ref{tab:adi}. We also include HD 100546, HD 139614 and CQ Tau for completeness, although no clear gaps or simulated planets creating the spiral pattern exist. }
\label{fig:drawing}
\end{figure*}

\clearpage

\section{Deprojected mass sensitivity curves}
\label{appendix_B}

Here we show the detection limits in mass for all the targets analysed in this work, with three different initial luminosity conditions. Deprojected separations have been obtained as explained in Section \ref{sec:data}. The location of the various substructures observed in both scattered light and mm continuum (see Appendix \ref{appendix_A}) are also overplotted.

\begin{figure}[H]
\centering
\setlength{\unitlength}{\textwidth}
\hbox{\hspace{-0.1cm}\includegraphics[width=0.49\textwidth]{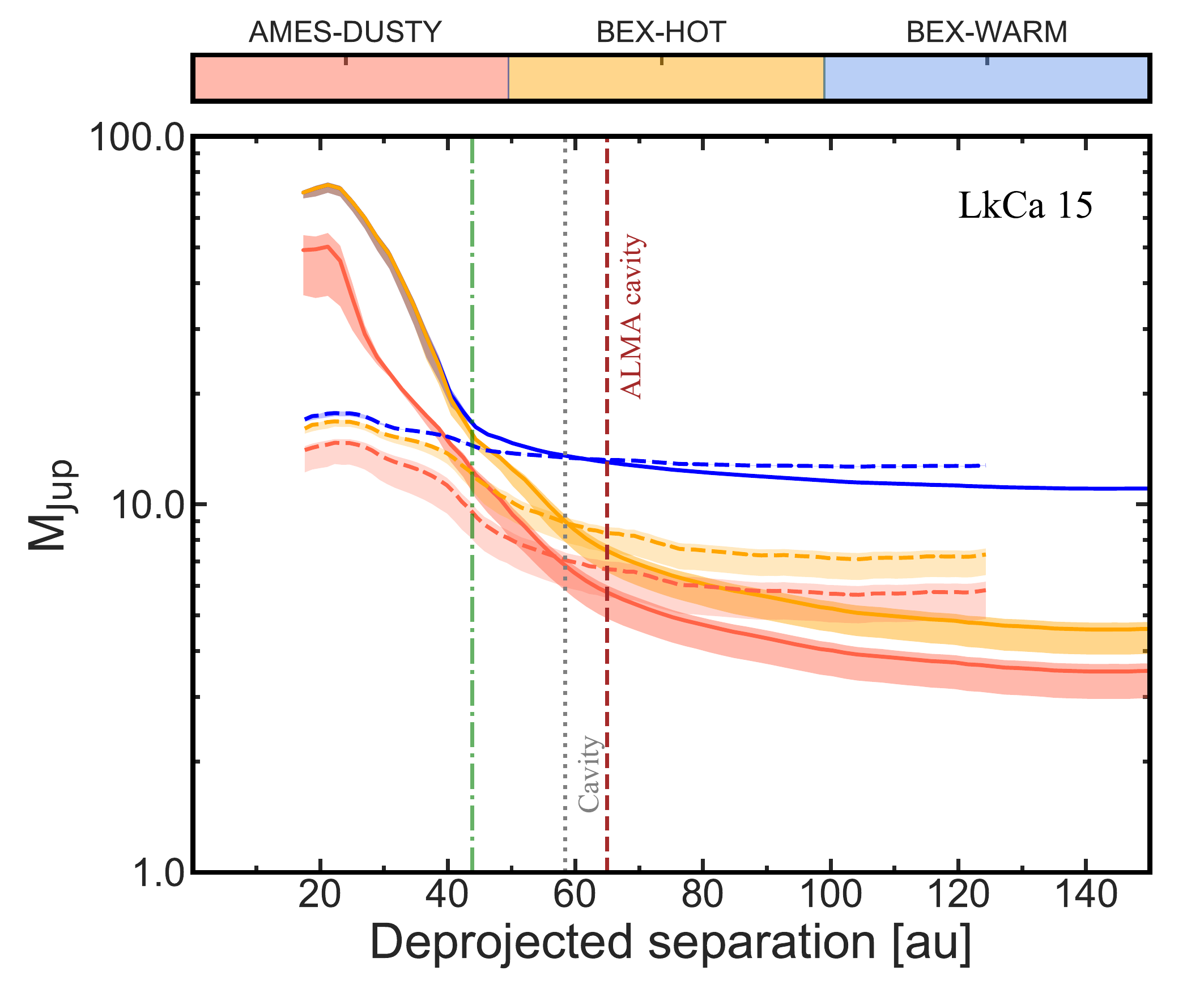}}
\caption{5$\sigma$ sensitivity to planet masses in LkCa 15. The different initial luminosities for the derived age in Table \ref{tab:host_param} are shown as red, orange and blue curves, respectively for the AMES-DUSTY, BEX-HOT and BEX-WARM models. Thick curves correspond to the ANDROMEDA IRDIS limits, and dashed curves show the IFS-ASDI performance using the SPHERE mode of Table \ref{tab:adi}.  Shaded regions encompass the upper and lower limit of the age uncertainty. SPHERE PDI substructures are shown in gray as vertical dotted lines, and the observed peaks in ALMA dust continuum are shown as brown dashed lines. Only ALMA rings with correspondence to scattered light substructures are shown. The green dashed-dotted vertical lines indicate reported gap locations in PDI.}
\label{fig:LkCa_15}
\end{figure}

\begin{figure}[H]
\centering
\setlength{\unitlength}{\textwidth}
\hbox{\hspace{-0.1cm}\includegraphics[width=0.49\textwidth]{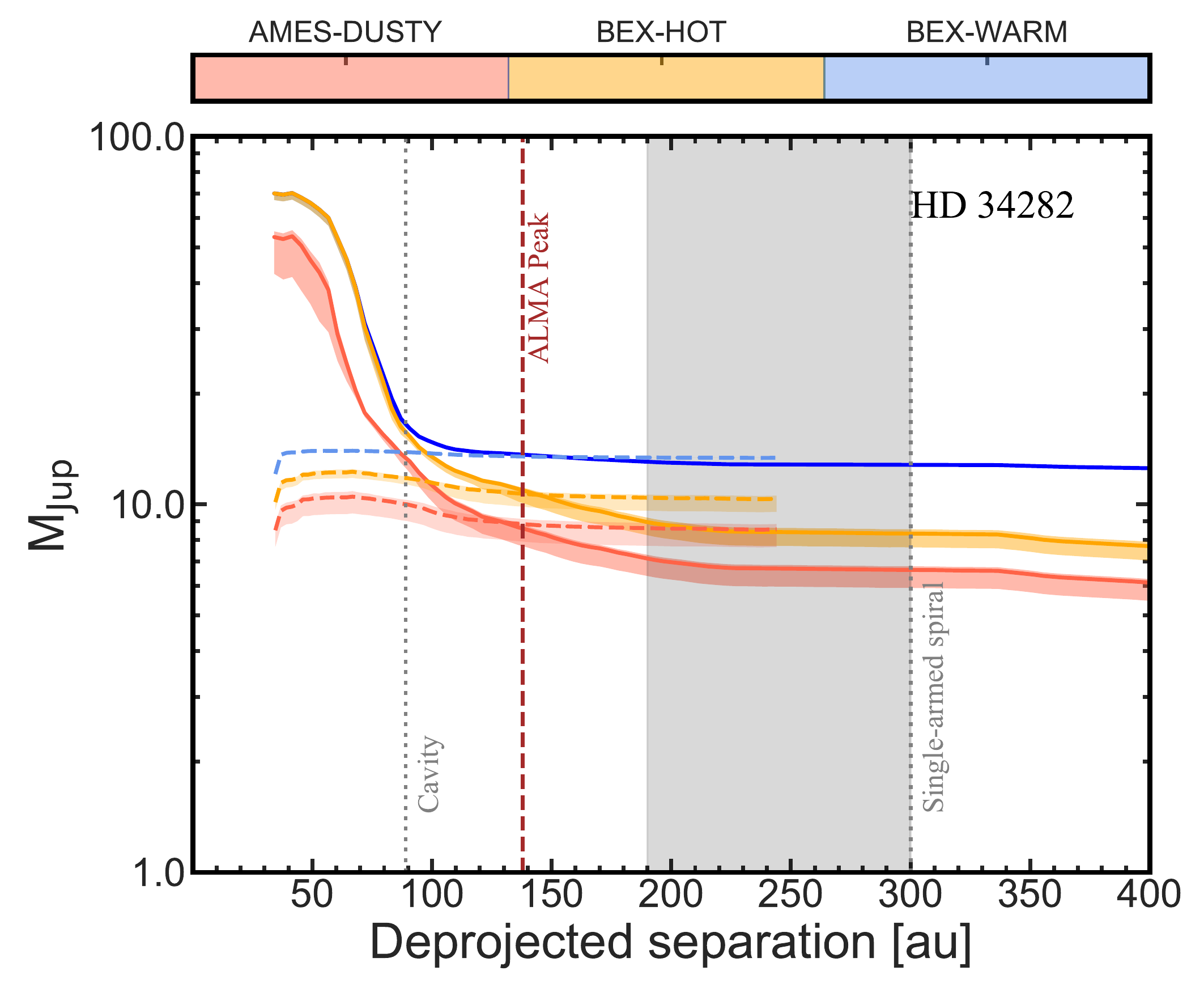}}
\caption{5$\sigma$ sensitivity to planet masses in HD 34282 (see Figure \ref{fig:LkCa_15}). The vertical gray-shaded region corresponds to the spatial extension of the spiral feature.}
\label{fig:HD_34282}
\end{figure}

\begin{figure}[H]
\centering
\setlength{\unitlength}{\textwidth}
\hbox{\hspace{-0.1cm}\includegraphics[width=0.49\textwidth]{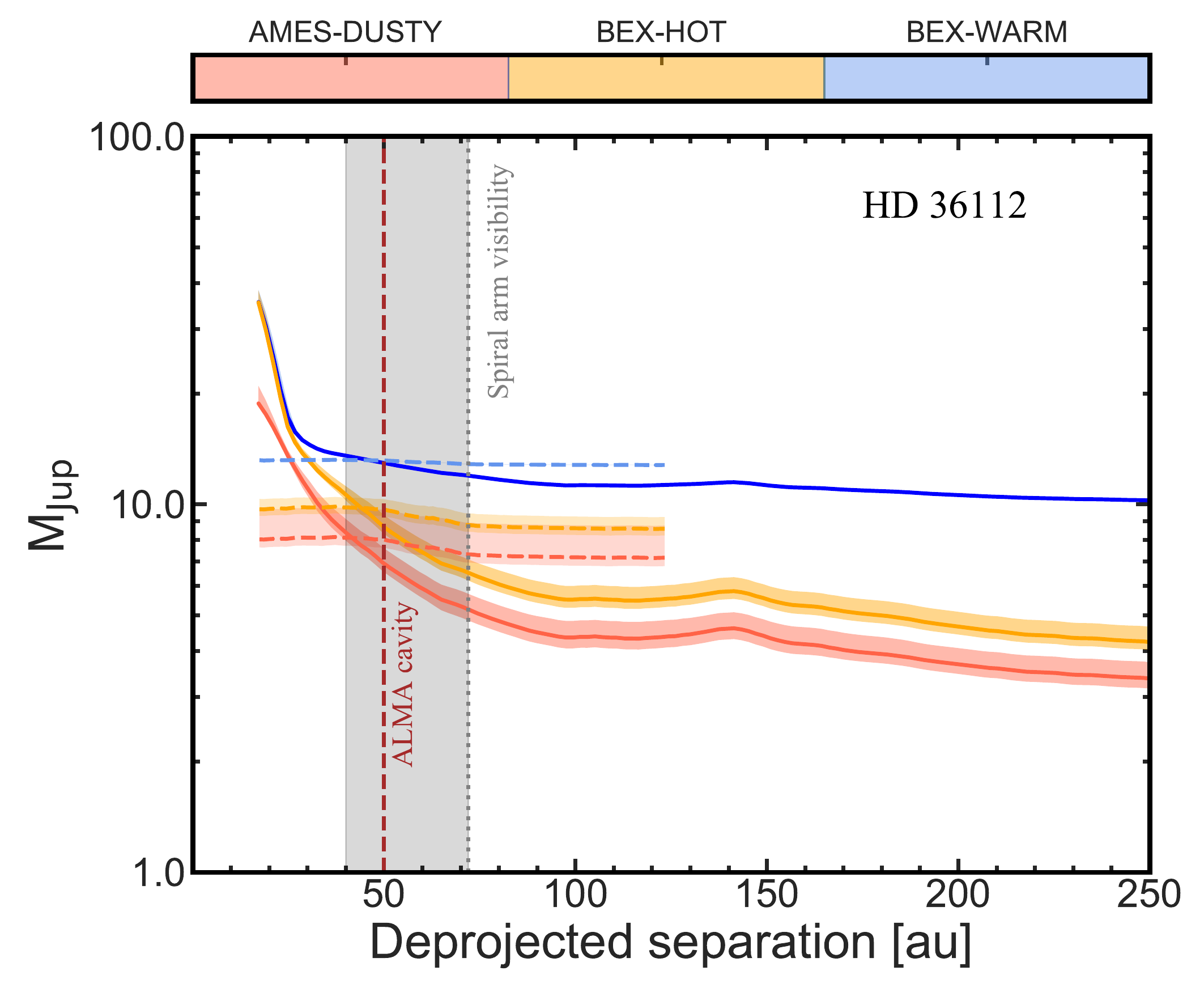}}
\caption{5$\sigma$ sensitivity to planet masses in HD 36112 (see Figure \ref{fig:HD_34282}).}
\label{fig:HD_36112}
\end{figure}

\begin{figure}[H]
\centering
\setlength{\unitlength}{\textwidth}
\hbox{\hspace{-0.1cm}\includegraphics[width=0.49\textwidth]{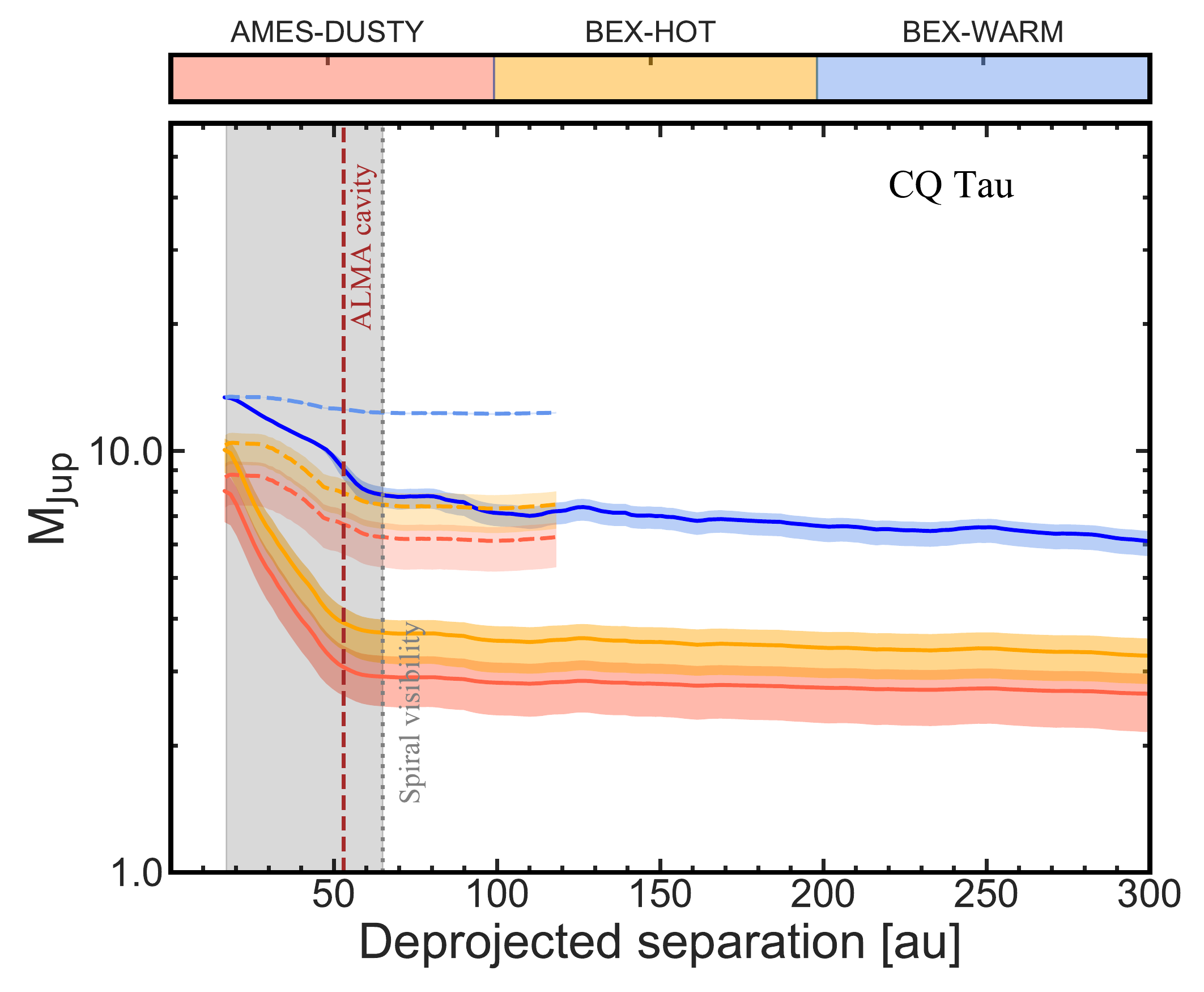}}
\caption{5$\sigma$ sensitivity to planet masses in CQ Tau (see Figure \ref{fig:HD_34282}).}
\label{fig:CQ_Tau}
\end{figure}

\begin{figure}[H]
\centering
\setlength{\unitlength}{\textwidth}
\hbox{\hspace{-0.1cm}\includegraphics[width=0.49\textwidth]{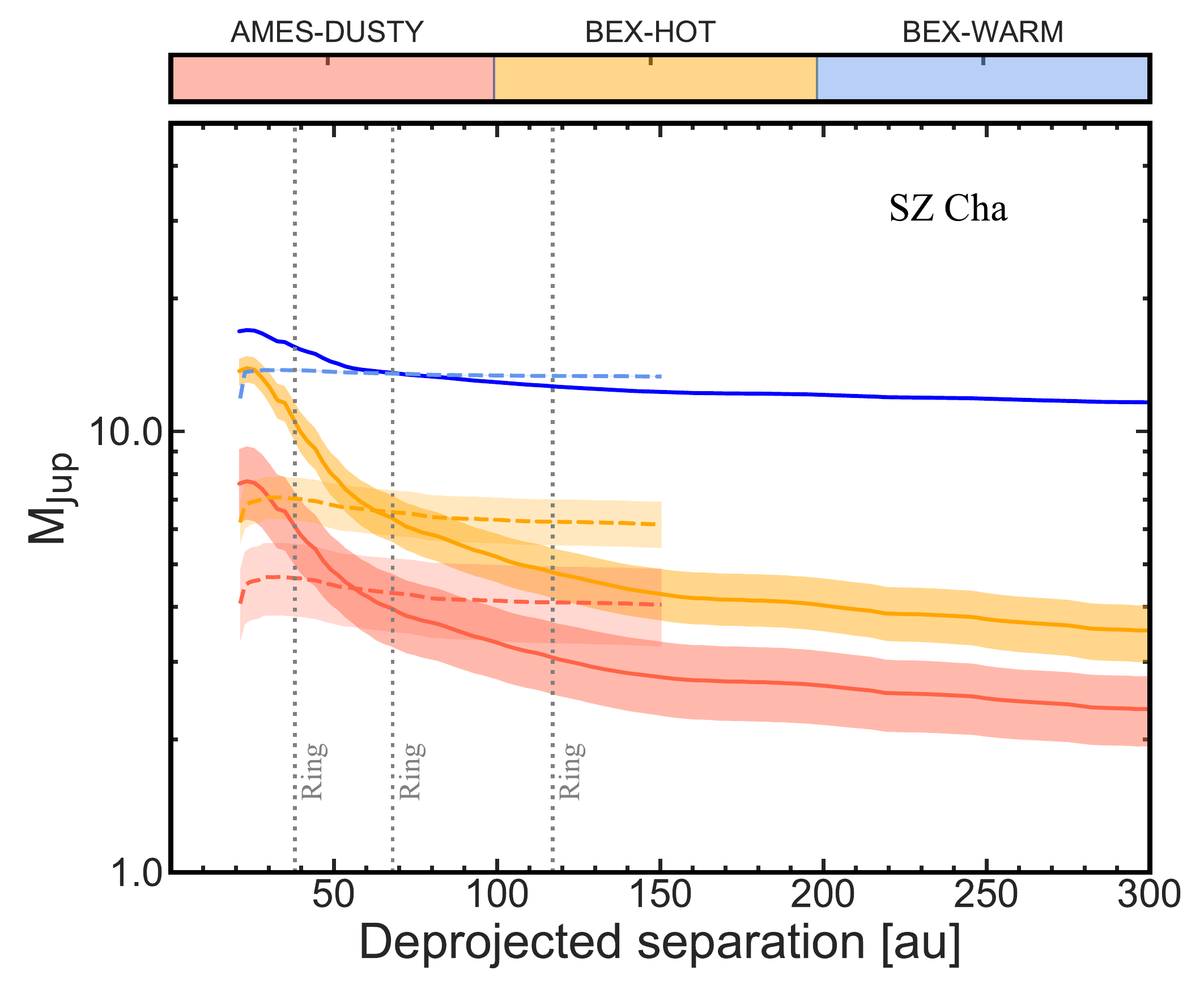}}
\caption{5$\sigma$ sensitivity to planet masses in Sz Cha (see Figure \ref{fig:LkCa_15}).}
\label{fig:SZ_CHA}
\end{figure}

\begin{figure}[H]
\centering
\setlength{\unitlength}{\textwidth}
\hbox{\hspace{-0.1cm}\includegraphics[width=0.49\textwidth]{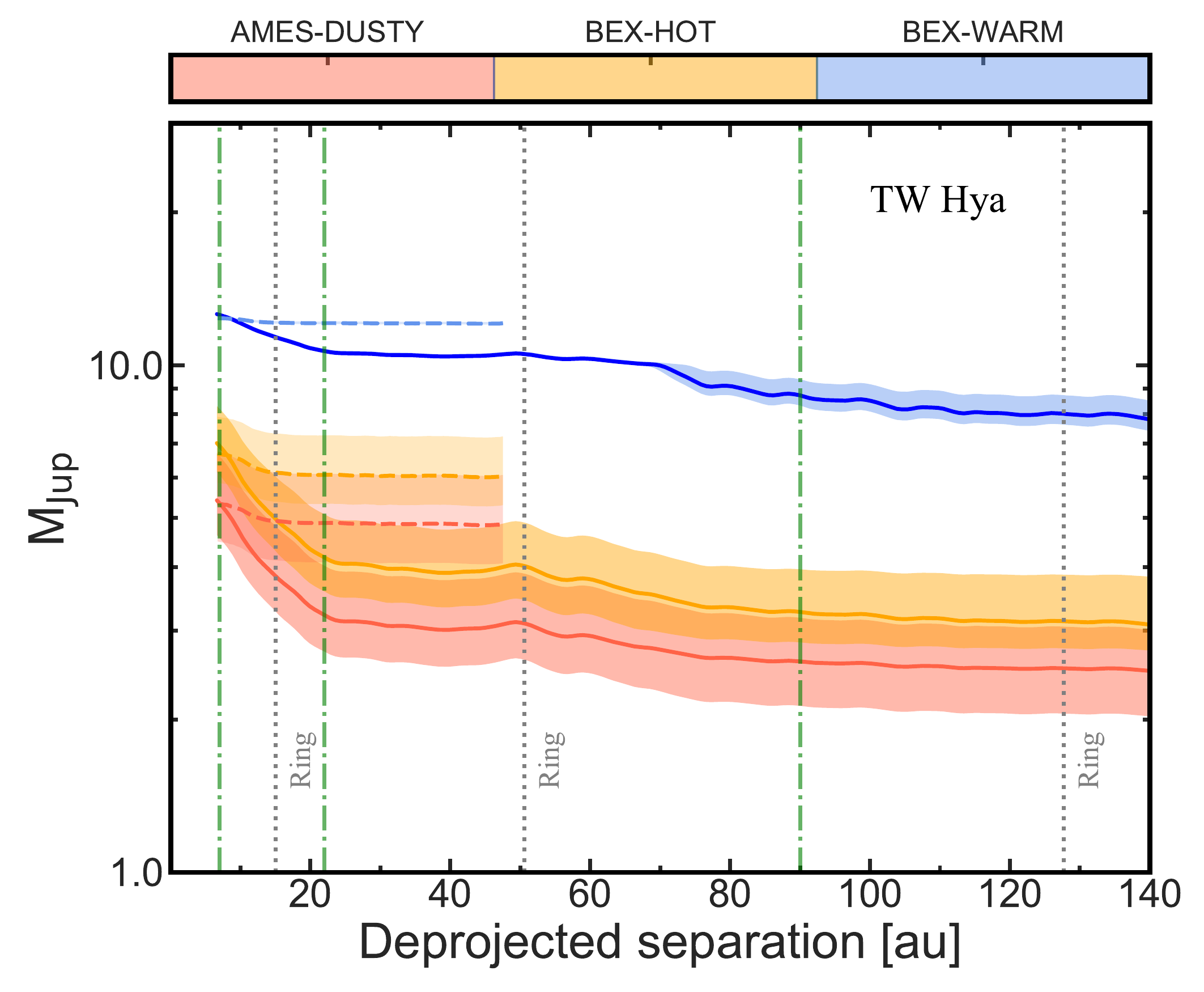}}
\caption{5$\sigma$ sensitivity to planet masses in TW Hya (see Figure \ref{fig:LkCa_15}).}
\label{fig:TW_Hya}
\end{figure}

\begin{figure}[H]
\centering
\setlength{\unitlength}{\textwidth}
\hbox{\hspace{-0.1cm}\includegraphics[width=0.49\textwidth]{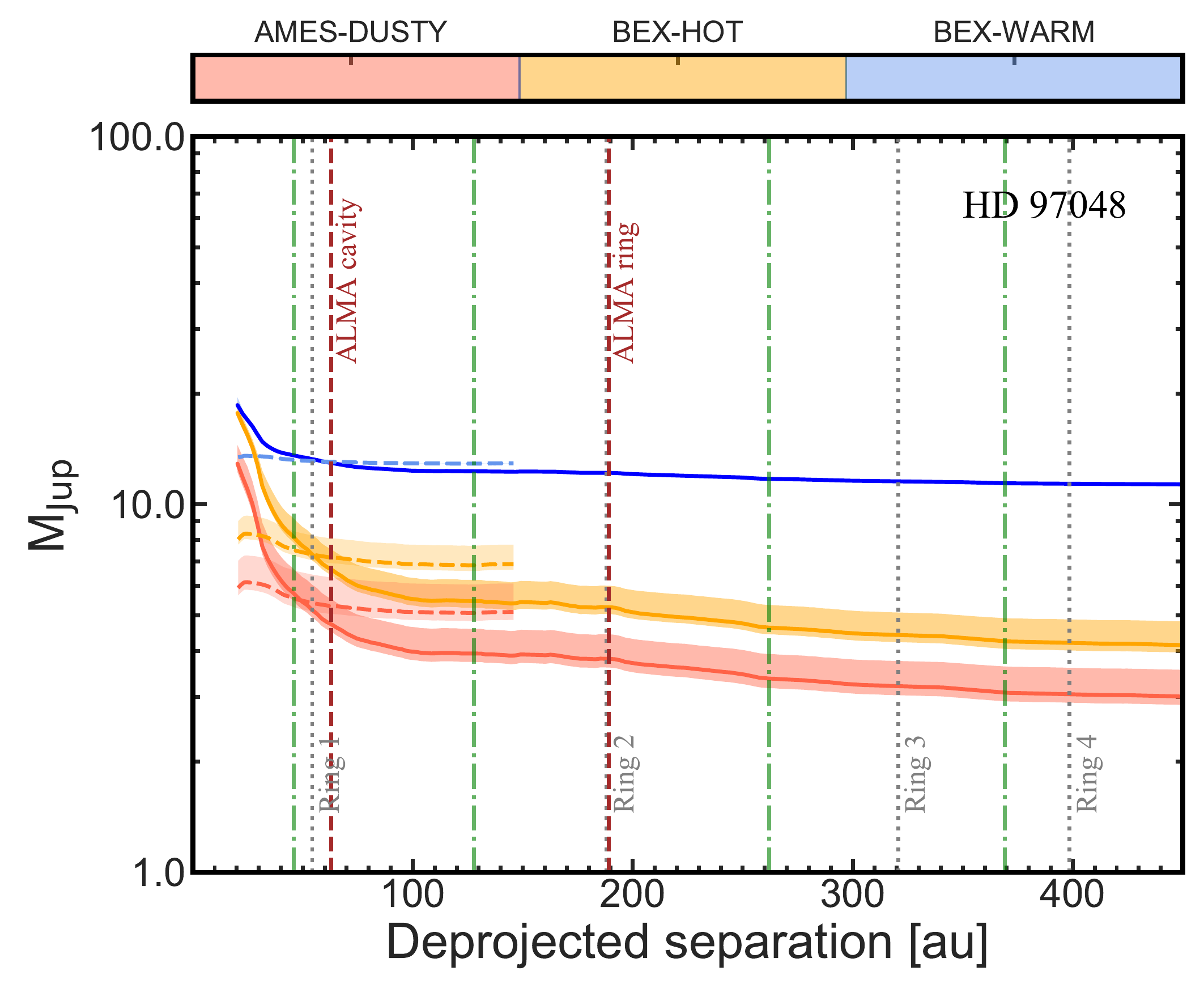}}
\caption{5$\sigma$ sensitivity to planet masses in HD 97048 (see Figure \ref{fig:LkCa_15}). }
\label{fig:HD_97048}
\end{figure}

\begin{figure}[H]
\centering
\setlength{\unitlength}{\textwidth}
\hbox{\hspace{-0.1cm}\includegraphics[width=0.49\textwidth]{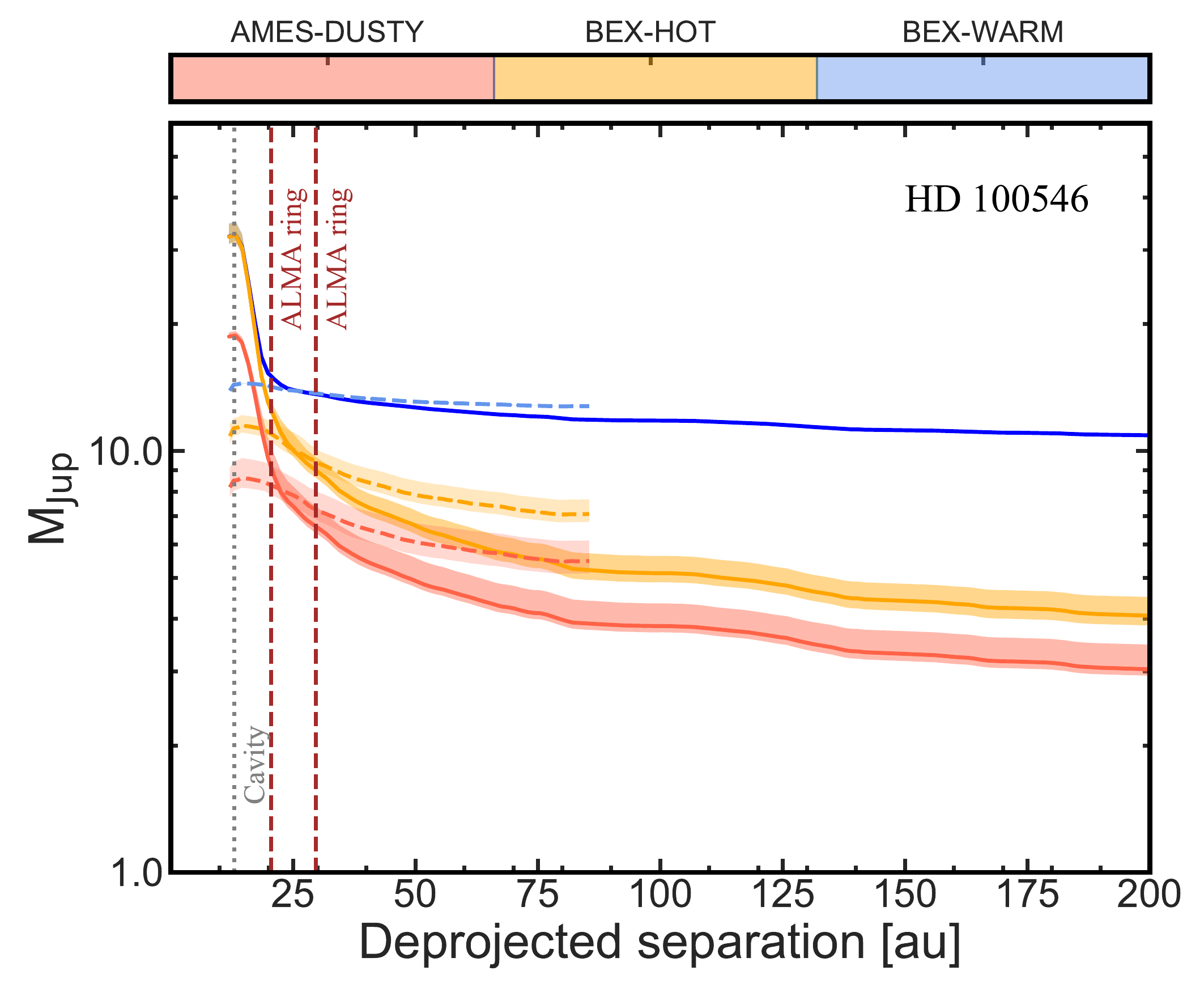}}
\caption{5$\sigma$ sensitivity to planet masses in HD 97048 (see Figure \ref{fig:LkCa_15}).}
\label{fig:HD_100546}
\end{figure}

\begin{figure}[H]
\centering
\setlength{\unitlength}{\textwidth}
\hbox{\hspace{-0.1cm}\includegraphics[width=0.49\textwidth]{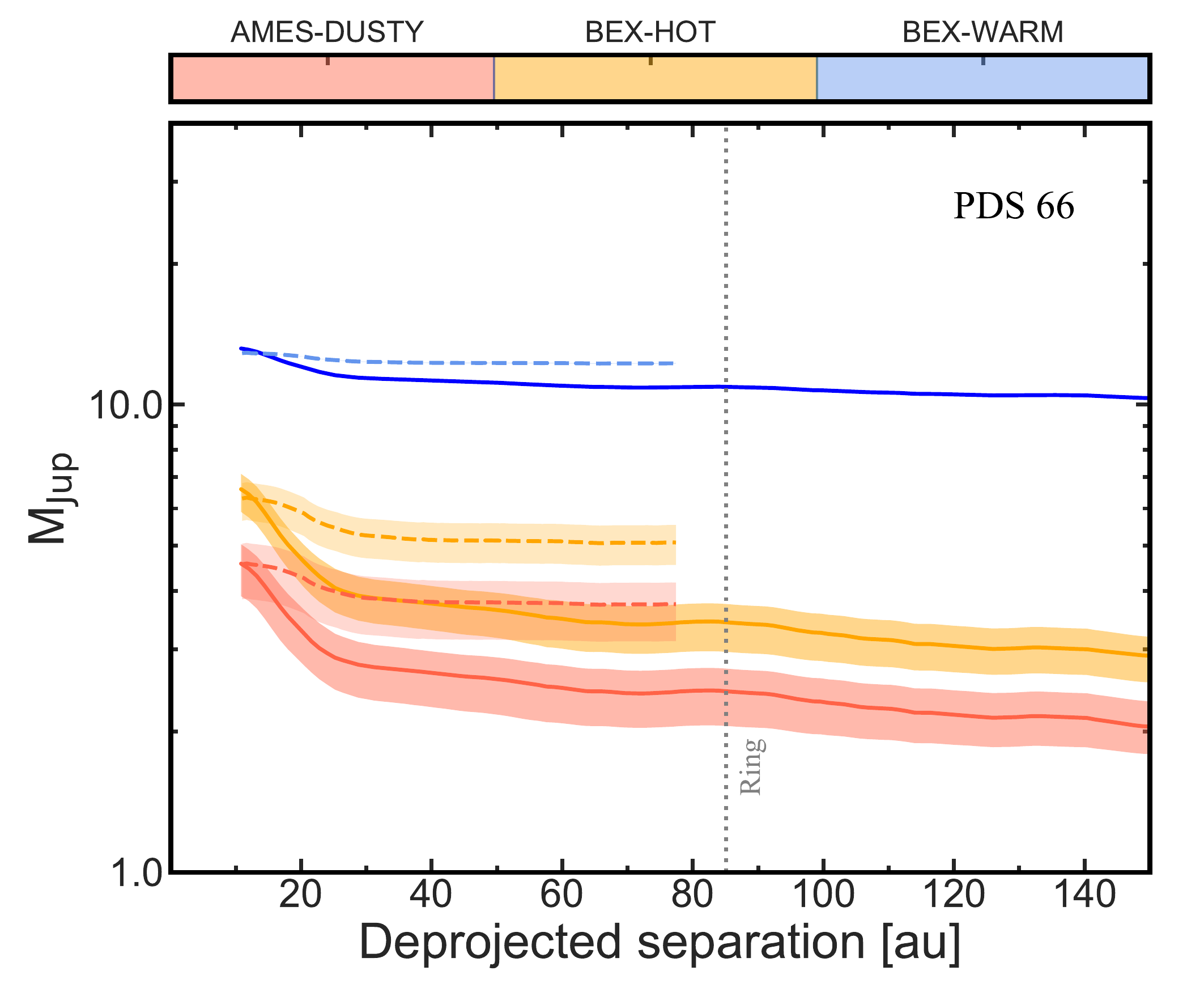}}
\caption{5$\sigma$ sensitivity to planet masses in PDS 66 (see Figure \ref{fig:LkCa_15}).}
\label{fig:PDS_66}
\end{figure}

\begin{figure}[H]
\centering
\setlength{\unitlength}{\textwidth}
\hbox{\hspace{-0.1cm}\includegraphics[width=0.49\textwidth]{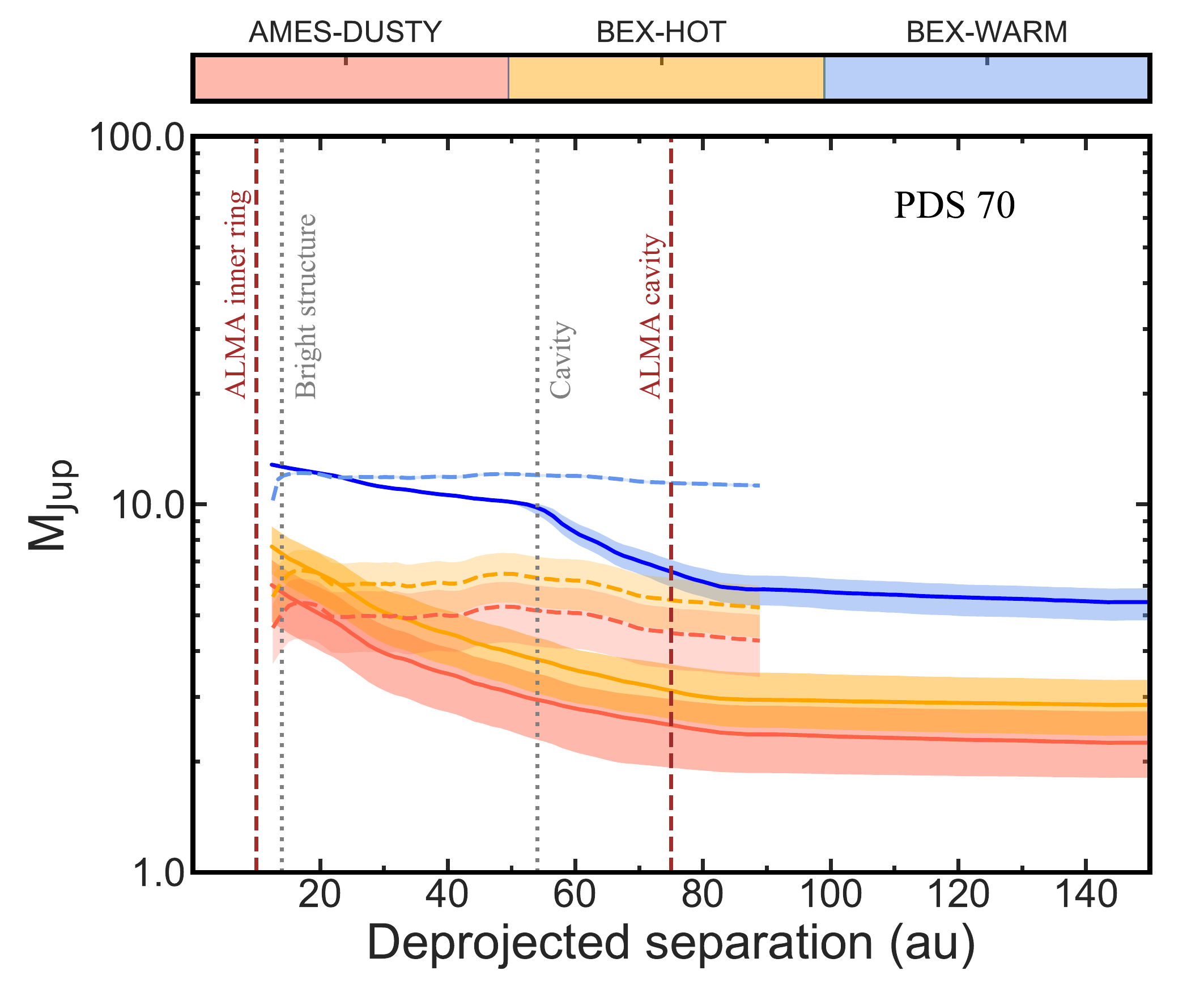}}
\caption{5$\sigma$ sensitivity to planet masses in PDS 70 (see Figure \ref{fig:LkCa_15}).}
\label{fig:PDS_70}
\end{figure}

\begin{figure}[H]
\centering
\setlength{\unitlength}{\textwidth}
\hbox{\hspace{-0.1cm}\includegraphics[width=0.49\textwidth]{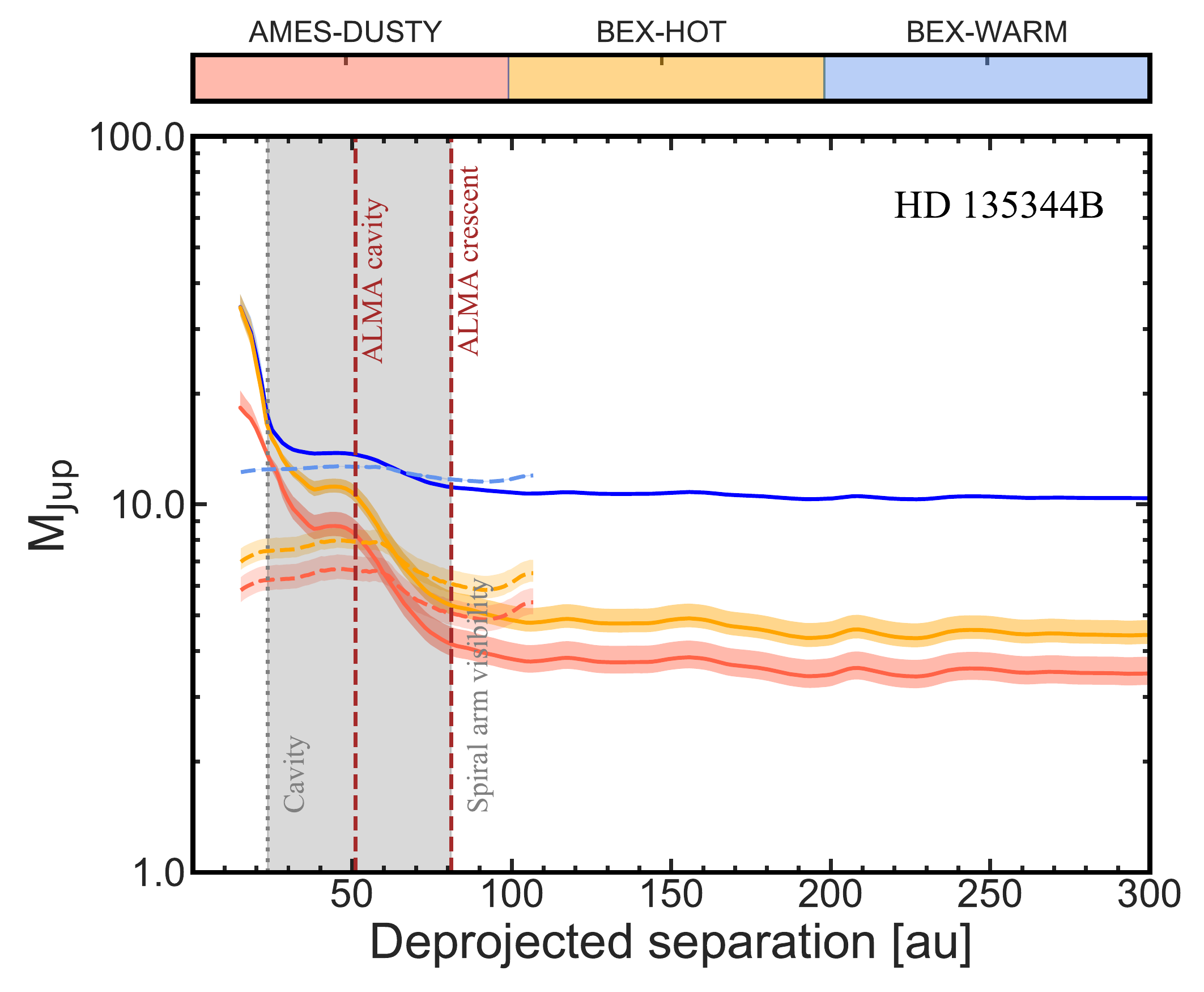}}
\caption{5$\sigma$ sensitivity to planet masses in PDS 66 (see Figure \ref{fig:HD_34282}). }
\label{fig:HD_135344B}
\end{figure}

\begin{figure}[H]
\centering
\setlength{\unitlength}{\textwidth}
\hbox{\hspace{-0.1cm}\includegraphics[width=0.49\textwidth]{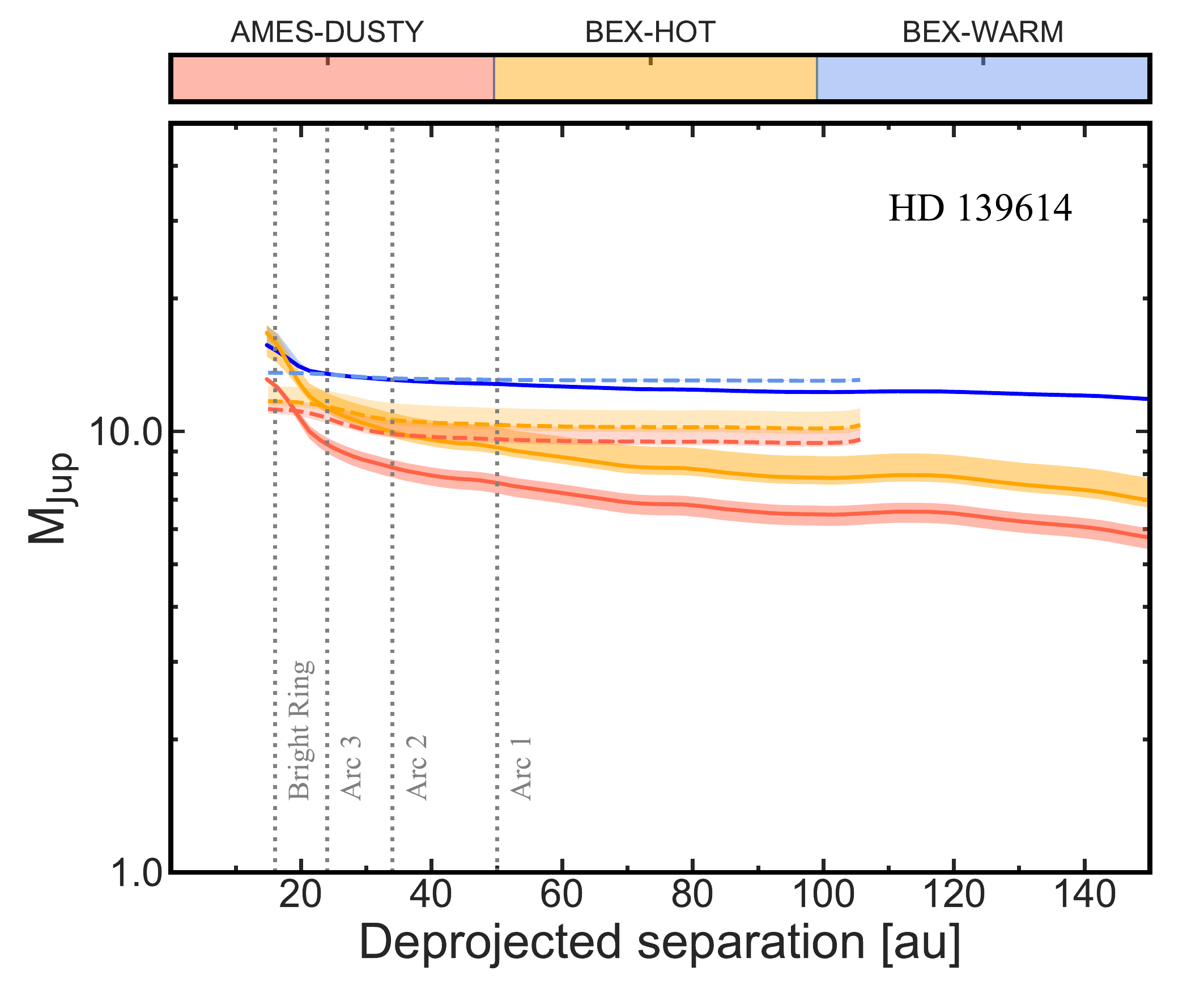}}
\caption{5$\sigma$ sensitivity to planet masses in HD 139614 (see Figure \ref{fig:LkCa_15}).}
\label{fig:HD_139614}
\end{figure}

\begin{figure}[H]
\centering
\setlength{\unitlength}{\textwidth}
\hbox{\hspace{-0.1cm}\includegraphics[width=0.49\textwidth]{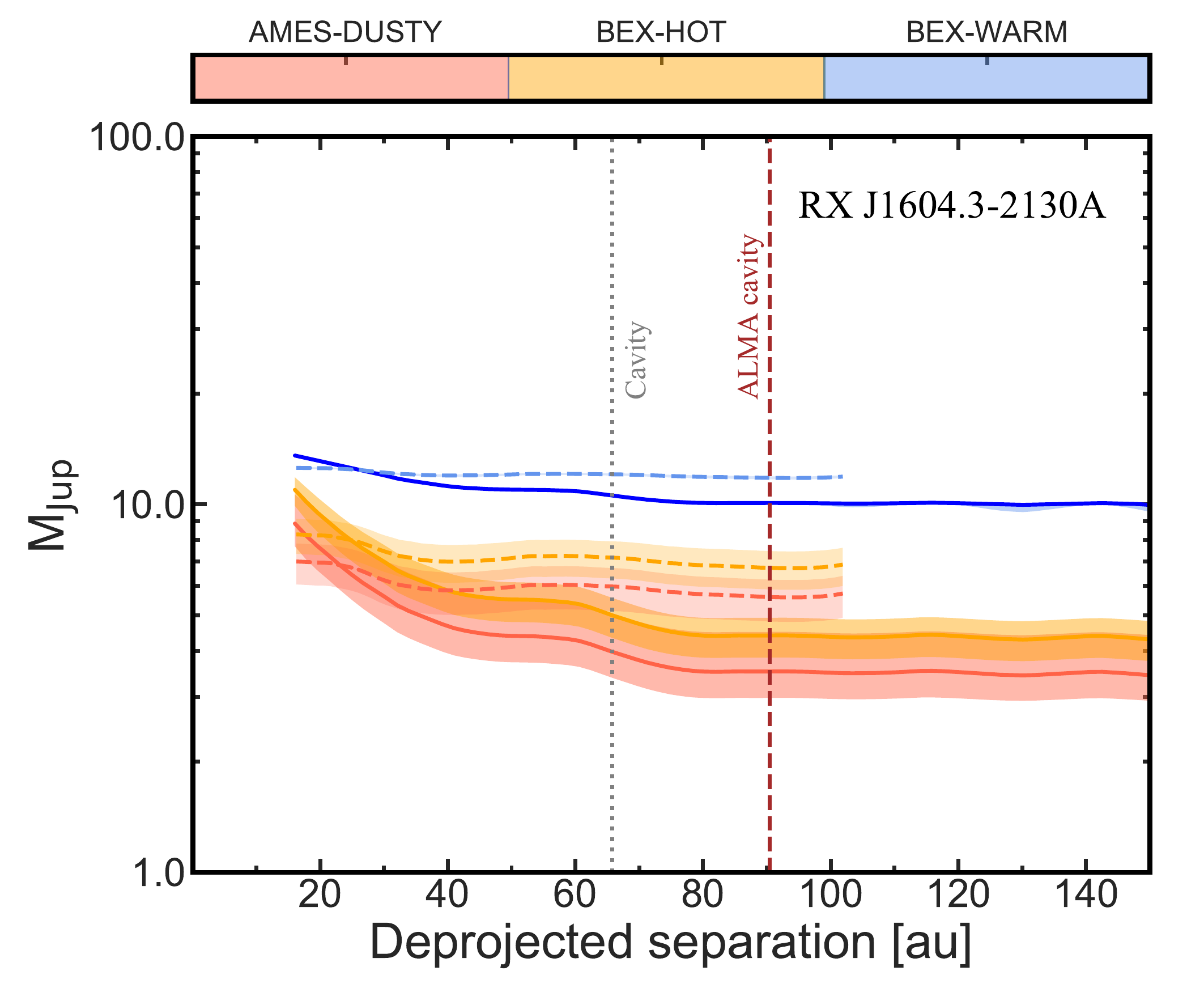}}
\caption{5$\sigma$ sensitivity to planet masses in RXJ1604 (see Figure \ref{fig:LkCa_15}).}
\label{fig:RX_J1604}
\end{figure}

\begin{figure}[H]
\centering
\setlength{\unitlength}{\textwidth}
\hbox{\hspace{-0.1cm}\includegraphics[width=0.49\textwidth]{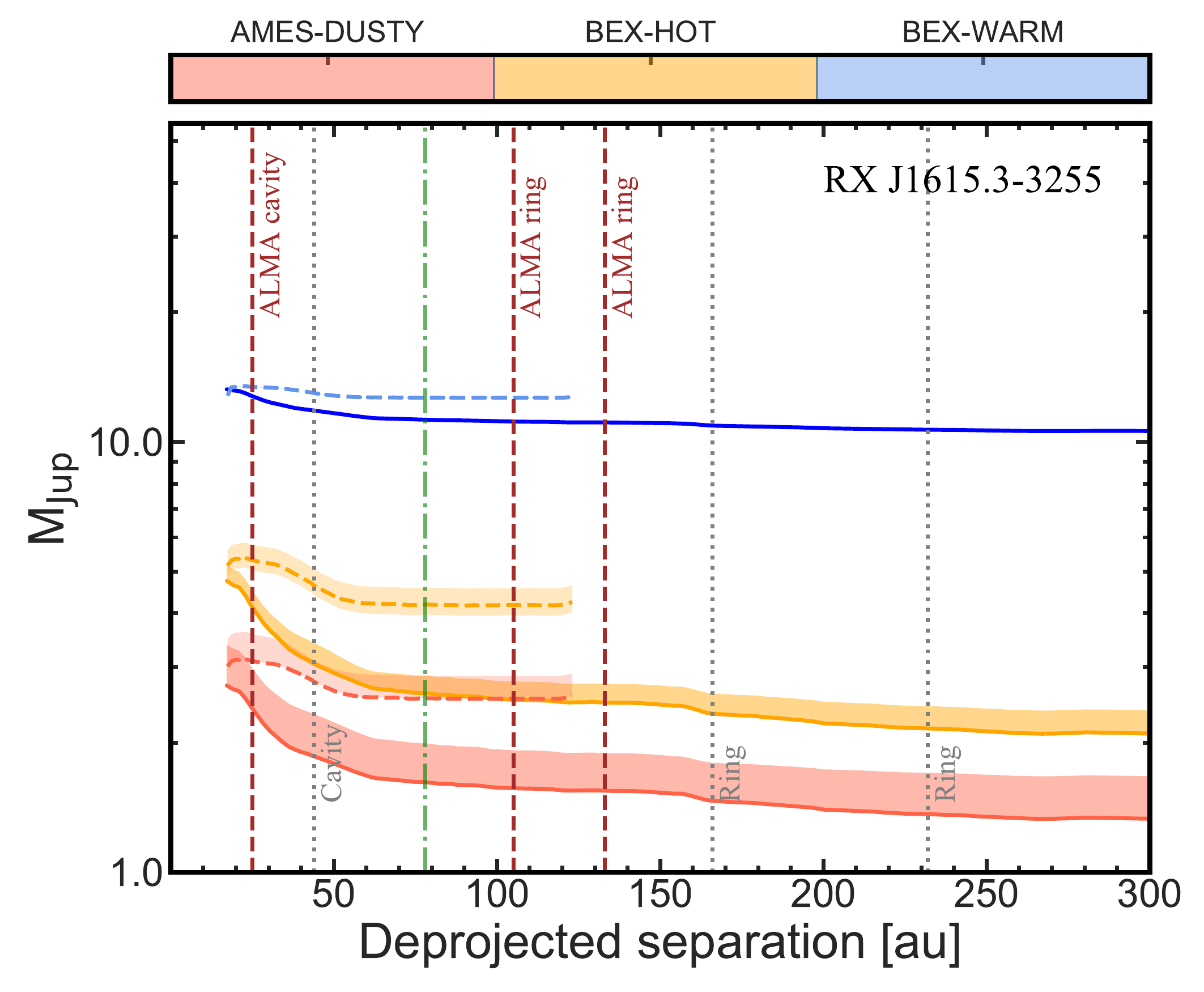}}
\caption{5$\sigma$ sensitivity to planet masses in RXJ1615 (see Figure \ref{fig:LkCa_15}). As the minimum age for this system is below the AMES-DUSTY lower limit, the minimum mass estimate for this model is not included here.  }
\label{fig:RX_J1615}
\end{figure}

\begin{figure}[H]
\centering
\setlength{\unitlength}{\textwidth}
\hbox{\hspace{-0.1cm}\includegraphics[width=0.49\textwidth]{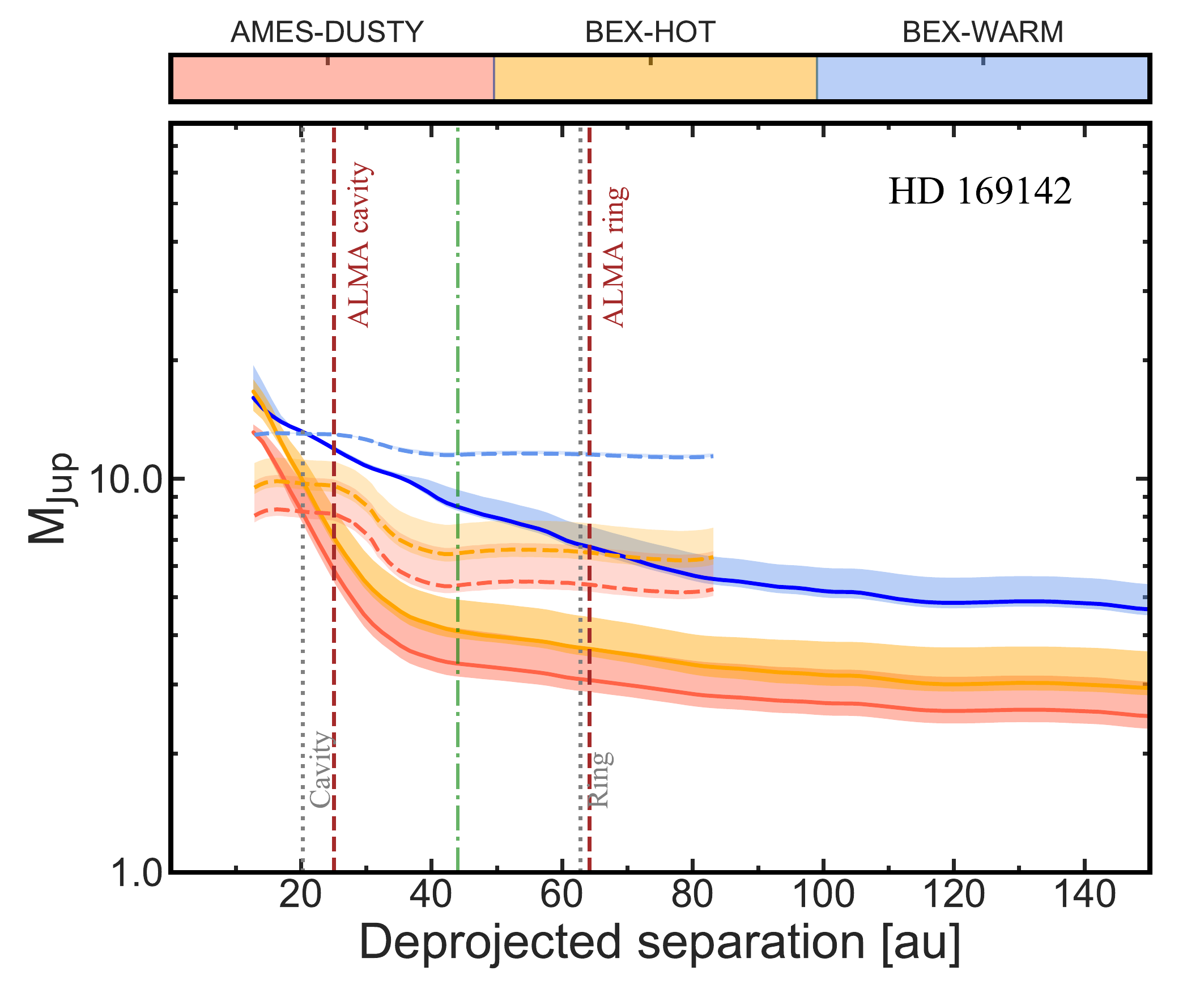}}
\caption{5$\sigma$ sensitivity to planet masses in HD 169142 (see Figure \ref{fig:LkCa_15}).}
\label{fig:HD_169142}
\end{figure}

\section{Detection probability maps}
\label{appendix_C}

Here we present the detection probability maps to low-mass companions in the PPDs present in our sample (see Table \ref{tab:host_param}). These maps represent the companion mass-semimajor axis parameter space of the SPHERE pupil-tracking data. To obtain them, we first converted the projected 2D limiting magnitude maps from the reduced IRDIFS data to planet masses, using the AMES-DUSTY \citep{chabrier2000}, BEX-HOT and BEX-WARM \citep{marleau2019} models. We then ran the MESS code, a Monte Carlo tool for the predictions of exoplanet search results \citep{bonavita2012} for each PPD system, which calculates the probability of detecting a planet of a given mass at a given separation. This is achieved  by the injection of a set of test companions on different orbits, which parameters are drawn from probability distributions. The resulting projected separation of each companion is then calculated and compared to the mass sensitivity of our data at the same location, determining whether or not the planet would have been found by SPHERE. For each target we generated a uniform grid of mass and semi-major axis in the interval [0.5, 200]\,$M\rm_{Jup}$ and [1, 1000]\,au with a sampling of 0.5\,$M\rm_{Jup}$ and 0.5\,au, respectively. We then generated $10^{4}$ orbits for each point in the grid, randomly oriented in space from uniform distributions in  $\omega$, $e$ and $M$, corresponding to the argument of periastron with respect to the line of nodes, eccentricity, and mean anomaly, respectively,  while the inclination and position angles were fixed to be in the disk plane (Table \ref{tab:adi}).

\setlength{\unitlength}{\textwidth}
\begin{figure*}
\includegraphics[width=1\textwidth]{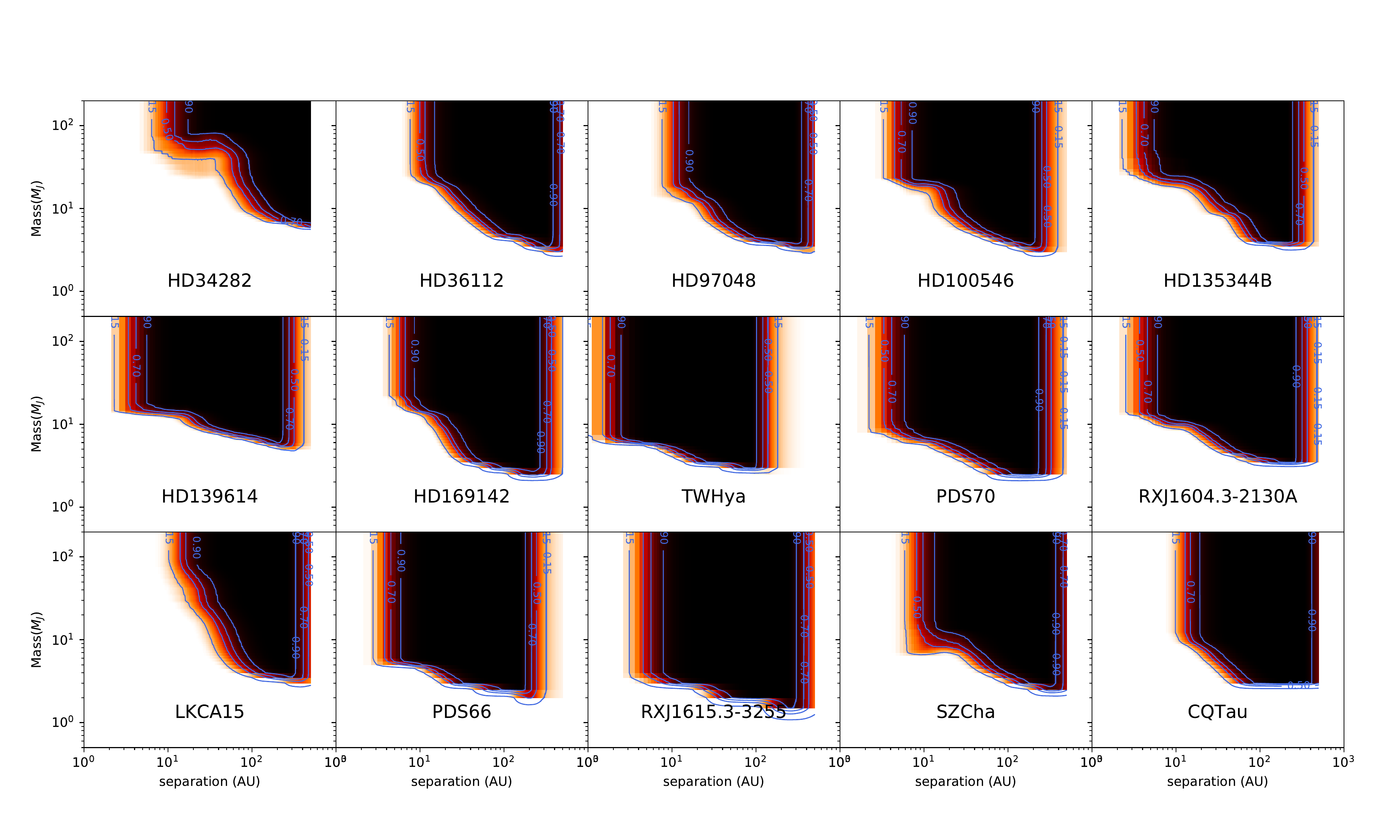}
\caption{SPHERE/IRDIS detection probability maps for the sample of PPD systems considered in this work. The nominal ages of Table \ref{tab:host_param} and AMES-DUSTY conditions have been assumed to convert ANDROMEDA contrast limits to companion masses. The blue curves circumscribe the probability of detecting a planet of a given mass-semimajor axis space. }
\label{fig:det_map_IRD_Dusty}
\end{figure*}

\setlength{\unitlength}{\textwidth}
\begin{figure*}
\includegraphics[width=1\textwidth]{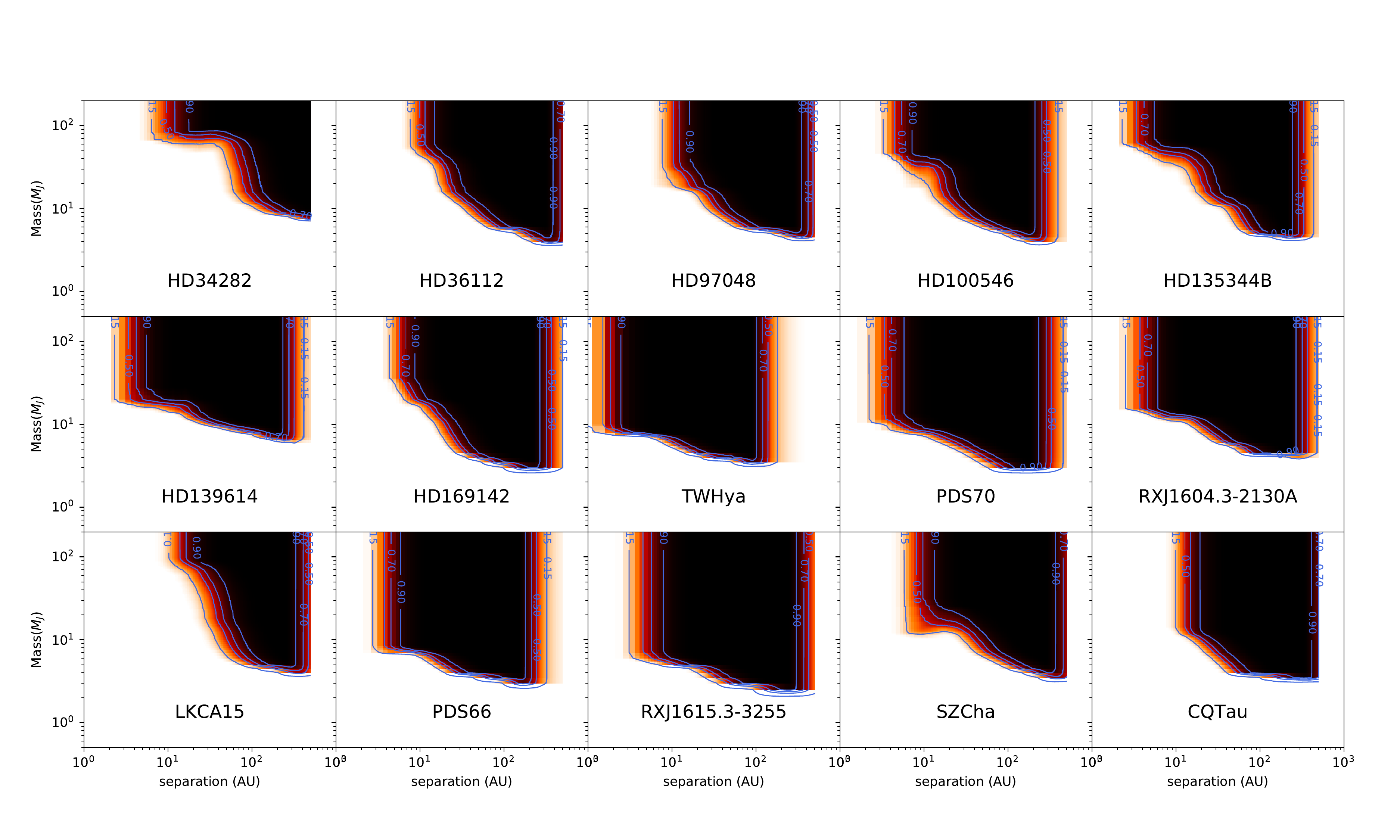}
\caption{Same as \ref{fig:det_map_IRD_Dusty} but for BEX-HOT conditions.}
\label{fig:det_map_IRD_Hot}
\end{figure*}

\setlength{\unitlength}{\textwidth}
\begin{figure*}
\includegraphics[width=1\textwidth]{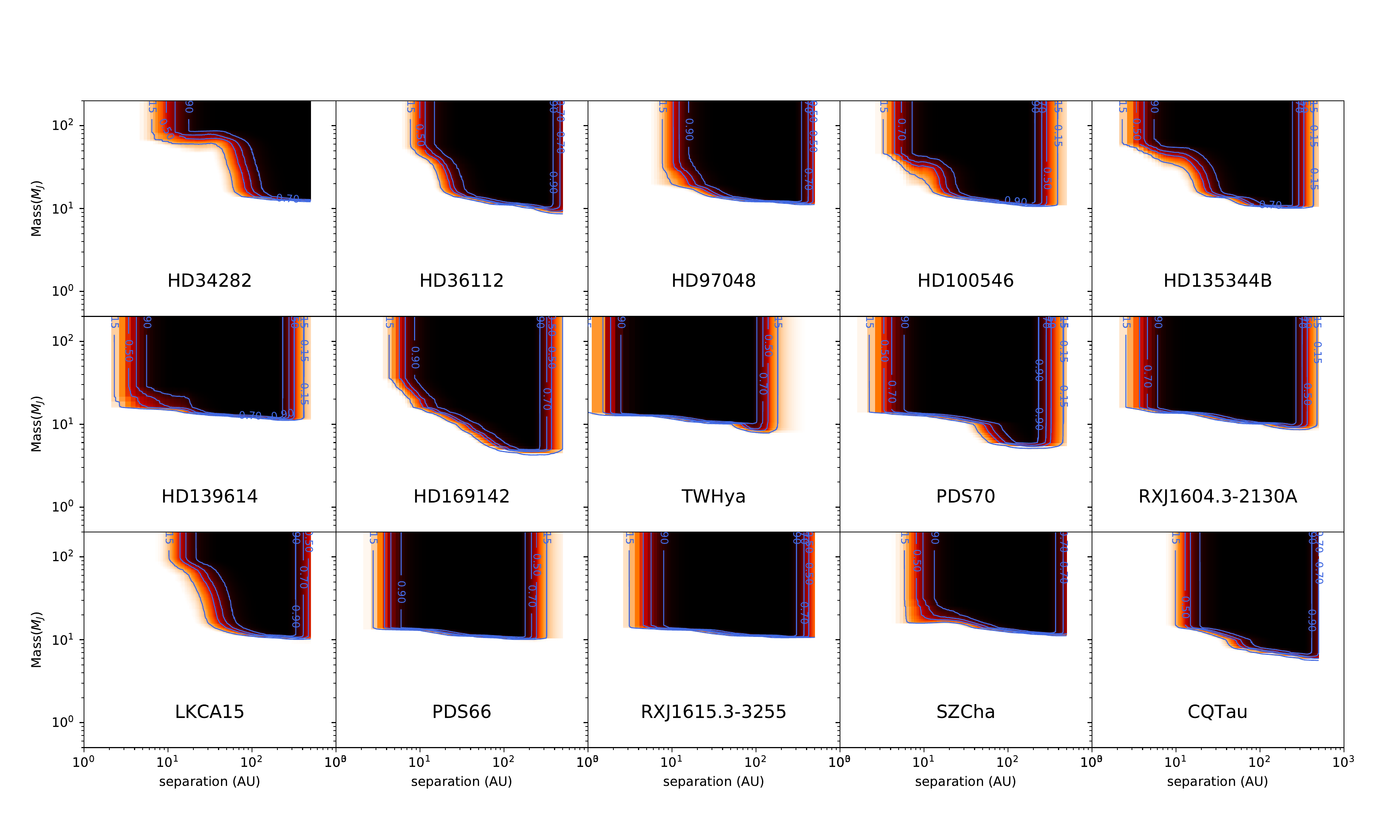}
\caption{Same as \ref{fig:det_map_IRD_Dusty} but for BEX-WARM conditions.}
\label{fig:det_map_IRD_Warm}
\end{figure*}

\setlength{\unitlength}{\textwidth}
\begin{figure*}
\includegraphics[width=1\textwidth]{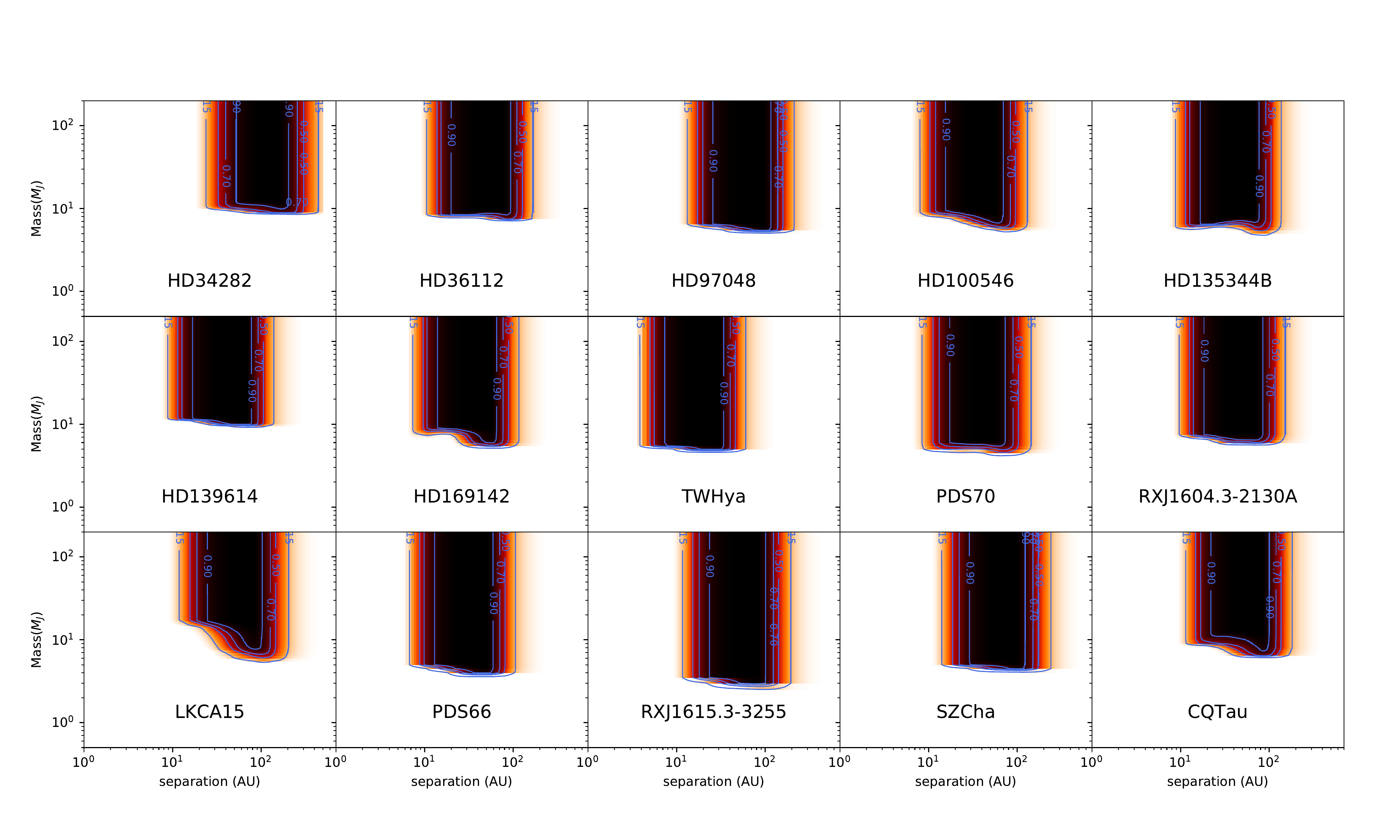}
\caption{SPHERE/IFS detection probability maps for the sample of PPD systems considered in this work. The nominal ages of Table \ref{tab:host_param} and AMES-DUSTY conditions have been assumed to convert the ASDI contrast limits to companion masses. The blue curves circumscribe the probability of detecting a planet of a given mass-semimajor axis space. }
\label{fig:det_map_IFS_Dusty}
\end{figure*}

\setlength{\unitlength}{\textwidth}
\begin{figure*}
\includegraphics[width=1\textwidth]{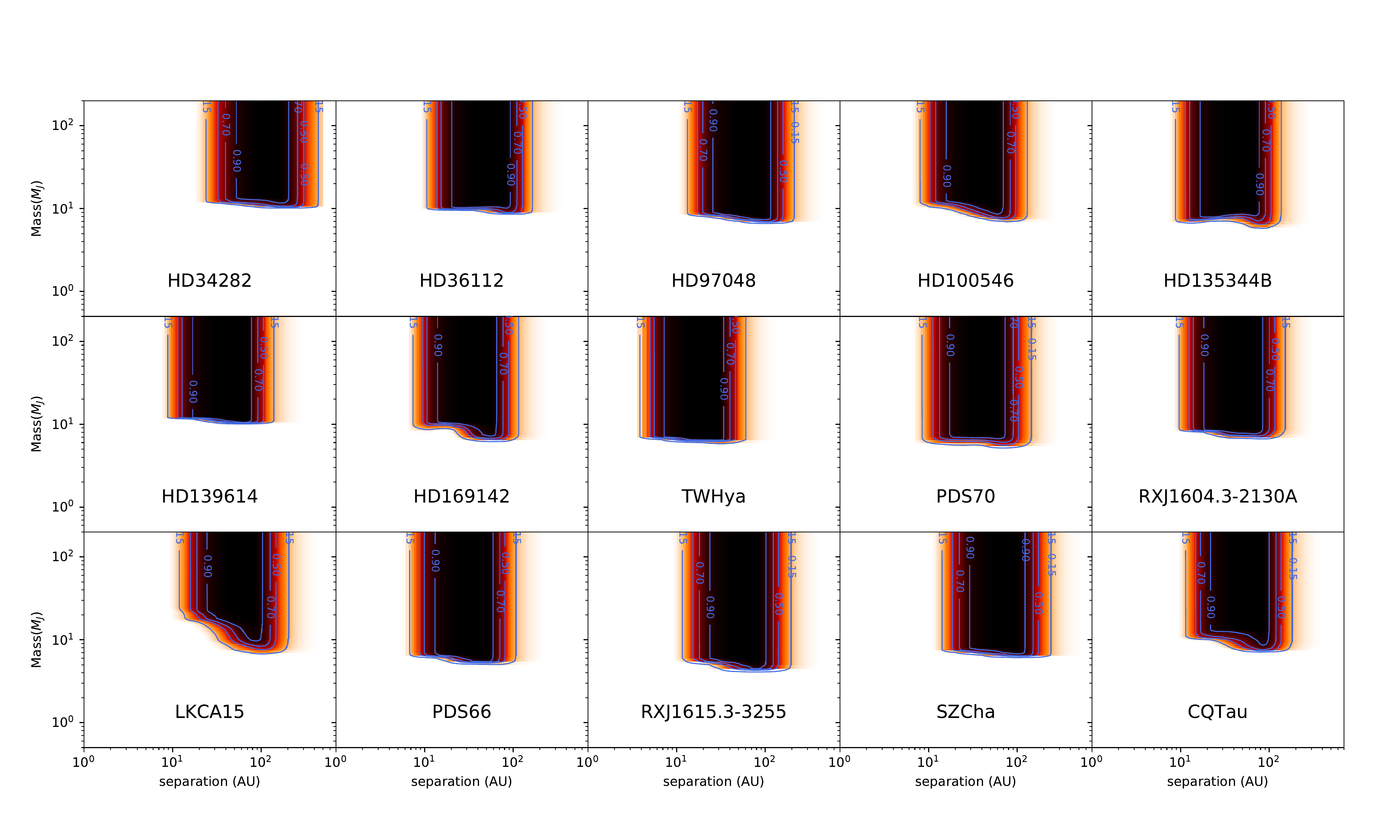}
\caption{Same as \ref{fig:det_map_IFS_Dusty} but for BEX-HOT conditions.}
\label{fig:det_map_IFS_Hot}
\end{figure*}

\setlength{\unitlength}{\textwidth}
\begin{figure*}
\includegraphics[width=1\textwidth]{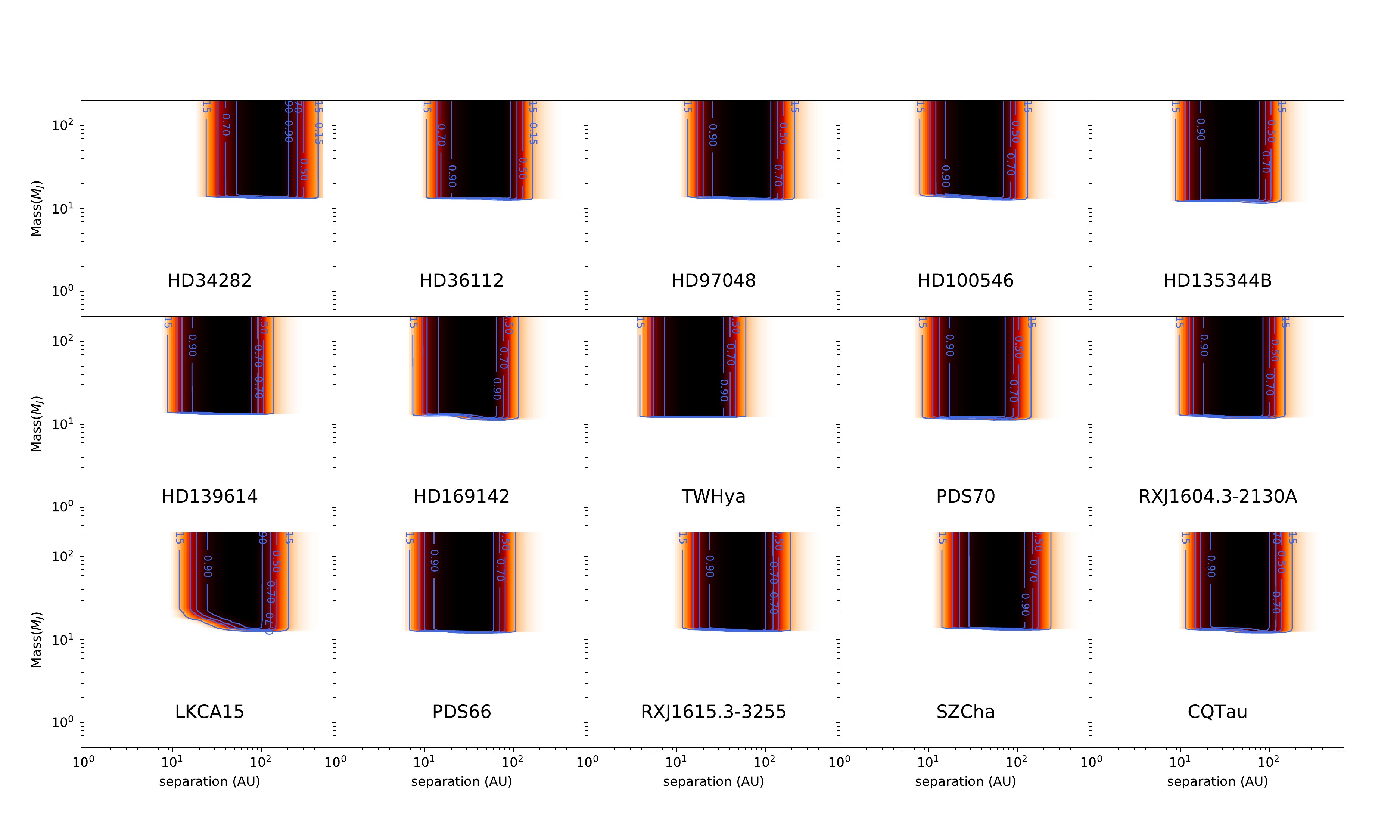}
\caption{ Same as \ref{fig:det_map_IFS_Dusty} but for BEX-WARM conditions.}
\label{fig:det_map_IFS_Warm}
\end{figure*}

\end{appendix}

\end{document}